\documentclass[reprint, superscriptaddress, nofootinbib, amsmath, amssymb, aps, onecolumn, 12pt,longbibliography, floatfix, prd]{revtex4-2}

\usepackage[hmargin=.7in,vmargin=1.1in]{geometry}
\usepackage{framed}

\usepackage{fancyhdr} 
\usepackage{setspace}
\usepackage[medium]{titlesec}

\usepackage[normalem]{ulem}
\usepackage{isodateo}
\usepackage{indentfirst}

\usepackage{enumitem}
\usepackage{graphicx}
\usepackage{dcolumn}
\usepackage{bm}
\usepackage{hyperref}
\usepackage{comment}
\usepackage{slashed}
\usepackage{xspace}
\usepackage{multirow}
\usepackage{xcolor}
\usepackage{bookmark}

\newcommand{\figref}[1]{Fig.~\ref{fig:#1}}
\newcommand{\CP}{\ensuremath{{\cal CP}}\xspace}
\newcommand{\gev}{\;\text{GeV}\xspace}

\begin{document}

\title{Tripling down on the \texorpdfstring{$W$}{W} boson mass}

\author{Henning Bahl}
\email{hbahl@uchicago.edu}
\affiliation{Department of Physics, University of Chicago, Chicago, IL 60637, USA}
\affiliation{Enrico Fermi Institute, University of Chicago, Chicago, IL 60637, USA}
 
\author{Wen Han Chiu}
\email{wenhan@uchicago.edu}
\affiliation{Department of Physics, University of Chicago, Chicago, IL 60637, USA}
\affiliation{Enrico Fermi Institute, University of Chicago, Chicago, IL 60637, USA}

\author{Christina~Gao}
\email{yanggao@fnal.gov}
\affiliation{
Theoretical Physics Department, Fermi National Accelerator Laboratory, Batavia, IL, 60510, USA
}%
\affiliation{University of Illinois at Urbana-Champaign, Urbana, IL 61801, USA}

\author{Lian-Tao Wang}
\email{liantaow@uchicago.edu}
\affiliation{Department of Physics, University of Chicago, Chicago, IL 60637, USA}
\affiliation{Enrico Fermi Institute, University of Chicago, Chicago, IL 60637, USA}
\affiliation{Kavli Institute for Cosmological Physics, University of Chicago, Chicago, IL 60637, USA}

\author{Yi-Ming Zhong}
\email{ymzhong@kicp.uchicago.edu}
\affiliation{Kavli Institute for Cosmological Physics, University of Chicago, Chicago, IL 60637, USA}

\date{\today}

\begin{abstract}

A new precision measurement of the $W$ boson mass has been announced by the CDF collaboration, which strongly deviates from the Standard Model prediction. In this article, we study the implications of this measurement on the parameter space of the $SU(2)_L$ triplet extension (with hypercharge $Y=1$) of the Standard Model Higgs sector, focusing on a limit where the new triplet is approximate $\mathbb{Z}_2$-odd while the SM is $\mathbb{Z}_2$-even. We study the compatibility of the triplet spectrum preferred by the $W$ boson mass measured by the CDF-II experiment with other electroweak precision observables and Higgs precision data. We comprehensively consider the signals of new Higgs states at the LHC and highlighted the promising search channels. In addition, we also investigate the cosmological implications of the case in which the lightest new Higgs particle is either late decaying or cosmologically stable. 

\end{abstract}

\maketitle


\section{Introduction}

Recently, the CDF-II experiment~\cite{CDF:2022hxs} measured the $W$ boson mass to be
\begin{equation}
m_{W,\text{CDF-II}} = 80.4335 \pm 0.0094~\text{GeV}.
\end{equation}
This suggests a 7$\sigma$ derivation from the Standard Model (SM) prediction~\cite{ParticleDataGroup:2020ssz},
\begin{equation}
m_{W,\text{SM}} = 80.357 \pm 0.006~\text{GeV}.   
\end{equation} 
The CDF-II measurement of $m_W$ is also in tension with the measurements from the previous collider experiments at $\sim 2.6\sigma$~\cite{ParticleDataGroup:2020ssz,Dorigo}. The discrepancy might be due to some unknown experimental systematical uncertainties, but it could also be a hint for new physics~\cite{Lu:2022bgw,DiLuzio:2022xns,Song:2022xts,Sakurai:2022hwh,Cheng:2022jyi,Bahl:2022xzi,Heo:2022dey,Biekotter:2022abc,Du:2022brr,Han:2022juu,Ahn:2022xeq,FileviezPerez:2022lxp,Ghoshal:2022vzo,Kanemura:2022ahw,Popov:2022ldh,Arcadi:2022dmt,Ghorbani:2022vtv,Lee:2022gyf,Heeck:2022fvl,Abouabid:2022lpg,Benbrik:2022dja,Kim:2022hvh,Atkinson:2022qnl,Strumia:2022qkt,deBlas:2022hdk,Yang:2022gvz,Yuan:2022cpw,Athron:2022qpo,Fan:2022dck,Babu:2022pdn,Heckman:2022the,Gu:2022htv,Athron:2022isz,Asadi:2022xiy,Paul:2022dds,Bagnaschi:2022whn,Lee:2022nqz,Liu:2022jdq,Fan:2022yly,Balkin:2022glu,Endo:2022kiw,Crivellin:2022fdf,Han:2022juu,Blennow:2022yfm,Cacciapaglia:2022xih,Tang:2022pxh,Zhu:2022tpr,Zheng:2022irz,Krasnikov:2022xsi,Arias-Aragon:2022ats,Du:2022pbp,Kawamura:2022uft,Nagao:2022oin,Zhang:2022nnh,Carpenter:2022oyg,Senjanovic:2022zwy,Chowdhury:2022moc,Borah:2022obi,Zeng:2022lkk,Du:2022fqv,Bhaskar:2022vgk,Baek:2022agi,Cao:2022mif,Borah:2022zim,Batra:2022org,Almeida:2022lcs,Cheng:2022aau,Batra:2022pej,Benbrik:2022dja,Cai:2022cti,Zhou:2022cql,Gupta:2022lrt,Wang:2022dte,Barman:2022qix,Kim:2022xuo,Dcruz:2022dao,Isaacson:2022rts,Chowdhury:2022dps,Kim:2022zhj,Gao:2022wxk,Lazarides:2022spe,Rizzo:2022jti,VanLoi:2022eir,YaserAyazi:2022tbn,Chakrabarty:2022voz,CentellesChulia:2022vpz,Nagao:2022dgl}. A class of new physics solutions contain extensions to the Standard Model (SM) Higgs sector, whereby the new Higgs states provide additional sources of custodial symmetry breaking~\cite{Lu:2022bgw,DiLuzio:2022xns,Song:2022xts,Sakurai:2022hwh,Cheng:2022jyi,Bahl:2022xzi,Heo:2022dey,Biekotter:2022abc,Du:2022brr,Han:2022juu,Ahn:2022xeq,FileviezPerez:2022lxp,Ghoshal:2022vzo,Kanemura:2022ahw,Popov:2022ldh,Arcadi:2022dmt,Ghorbani:2022vtv,Lee:2022gyf,Heeck:2022fvl,Abouabid:2022lpg,Benbrik:2022dja,Kim:2022hvh,Atkinson:2022qnl}. In a particular class of models, the correction to the $W$ mass from new physics enters at the one-loop level. The new physics scale is then predicted to be around a few hundreds of GeV. This is particularly interesting since it could give rise to signals in LHC new physics searches. The new physics in this class are generically in some $SU(2)_L$ multiplet. The correction to the $W$ mass requires that the masses of the different members of the multiplet receive different custodial symmetry breaking contributions from the electroweak symmetry breaking. Hence, some of the couplings between the new physics and the Higgs need to be sizable, and there need to be significant mass splittings within the multiplet. Both of these features have interesting phenomenological consequences. 

In this paper, we explore the phenomenology of the Higgs Triplet Model (HTM), with hypercharge $Y=1$, in the context of electroweak precision measurements, direct collider searches, and Higgs precision measurement in light of the CDF-II $W$ mass measurement. The prediction of this model for the $W$ mass has been investigated in Ref.~\cite{Kanemura:2022ahw}. We go beyond the existing works by investigating the compatibility of the $m_{W,{\rm CDF-II}}$ preferred triplet spectra with the measurements of the effective weak mixing angle and Higgs precision data as well as by providing a comprehensive analysis of possible signatures at the Large Hadron Collider (LHC). Furthermore, we explore the situation that the new Higgs triplet is approximately inert. This can be achieved naturally by imposing an approximate $\mathbb{Z}_2$ symmetry, which can be broken softly. In this case, its lightest neutral states can be candidates for a fraction of stable dark matter or decaying dark matter. We explore the CDF-II measurement's impact on those dark matter candidates.  

The paper is organized as follows: in Section~\ref{sec:HTM}, we briefly review the Higgs Triplet Model; in Section~\ref{sec:oneloop}, we calculate the HTM's correction to the $W$ mass at the one loop and give the preferred mass spectra for the new Higgses from the CDF-II measurement. We explore the phenomenology of this spectra in various aspects, including their contributions to the effective weak mixing angle in Section~\ref{sec:weakangle}, the compatibility with the Higgs precision measurement in Section~\ref{sec:precisionhiggs}, the bounds and discovery channels from the LHC direct searches in Section~\ref{sec:direct_searches}, and their cosmological implications in Section~\ref{sec:darkmatter}. We conclude in Section~\ref{sec:summary}. In the appendices, we give details of the self-energy corrections and the SM fitting formula and discuss the soft $\mathbb{Z}_2$ breaking limit, the unitarity and vacuum stability bounds, the Landau pole, and the decoupling limit of the HTM. 


\section{Higgs Triplet Model}
\label{sec:HTM}
In the HTM, the Higgs sector contains an isospin doublet $\Phi$ with hypercharge $Y=\frac12$ and an isospin triplet $\Delta$ with $Y=1$.\footnote{
Alternatively, one can also consider a adding a $Y=0$ triplet to the SM. In this model, $M_W$ receives a positive tree-level shift allowing to easily fit the CDF-II anomaly (see e.g.\ Refs.~\cite{Strumia:2022qkt,FileviezPerez:2022lxp}).}
They can be parameterized as
\begin{equation}
\begin{split}
    \Phi= \left(\begin{array}{c}
         G^+ \\
         \frac {v_{\phi}+h+iG^0}{\sqrt2}
    \end{array}
    \right)~,~
    \Delta={}& \left(\begin{array}{c c}
        \frac{H^+}{\sqrt2}&H^{++} \\
        \Delta^0&-\frac{H^+}{\sqrt2}
    \end{array}
    \right)~~{\rm with}~\Delta^0=\frac{v_\Delta+H+iA}{\sqrt 2},
    \end{split}
\end{equation}
where $v_\phi$ and $v_\Delta$ are the vacuum expectation values (vev's) of the doublet and triplet field obeying 
\begin{equation}
v^2\equiv v_\phi^2+2v_\Delta^2\approx (246\,{\rm GeV})^2.
\end{equation}
In addition to the SM-like Higgs boson, the scalar sector contains six new Higgs bosons (degrees of freedom): the \CP-even $H$ boson, the \CP-odd $A$ boson, the singly-charged $H^\pm$ bosons, and the doubly-charged $H^{\pm\pm}$ bosons.

In this model, the tree level $W$ and $Z$ boson masses are given by
\begin{equation}
    m_W^2=\frac{g^2}{4}v^2,~\quad m_Z^2=\frac{g^2}{4c_W^2}(v^2+2v_\Delta^2),
\end{equation}
where $c_W^2 \equiv \cos^2 \theta_W$ and $\theta_W$ is the weak mixing angle. If we take the $Z$ boson mass as an input, the expected $W$ boson mass is naively smaller than the SM prediction at the tree level
\begin{equation}
    m_W=m_{W,{\rm SM}}^\text{tree}\left(1-\frac{v_\Delta^2}{v^2}\right) + \Delta m_{W},
\end{equation}
where $\Delta m_{W}$ denotes loop corrections.

However, as we shall see below, a mass splitting between the new Higgs states can correct the $W$ mass at the loop level with an opposite sign compared to the tree level correction. 
To explain the CDF-II result, it is preferred that the 1-loop correction dominates over the tree level correction, i.e., $v_\Delta\ll v$. Assuming the difference between the CDF-II measurement and the SM prediction mainly comes from the loop correction $\Delta m_{W}$, i.e., 
\begin{equation}\label{eq:vDelta_estimate}
    \frac{v_\Delta^2}{v^2}\ll \frac{m_{W,{\rm CDF-II}}-m_{W,{\rm SM}}}{m_{W,{\rm SM}}}~,
\end{equation}
this restricts $v_\Delta \ll 7.6~ {\rm GeV}$. To be concrete, we assume that $v_\Delta < 1~{\rm GeV}$ in the rest of the paper.
For simplicity, we will work in the limit $v_\Delta=0$ for the calculation of the $W$ mass correction, effective weak mixing angle, the Higgs di-photon rate, and the trilinear Higgs coupling (see below). Note that deviations from this limit will be suppressed by powers of $v_\Delta^2/v^2 \lesssim 2\times10^{-5}$ and will be ignored.

\subsection{The Inert Triplet}

The limit of $v_\Delta=0$ can be realized in a strict sense by imposing a $\mathbb{Z}_2$ symmetry, under which $\Phi$ is $\mathbb{Z}_2$-even and $\Delta$ is $\mathbb{Z}_2$-odd. This $\mathbb{Z}_2$ can also be used to forbid the neutrino yukawa term typically seen in the Type-II seesaw model. The general gauge invariant potential is then given by 
\begin{equation}
\begin{split}
    V(\Phi,\Delta)={}&m^2\Phi^\dagger \Phi+M^2{\rm Tr}(\Delta^\dagger\Delta)\\
    &+\lambda_1\left(\Phi^\dagger \Phi\right)^2
    +\lambda_2\left[{\rm Tr}(\Delta^\dagger\Delta)\right]^2+\lambda_3{\rm Tr}\left[(\Delta^\dagger\Delta)^2\right]\\
    &+\lambda_4\left(\Phi^\dagger \Phi\right){\rm Tr}(\Delta^\dagger\Delta)
    +\lambda_5 \Phi^\dagger\Delta\Delta^\dagger \Phi,
    \end{split}
\end{equation}
where all the parameters in the potential can be taken to be real.
The minimization of the potential yields 
\begin{equation}
    m^2=-\lambda_1v^2,\quad v_{\Delta}=0.
\end{equation}
In terms of the physical states, the quadratic part of the Higgs potential is
\begin{equation}
\begin{split}
    V(\Phi,\Delta)\supset{}&\frac 12 (2\lambda_1 v^2) h^2 + \frac12 \left(M^2+\frac{\lambda_4 v^2}{2}\right)\left(A^2+H^2+ 2H^{++}H^{--}+ 2H^+H^-\right)\\
    &+\frac14  \lambda_5 v^2 \left(A^2+H^2+H^+H^-\right)~.
\end{split}
\end{equation}
Then, the mass spectrum is given by
\begin{equation}
\begin{split}
m_h^2&=2\lambda_1v^2,\\
    m_A^2={}&m_H^2=M^2+(\lambda_4+\lambda_5)\frac{v^2}2,\\
    m_{H^+}^2={}&M^2+\lambda_4\frac{v^2}2+\lambda_5\frac{v^2}4,\\
    m_{H^{++}}^2={}&M^2+\lambda_4\frac{v^2}2.
\end{split}
\end{equation}
We can substitute the Higgs potential parameters $m^2$, $M^2$, $\lambda_1$, $\lambda_5$ by $v$, $m_h^2$, $m_{A,H}^2=m_A^2=m_H^2$, $m_{H^+}^2$, and $m_{H^{++}}^2$, where $v= (\sqrt{2} G_F)^{-1/2}\simeq 246$~GeV and $m_h\simeq 125$~GeV. The free parameters in this model are thus given by
\begin{equation}
   m_{A,H},\quad m_{H^+},\quad m_{H^{++}},\quad \lambda_2,\quad \lambda_3,\quad \lambda_4
\end{equation}
with the condition that
\begin{equation}
    m_{H^{+}}^2 - m_{H^{++}}^2 = m_{A,H}^2 - m_{H^+}^2 = \frac{\lambda_5 v^2}{4}.
\end{equation}
I.e., the coupling $\lambda_5$ controls the splitting of the mass spectrum. For the rest of the discussion, the model with $\lambda_5 >0$ ($\lambda_5 <0$) will be referred to as the type-I (II) Higgs triplet model, which has a mass ordering of $m_{H^{++}} < m_{H^+} < m_{A,H}$ ($m_{A,H} < m_{H^+} < m_{H^{++}}$), respectively.


\section{One-loop corrected \texorpdfstring{$W$}{W} boson mass}
\label{sec:oneloop}

If $H^{++}, H^+$ and $H,A$ have sizable mass splittings, i.e., if $|\lambda_5|$ is large, the HTM provides additional sources of custodial symmetry breaking, therefore correcting the $W$ mass differently than the $Z$ mass. We summarize our results below. 
Note that we perform this calculation in the limit $v_\Delta=0$. Finite values for $v_\Delta$ compatible with the upper bound of Eq.~(\ref{eq:vDelta_estimate}) will only induce negligible small shifts of $m_W$. All necessary self-energy corrections are listed in App.~\ref{app:self_energies} (see also Ref.~\cite{Aoki:2012jj}).
Tab.~\ref{tab:input} lists all of the input parameters~\cite{ParticleDataGroup:2020ssz} used in the computation.

\begin{table}
    \centering
    \begin{tabular}{|c c c|}
    \hline
  $\alpha_\text{em}^{-1} =  137.035999084$,      & $m_Z =91.1876~\text{GeV}$, & $G_F = 1.166378\cdot 10^{-5}~\text{GeV}^{-2}$, \\
  $m_t = 172.76~\text{GeV}$, &
    $m_h = 125.09~\text{GeV}$,     & $\alpha_s(m_Z^2) = 0.1179$, \\ 
    $\Delta \alpha = 0.0591577$.& & \\
    \hline
    \end{tabular}
    \caption{Input parameters used in computation. Note that $\Delta \alpha$ is the sum of the hadronic contribution $ \Delta \alpha_\text{had}^{(5)} (m_Z^2) = 0.02766$ and the leptonic contribution $\Delta \alpha_\text{lept} =  0.031497687$~\cite{Steinhauser:1998rq}.}
    \label{tab:input}
\end{table}

To determine the $W$ mass from the measurement of the Fermi coupling constant $G_F$, we note that,
\begin{equation}
    G_F=\frac{\pi\alpha_{\text{em},0}}{\sqrt{2}m_{W,0}^2 s_{W,0}^2}\left(1+\frac{{\Pi_{WW}}(0)}{m_W^2}+\delta_{VB}\right),
\end{equation}
where terms with 0 subscripts are the bare parameters, $ \Pi_{WW}$ is the self-energy of the $W$, and $\delta_{VB}$ are the vertex and the box diagram corrections to the muon decay process. Rewriting this expression in terms of the physical parameters, one gets at the one-loop level that
\begin{equation}
   \begin{aligned}
    G_F=\frac{\pi\alpha_{\text{em}}}{\sqrt{2}m_{W}^2 s_{W}^2}\left(1+\frac{\delta\alpha_\text{em}}{\alpha_\text{em}}-\frac{\delta m_W^2}{m_W^2}-\frac{\delta s_W^2}{s_W^2}+\frac{{\Pi}_{WW}(0)}{m_W^2}+\delta_{VB}\right)
    \equiv\frac{\pi\alpha_{\text{em}}}{\sqrt{2}m_{W}^2 s_{W}^2}(1+\Delta r),
   \end{aligned}\label{eq:GF_corr}
\end{equation}
where $\delta\alpha_\text{em}$, $\delta m_W^2$, and $\delta s_W^2$ are the counterterms for the fine-structure constant $\alpha_\text{em}$, the $W$ mass, and weak mixing angle $s_W\equiv \sin \theta_W$, respectively. The counterterm of $s^2_W$ can be expressed in terms of the $W$ and $Z$ mass counterterms, which we define in the on-shell scheme, 
\begin{equation}
    \delta{s^2_W}=-\delta{c^2_W}=-c^2_W\left(\frac{\delta m_W^2}{m_W^2}-\frac{\delta m_Z^2}{m_Z^2}\right).
\end{equation}
The $\alpha_\text{em}$ counterterm, which we also define in the on-shell scheme, is given by
\begin{equation}
    \frac{\delta\alpha_\text{em}}{\alpha_\text{em}}=\Pi'_{\gamma\gamma}(0)+
     2\frac{{c}_W}{{s}_W}\frac{\Pi_{Z\gamma}^{\rm 1PI}(0)}{m_Z^2},
\end{equation}
where $\Pi'_{\gamma\gamma}(0)\equiv d \Pi_{\gamma\gamma}(p^2)/dp^2|_{p^2=0}$. Combining everything, $\Delta r$ is at the one-loop level given by
\begin{align}
    \Delta r ={}&\Pi'_{\gamma\gamma}(0)
    +\frac{\Pi^{\rm 1PI}_{WW}(0)-{\rm Re}\Pi^{\rm 1PI}_{WW}(m_W^2)}{m_W^2}+\frac{{c}_W^2}{{s}_W^2}\left(\frac{{\rm Re}\Pi^{\rm 1PI}_{WW}(m_W^2)}{m_W^2}-\frac{{\rm Re}\Pi^{\rm 1PI}_{ZZ}(m_Z^2)}{m_Z^2}
    \right)\nonumber\\
    &+2\frac{{c}_W}{{s}_W}\frac{\Pi_{Z\gamma}^{\rm 1PI}(0)}{m_Z^2}+\delta_{VB}.
 \label{eq:master}
\end{align}
The vertex and box diagram corrections to the muon decay, $\delta_{VB}$, are given by (see e.g.\ Ref.~\cite{Bohm:2001yx})
\begin{equation}
\delta_{VB} = \frac{\alpha_\text{em}}{4\pi {s}_W^2}\left(6+\frac{7c_W^2+3 {s}_W^2}{2s_W^2}\ln c_W^2\right)    ,
\end{equation}
where we neglected the contributions proportional to the electron and muon Yukawa couplings.

Based on Eq.~(\ref{eq:GF_corr}), we can then write
\begin{align}
   {m_W^2}=m_Z^2 \times  \left(
    \frac{1}{2}+\sqrt{\frac{1}{4}-\frac{\pi\alpha_{\rm em}}{\sqrt2G_Fm_Z^2}\left(1+\Delta r(m_{W})\right)}
    \right),
\label{eq:iter}
\end{align}
which we can iterate to solve for $m_W$.

For the numerical implementation, we follow the procedure outlined e.g.\ in \cite{Hessenberger:2018xzo}. We split $\Delta r$ into three parts: the one-loop SM contributions that depend on $m_W$ as an input ($\Delta r_{\text{SM}, W}$), the remaining one-loop and higher-order SM contributions ($\Delta r_{\text{SM, rest}}$), and the beyond-the-Standard-Model (BSM) contributions ($\Delta r_\text{BSM}$),
\begin{equation}
\Delta r =  \Delta r_{\text{SM}, W}(m_{W, \text{BSM}}) + \Delta r_{\text{SM, rest}}  + \Delta r_\text{BSM}.
\label{eq:key1}
\end{equation}
The quantity $\Delta r_\text{SM, rest}$ is given by
\begin{equation}
    \Delta r_{\text{SM, rest}} =\Delta r_\text{SM}- \Delta r_{\text{SM}, W}(m_{W, \text{SM}}),
    \label{eq:key2}
\end{equation}
where $m_{W,\text{SM}}$ is computed from the fitting formula given in Ref.~\cite{Awramik:2003rn} (see App.~\ref{sec:fitting}). The fitting formula can also be used to obtain a number for $\Delta r_\text{SM}$ (i.e., $\Delta r_\text{SM} \simeq 0.03807$).
Combining the Eqs.~(\ref{eq:key1}) and~(\ref{eq:key2}) together yields
\begin{align}
    \Delta r (m_{W, \text{BSM}}) = \Delta r_\text{SM}  - \Delta r_{\text{SM}, W}(m_{W, \text{SM}}) + \Delta r_{\text{SM}, W}(m_{W, \text{BSM}}) + \Delta r_\text{BSM},
   \label{eq:key3}
\end{align}
This equation consistently combines the full HTM one-loop corrections with the SM higher-order corrections, which are crucial for a precise result. 

\begin{figure*}[t]
    \centering
    \includegraphics[width=0.48\textwidth]{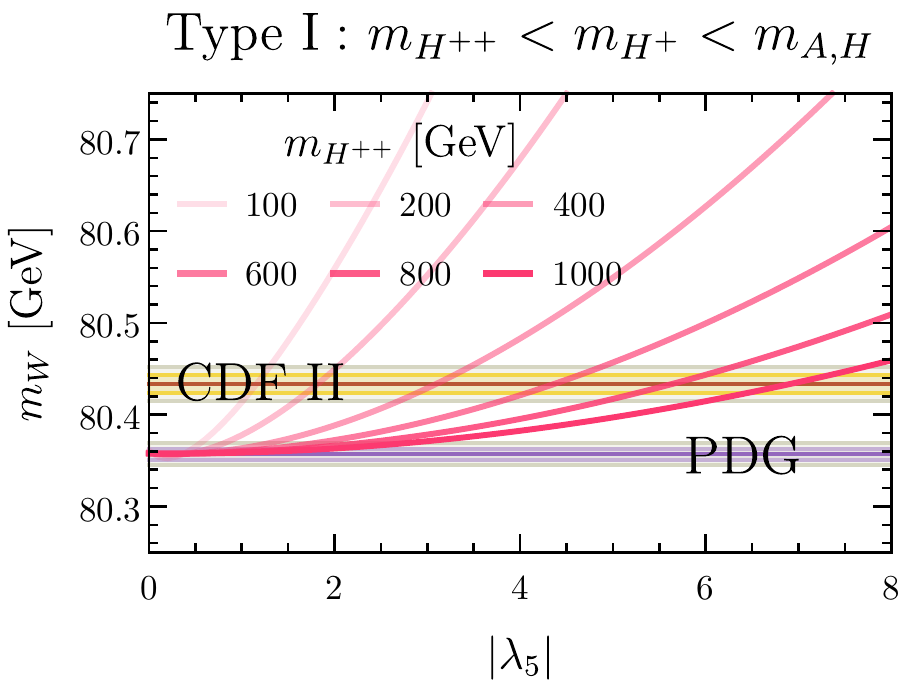}\quad\includegraphics[width=0.48\textwidth]{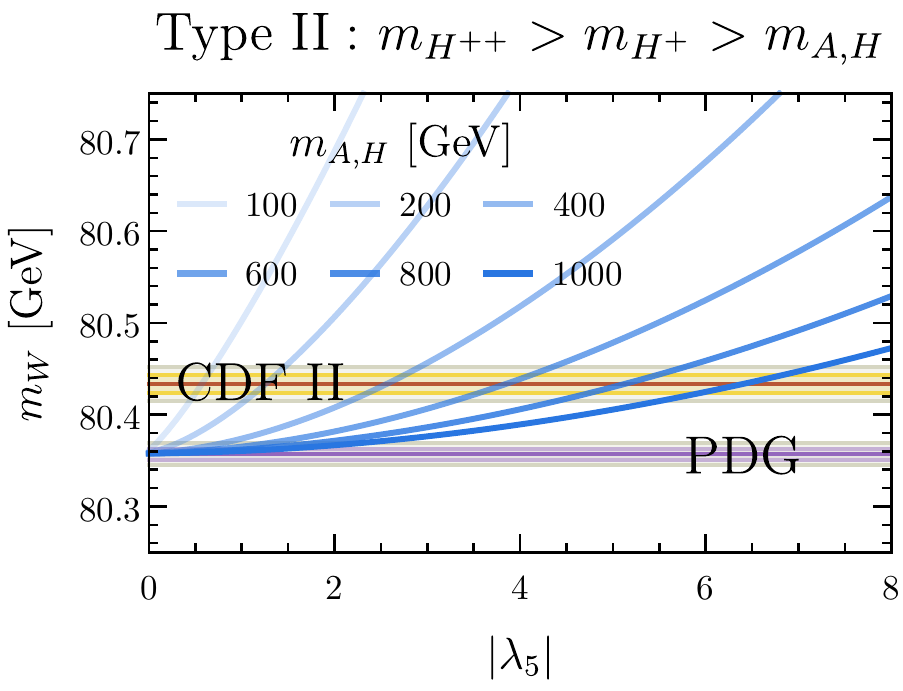}
    \caption{One-loop corrected $W$ boson mass $m_W$ as a function of  the coupling $|\lambda_5| = 4|m_{H^+}^2 - m_{H^{++}}^2|/v^2$ for various masses of the lightest state in the HTM. We assume the mass hierarchy of the new states following $m_{H^{++}}< m_{H^+}< m_A$ ($m_{H^{++}}> m_{H^+}> m_A$) in the left (right) panel. Different curves in each panel represent different masses for the lightest state. The brown (dark purple) line represent the CDF II measured (PDG) value and the yellow (purple)/gray band shows the $1\sigma$/$2\sigma$ intervals.}
    \label{fig:fig1}
\end{figure*}

In the left (right) panel of~\figref{fig1}, we show the resulting numerical value for $m_W$ as a function of $|\lambda_5|=\lambda_5$ ($|\lambda_5|=-\lambda_5$) and $m_\text{lightest}=m_{H^{++}}$ ($m_\text{lightest}=m_{A,H}$) for the type-I (II) HTM.\footnote{A plot showing $m_W$ as a function of the mass difference between the doubly- the singly-charged Higgs bosons alongside a discussion of the decoupling of the BSM states can be found in App.~\ref{app:decoupling}.} In both panels, we depict the CDF measured (PDG) value as a brown (dark purple) line, the 1$\sigma$ region as a yellow (purple) band, and the 2$\sigma$ region as gray bands. For a fixed value of the lightest BSM state $m_\text{lightest}$, the one-loop corrected $W$ boson mass increases with $|\lambda_5|$. For a fixed shift in the $W$ mass, a heavier $m_\text{lightest}$ requires a larger value of $|\lambda_5|$. With the same $m_\text{lightest}$, the type-I model needs a larger value of $|\lambda_5|$ to obtain the same $W$ mass shift compared to the type II.

The largish value of $|\lambda_5|$ required for large choices of $m_\text{lightest}$ could potentially cause the appearance of a Landau pole close to the electroweak scale. As we show in App.~\ref{app:landau}, no Landau pole appears below $\sim 10 \text{ TeV}$. 
This makes the additional contribution from the UV completion above the Landau pole subleading in comparison to those considered here.

\begin{figure*}[t]
    \centering
    \includegraphics[width=\textwidth]{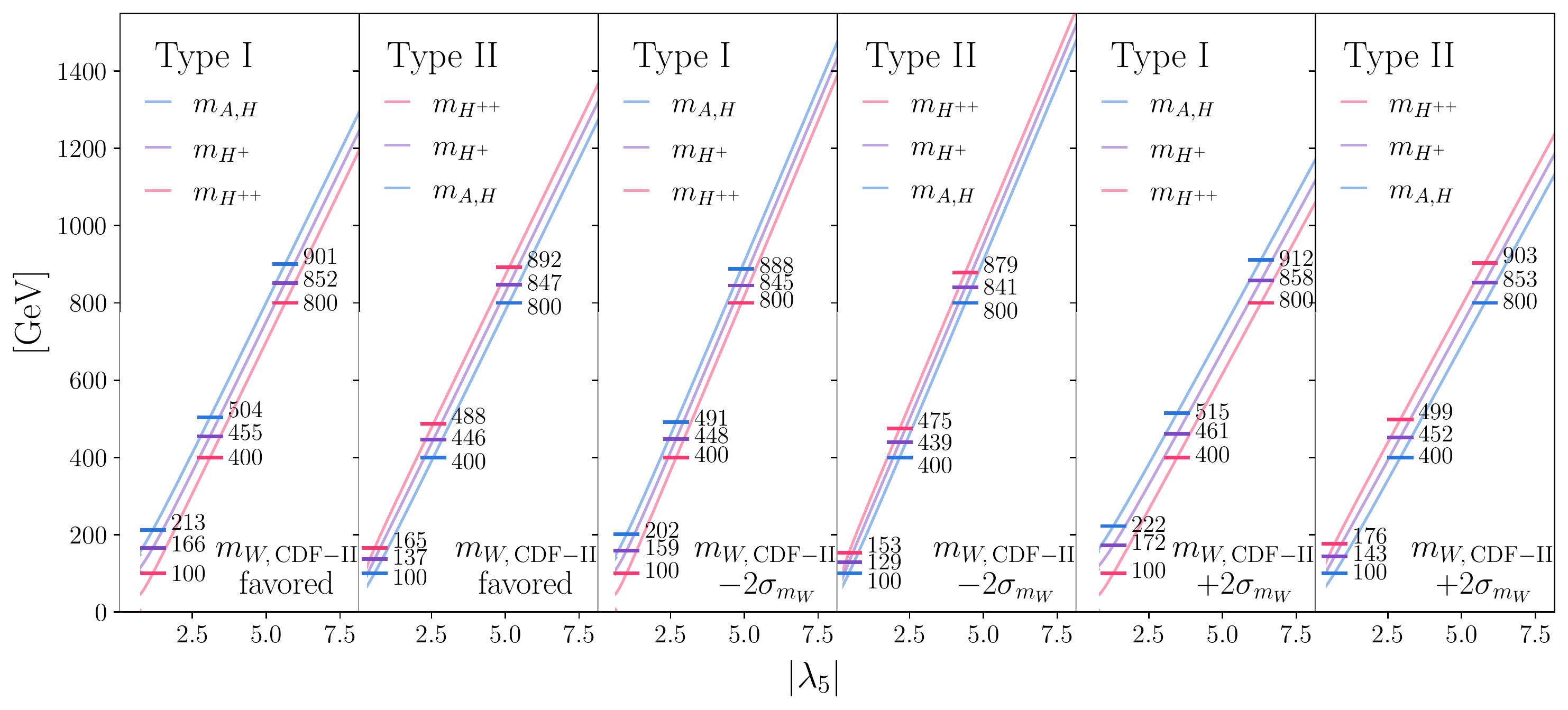}
    \caption{Mass spectrum of the new Higgs states for a given $|\lambda_5|$ that yields the CDF-II measured central values of $m_W$,  $m_W-2\sigma_{m_W}$, or $m_W+2\sigma_{m_W}$ for the type-I and type-II HTM. We exclude the mass spectrum that corresponds $m_{H^{++}} < 42.9\,\text{GeV}$ (excluded by $Z$ decays~\cite{Kanemura:2014goa}) for the type-I model and that corresponds to $m_{A,H} < m_h/2 = 62.5\,\text{GeV}$ (excluded by Higgs precision measurement, see Sec.~\ref{sec:exotichiggsdecays}) for the type-II model. In each panel, we explicitly show three sets of benchmark values.}
    \label{fig:fig3}
\end{figure*}

Furthermore, we scan $m_\text{lightest}$ and $\lambda_5$ to pinpoint the parameter regions predicting a $m_W$ value close to the CDF-II measurement. The resulting mass spectra for the new Higgs states are shown in~\figref{fig3}. The first, third, and fifth (second, fourth, and sixth) panels respectively represent the spectra for the type-I (II) HTM that yield the $m_W$ value measured by CDF-II, the CDF-II value plus two times the experimental one-sigma uncertainty, and the CDF-II value minus two times the experimental one-sigma uncertainty. The blue, purple, and red lines in each panel represent the corresponding values of $m_{A,H}$, $m_{H^{+}}$, and $m_{H^{++}}$, respectively. We also explicitly show three sets of benchmark values for each scenario. For the type-I HTM, we do not show the mass spectrum corresponding to $m_{H^{++}} < 42.9\,\text{GeV}$ since it is excluded by the measurement of $Z$ boson decays~\cite{Kanemura:2014goa}. For the type-II HTM, we do not show the mass spectrum corresponding to $m_{H,A} < m_h/2 =62.5\,\text{GeV}$ given it is excluded by the precision measurement of exotic Higgs decays as we discuss in Sec.~\ref{sec:exotichiggsdecays}. Note that there are stronger yet model-dependent constraints on the HTM mass spectrum from direct collider searches. We will summarize them in detail in Sec.~\ref{sec:direct_searches}.


\section{Effective weak mixing angle}
\label{sec:weakangle}

After obtaining the preferred spectra of the HTM, we assess whether these are compatible with the electroweak precision data by computing the effective weak mixing angle, $\sin^2\theta_\text{eff}$. In this computation, $\alpha_\text{em},~M_Z,$ and $G_F$ are chosen as inputs. Experimentally, $\sin^2\theta_\text{eff}$ is defined as the ratio of the leptonic vector current to the leptonic axial current at the $Z$ pole. The deviation from the tree-level value of the mixing angle, $s_W^2$, can be parameterized by $\Delta\kappa$, where 
\begin{equation}
    \sin^2\theta_\text{eff}=s_W^2(1+\Delta\kappa).
\end{equation}
At one-loop, $\Delta\kappa$ obtains contributions from $A-Z$ mixing, corrections to the weak mixing angle, and corrections to the axial/vector vertices,
\begin{align}
\Delta\kappa={}& -\frac{c_W}{s_W}\left(\frac{\text{Re}\Pi^{1\text{PI}}_{Z\gamma}(m_Z^2)}{m_Z^2}\right)-\frac{{c}_W^2}{{s}_W^2}\left(\frac{{\rm Re}\Pi^{\rm 1PI}_{WW}(m_W^2)}{m_W^2}-\frac{{\rm Re}\Pi^{\rm 1PI}_{ZZ}(m_Z^2)}{m_Z^2}
    \right) \nonumber\\
    & +\frac{v_l}{v_l-a_l}\left(\frac{F^l_V(m_z^2)}{v_l}-\frac{F_A^l(m_z^2)}{a_l}\right),
    \label{eq:delta_kappa}
\end{align}
where $v_l$ and $a_l$ are the tree-level vector and axial couplings respectively, and $F_{V,A}^l$ are the form-factors for the leptonic vector/axial currents. Since the extra Higgs bosons do not couple to the SM fermions, they do not contribute to $F_{V,A}^l$.

\begin{figure*}[t]
    \centering
    \includegraphics[width=0.48\textwidth]{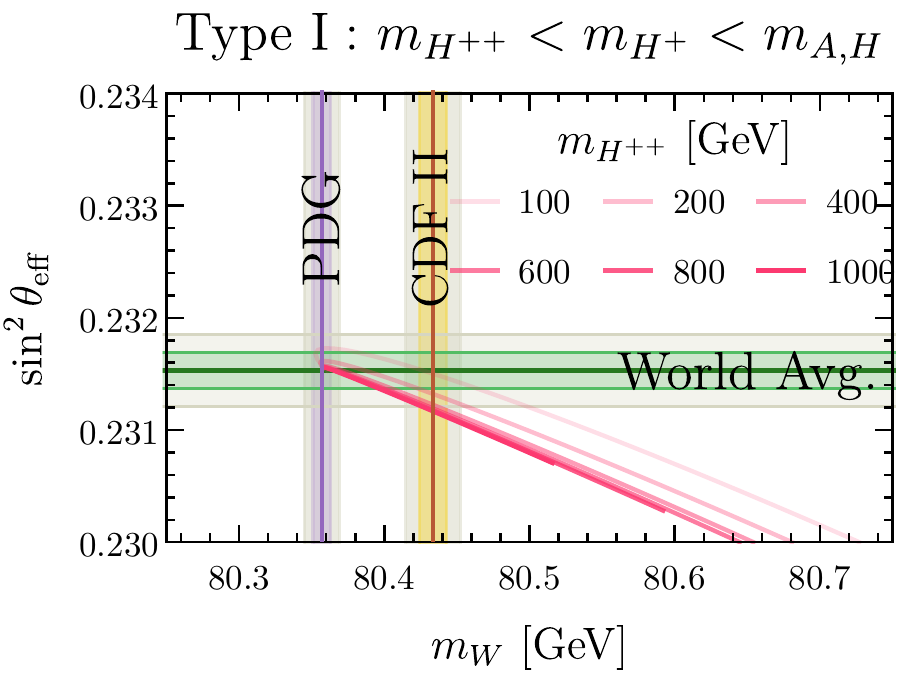}\quad\includegraphics[width=0.48\textwidth]{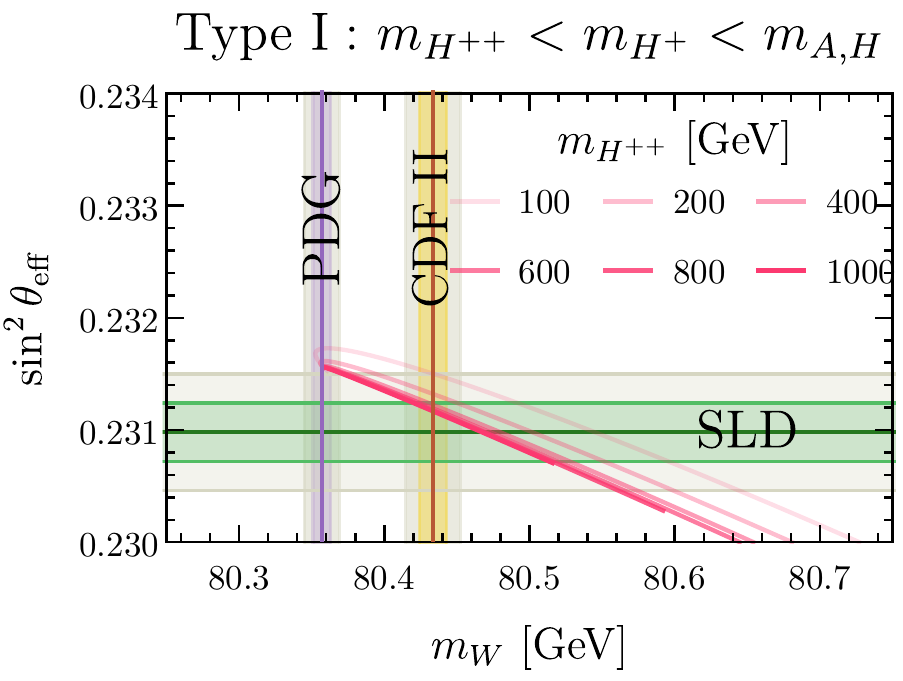}\\
    \includegraphics[width=0.48\textwidth]{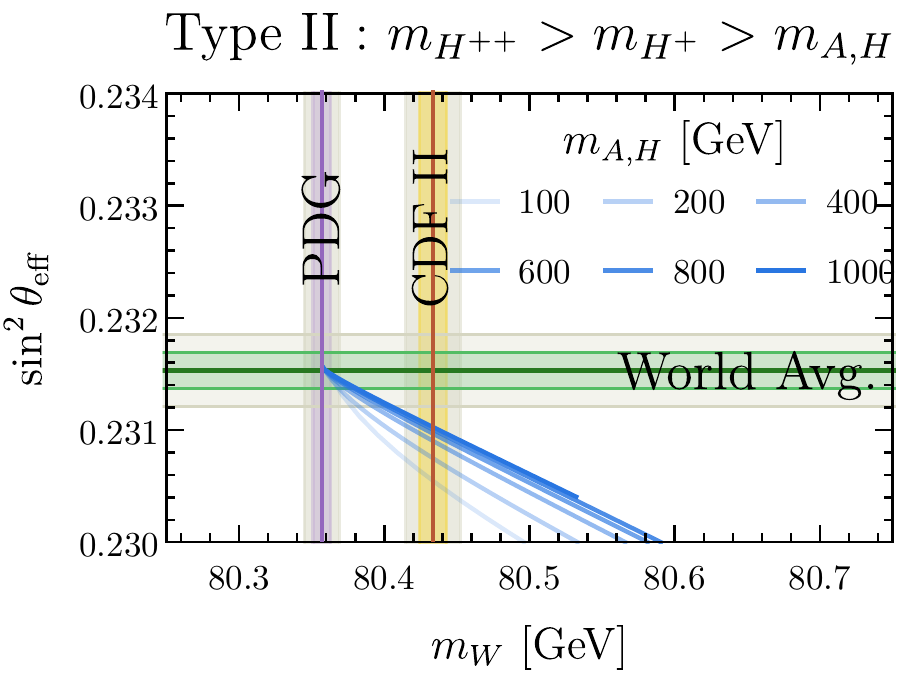}\quad
    \includegraphics[width=0.48\textwidth]{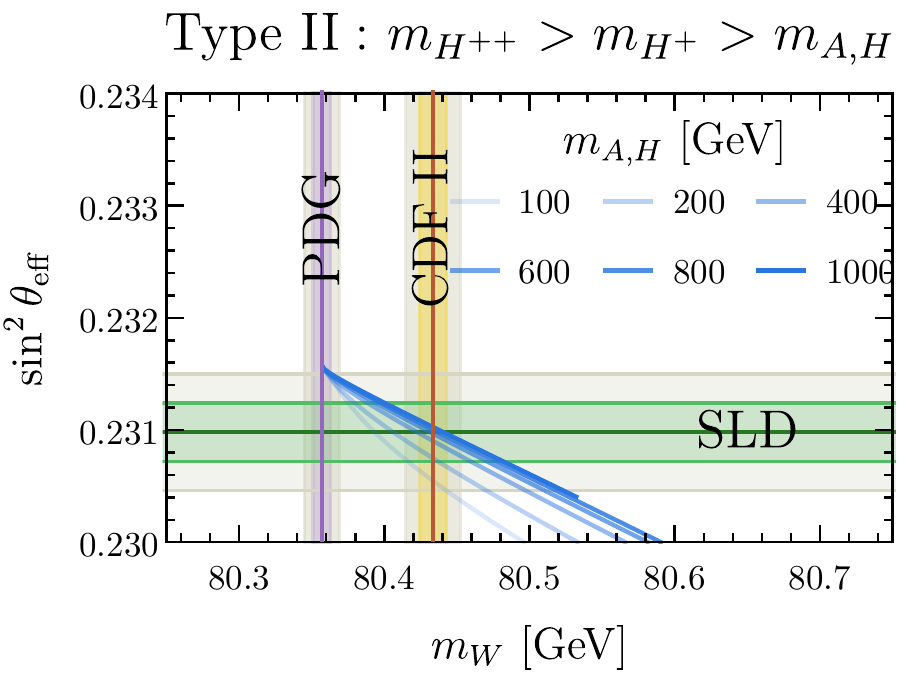}
    \caption{Effective weak mixing angle vs.\ $W$ boson mass for the type I (upper row) and II (lower row) mass hierarchies. For each panel, the different curves represent models with different mass values for the lightest state. For each curve, we vary $|\lambda_5|$ from 0 (corresponds to the SM values) to 10. This finite range of $|\lambda_5|$ scanned results in the endpoints of the contours. The brown line represent the CDF II measured $W$ boson mass and the yellow/gray band shows $1\sigma$/$2\sigma$ range. The dark purple line represent the PDG value for the $W$ boson mass with the purple/gray band showing the $1\sigma$/$2\sigma$ range. The dark green line in the left (right) column represent the world averaged value $0.23153\pm 0.00016$~\cite{ALEPH:2005ab,ParticleDataGroup:2020ssz} (SLD measured value $0.23098\pm 0.00026$~\cite{ALEPH:2005ab}) of $\sin^2 \theta_\text{eff}$ with the green/gray band shows $1\sigma$/$2\sigma$ range.}
    \label{fig:fig2}
\end{figure*}

We compute the SM contribution to $\sin^2 \theta_\text{eff}$ with the help of the SM fitting formula. Similar to the treatment of $\Delta r$, we split $\Delta \kappa$ into three pieces,
\begin{equation}
    \Delta \kappa = \Delta \kappa_{\text{SM},W} (m_{W, \text{BSM}})+ \Delta \kappa_\text{SM, rest} + \Delta \kappa_\text{BSM}~.
\end{equation} 
$ \Delta \kappa_\text{SM, rest}$ is determined via
\begin{equation}
    \Delta \kappa_\text{SM} = \Delta \kappa_{\text{SM},W} (m_{W, \text{SM}})+ \Delta \kappa_\text{SM, rest}, 
\end{equation}
where $\Delta \kappa_\text{SM} \simeq 0.03640$ is computed from the fitting formula given in Ref.~\cite{Awramik:2006uz} (see App.~\ref{sec:fitting}) and $\Delta \kappa_{\text{SM},W} (m_{W, \text{SM}}) \simeq 0.03628$ is Eq.~(\ref{eq:delta_kappa}) restricted to the SM contribution only, which explicitly depends on $m_W$.

In \figref{fig2}, we check whether the parameter space of the HTM that predicts $m_W$ close to the CDF-II value is compatible with the measured effective mixing angle. The upper (lower) row of \figref{fig2} shows the resulting $\sin^2 \theta_\text{eff}$ vs.\ $m_W$ plot for a given $m_\text{lightest}$ in the interval  $[100~\text{GeV}, 1000~\text{GeV}]$ and $|\lambda_5|\in[0, 10]$ for the type-I (II) HTM. For each panel, we highlight the CDF-II (PDG) $m_W$ value as the brown (dark purple) vertical lines while the yellow (purple) and gray vertical bands show the $1\sigma$ and $2\sigma$ ranges, respectively. The dark green horizontal lines in the left column represent the world-average value for the effective weak mixing angle~\cite{ALEPH:2005ab,ParticleDataGroup:2020ssz} while the green and gray horizontal band shows 1$\sigma$ and 2$\sigma$ range respectively. For comparison, we show in the right column the value of the single most precise effective weak mixing angle measurement obtained by the SLD collaboration~\cite{ALEPH:2005ab}.

In the limit of $|\lambda_5|=0$, the type-I/II HTM predicts a $W$ boson mass that agrees well with the world-averaged value. The effective weak mixing angle also agrees well with its world-average. As $|\lambda_5|$ increases, the resulting $m_W$ increases while the resulting $\sin^2 \theta_\text{eff}$ decreases.\footnote{In the limit of small $|\lambda_5|$, the correction to both the $W$ mass and effective mixing angle is sensitive to the sign of $\lambda_5$. In particular, this results in the turning behavior seen for the type-I HTM.} On the other hand, a change in $m_\text{lightest}$ has a less significant impact (at least for $m_\text{lightest}\gtrsim 400\,\text{GeV}$). Note that a heavier $m_\text{lightest}$ yields a larger deviation from the world average for $\sin^2\theta_\text{eff}$ for the type-I model while it yields a smaller departure for type II. For the type-I model, the parameter space that explains $m_{W, \text{CDF-II}}$ is consistent with the world averaged value of $\sin^2\theta_\text{eff}$ within $2\sigma$ level. For the type-II model, the two measurements are inconsistent at the $2\sigma$ level for the $m_\text{lightest}$--$|\lambda_5|$ parameter space that we scanned. If we instead compare $\sin^2 \theta_\text{eff}$ to the value measured by the SLD collaboration~\cite{ALEPH:2005ab}, we find that the parameter space explaining $m_{W, \text{CDF-II}}$ is consistent with the measured $\sin^2 \theta_\text{eff}$ within the $2\sigma$ level for both type-I and -II mass hierarchies. 

In the Two-Higgs-doublet model (2HDM), for which also large upwards shift of $m_W$ with respect to the SM prediction can be realized, a quite similar correlation between the predictions for $m_W$ and $\sin^2\theta_\text{eff}$ is known to exist (see e.g.\ Ref.~\cite{Bahl:2022xzi}). In comparison to the 2HDM, the type-I triplet model provides a slightly better fit of the effective weak mixing angle measurements if the the lightest BSM state is close to the electroweak scale; in contrast, the type-II triplet model provides a slightly worse fit if the lightest BSM state is close to the electroweak scale.


\section{Precision measurement of the Standard Model Higgs}
\label{sec:precisionhiggs}

For $v_\Delta \simeq 0$, the tree-level couplings of the SM-like Higgs boson are only modified negligibly with respect to the SM. Significant effects can, however, occur a the loop level or through the presence of new exotic decay modes.

\subsection{Higgs-photon coupling and Higgs self-coupling}

We define the ratio of the coupling between the SM-like Higgs boson and photon to the SM predicted coupling by $$\kappa_\gamma^2 \equiv \frac{\Gamma_{H\to\gamma\gamma}}{\Gamma^\text{SM}_{H\to\gamma\gamma}}.$$ For the triplet model, it is given by
\begin{align}
    \kappa_\gamma^2  &= \frac{\left|\frac{4}{3}F_{1/2}(\tau_t)+\ldots+Q_{H^{\pm\pm}}^2\frac{v^2\lambda_{h H^{\pm\pm}H^{\mp\mp}}}{m_h^2} F_{\pm}(\tau_{H^{\pm\pm}})+Q_{H^{\pm}}^2\frac{v^2\lambda_{h H^{\pm}H^{\mp}}}{m_h^2} F_{\pm}(\tau_{H^{\pm}})\right|^2}{\left|\frac{4}{3}F_{1/2}(\tau_t)+\ldots\right|^2},
\end{align}
where $Q$ denotes the electric charge; $\tau_f \equiv m_h^2/(4 m_f^2)$; and the ellipsis denotes subleading SM contributions. The scalar couplings are given by
\begin{align}
    \lambda_{h H^{\pm\pm}H^{\mp\mp}} &= - v\lambda_4, \hspace{.5cm} \lambda_{h H^{\pm}H^{\mp}} = -v(\lambda_4 + \lambda_5/2).
    \label{eq:higgscoupling}
\end{align}
The loop functions $F_{1/2}$ and $F_\pm$ have the form
\begin{align}
    F_{1/2}(\tau) &= \frac{(\tau - 1) f(\tau) + \tau}{\tau^2},\hspace{.4cm}F_{\pm}(\tau) = \frac{\tau - f(\tau)}{\tau}
\end{align}
with
\begin{align}
    f(\tau) = 
    \begin{cases}
        \arcsin^2(\sqrt{\tau}) & \text{if } \tau\le 1, \\
        -\frac{1}{4}\left(\ln\frac{1+\sqrt{1-1/\tau}}{1-\sqrt{1-1/\tau}} - i\pi\right)^2 & \text{if } \tau > 1.
    \end{cases}
\end{align}
We evaluate the LHC constraints set on the triplet couplings through modifications of the $H\to\gamma\gamma$ rate by employing \texttt{HiggsSignals}~\cite{Bechtle:2013xfa,Bechtle:2020uwn}.

In addition to the di-photon rate, we also evaluate loop corrections to the trilinear Higgs self-coupling, which can receive large quantum corrections in the presence of large scalar couplings potentially excluding otherwise unconstrained parameter space (see e.g.\ Ref.~\cite{Bahl:2022jnx}).

We compute the one-loop correction using \texttt{FeynArts}~\cite{Hahn:2000kx} and \texttt{FormCalc}~\cite{Hahn:1998yk} with the necessary model file derived using \texttt{FeynRules}~\cite{Christensen:2008py,Alloul:2013bka}. For this calculation, we renormalize the SM-like Higgs boson mass in the on-shell scheme. The SM-like vev is also renormalized in the on-shell scheme by renormalizing the $W$ and $Z$ boson masses as well as the electric charge in the on-shell scheme.

We compare the predicted value for the trilinear Higgs self-coupling normalized to the SM tree-level value, $\kappa_\lambda$, to the strongest current bound of $-1.0\leq \kappa_\lambda \leq 6.6$~\cite{ATLAS:2021tyg} (at 95\% CL). This bound is based on searches for the production of two Higgs bosons and assumes that this production mechanism is only affected by a deviation of the trilinear Higgs self-coupling from its SM value. While quantum corrections to double-Higgs production are not only induced by corrections to the trilinear Higgs self-coupling, evaluating the one-loop corrections to the trilinear Higgs self-coupling takes into account all one-loop corrections to double Higgs production leading in powers of scalar couplings. Since the scalar couplings are responsible for the dominant deviation from the SM, this justifies applying the bound of Ref.~\cite{ATLAS:2021tyg}.

\begin{figure}[t]
    \centering
    \includegraphics[width=0.8\textwidth]{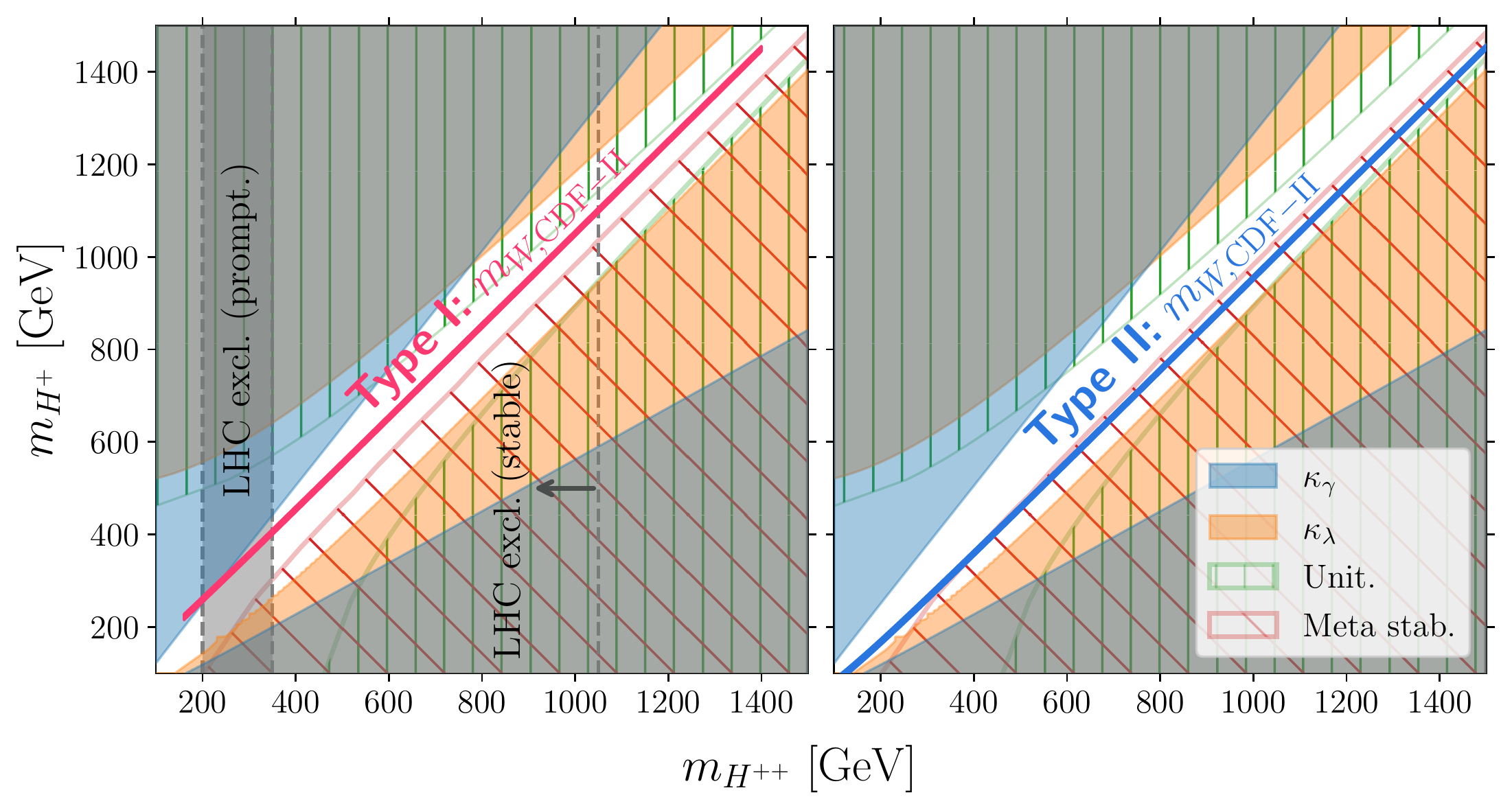}
    \caption{Constraints on $(m_{H^{++}},m_{H^{+}})$ parameter space for the HTM from the measurement of Higgs-photon coupling ($\kappa_\gamma$, blue shaded region), Higgs self-coupling ($\kappa_\lambda$, orange shaded region), perturbative unitarity (green hatched region), and meta-stability condition of the vacuum (red hatched region). We set $\lambda_4=0$  and $\lambda_\Delta=1$ for all panels. The left (right) column shows the $2\sigma$ favored parameter space that explained the measured $m_W$ by CDF-II for the type-I (-II) model as red (blue) narrow bands. Note that we do not show the parameters for $|\lambda_5| > 10$ in drawing the narrow bands. The parameter space with $m_{H^{++}} \gtrsim 250\,\text{GeV}$ ($m_{H^{++}} \lesssim 350\,\text{GeV}$) for the type-I (-II) HTM remains unconstrained. We also show the LHC constraints on $m_{H^{++}}$ for type-I if $H^{++}$ decays promptly (gray band) or if it is detector stable (left of the gray dash line). See Sec.~\ref{sec:detection} for more details.}
    \label{fig:figx}
\end{figure}

The constraints in the $(m_{H^{++}},m_{H^{+}})$ parameter plane due to modifications of $\kappa_\gamma$ and $\kappa_\lambda$ are shown in \figref{figx}. The blue shaded region shows the excluded parameter space by measurements of the Higgs di-photon rate (demanding compatibility at 95\% CL); the orange shaded region is excluded by the constraint on the trilinear Higgs couplings (at 95\% CL). Moreover, we show the constraints set by perturbative unitarity (green hashed region) and by the (meta-)stability of the electroweak vacuum (red hashed region), which we evaluate as detailed in App.~\ref{sec:unitarity_stability}. Note that we set $\lambda_4 = 0$ and $\lambda_2 = \lambda_3 = 1$ in drawing the plots.\footnote{Larger values for $\lambda_4$ tighten the constraints from $h\to\gamma\gamma$. Larger values for $\lambda_{2,3}$ tighten the perturbative unitarity constraint while relaxing the vacuum stability constraint.} 

For the left panel of \figref{figx}, we concentrate on the type-I hierarchy. Almost the complete lower right half of the parameter plane is excluded by requiring metastability of the electroweak vacuum. Perturbative unitarity excludes large differences between $m_{H^{++}}$ and $m_{H^+}$. Measurements of the Higgs to di-photon rate additionally exclude a portion of the parameter space around $m_{H^{++}}\sim 300\, \text{GeV}$ and $m_{H^{++}}\sim 450\, \text{GeV}$ unconstrained by perturbative unitary and vacuum stability. The experimental measurements of the Higgs trilinear coupling are, so far, not precise enough to probe parameter space unconstrained by perturbative unitarity and vacuum stability in the considered scenario. We find, the parameter space favored by the CDF-II measurement of $m_W$ (red narrow band) with $m_{H^{++}}\gtrsim 250\, \text{GeV}$, which lies close to the diagonal, to not lie in the parameter space excluded by the above mentioned constraints. In addition to the constraints discussed above, we also show the LHC constraints on $m_{H^{++}}$ if $H^{++}$ decays promptly (gray band) or if it is detector stable (gray dash line). These constraints are discussed in detail in Sec.~\ref{sec:detection} below.

For type II (see the right panel of \figref{figx}), the constraints set by the Higgs couplings, perturbative unitarity and vacuum stability are unchanged. The parameter space favored by the CDF-II $m_W$ measurements (blue narrow bands) is, however, shifted downwards with respect to the type-I hierarchy. As a result, the parameter space favored by the CDF-II $m_W$ measurements lies at the boundary of the region excluded by demanding vacuum stability. 
The parameter space favored by $m_{W,\text{CDF-II}}$  is  only accessible for $m_{H^{++}}\lesssim 350\,\text{GeV}$. Note, however, that the evaluating of the vacuum stability constraint relies on various assumptions (see App.~\ref{sec:unitarity_stability}).

\subsection{Exotic decays of the Higgs boson}
\label{sec:exotichiggsdecays}

For the type-I HTM, the branching ratio for $h\to H^{++} H^{--}$ depends on $\lambda_4$ (see Eq.~\ref{eq:higgscoupling}). This coupling needs to be small ($\lesssim 1$ for $m_\text{lightest} \sim \mathcal{O}(100)\gev$) in order to evade constraints from the di-photon decay rate of the SM-like Higgs boson. This leads to negligibly small exotic decay modes for the SM-like Higgs boson (if at all kinematically accessible). The situation is quite different for the type-II HTM. In this case, the exotic decay modes of the SM-like Higgs boson are mainly given by $h\to HH$, $h\to AA$, and $h\to H^+ H^-$ once they are kinetically accessible. Their branching ratios mostly depend on $\lambda_5$ (\ref{eq:higgscoupling}), which needs to be large to explain the CDF-II measurement of $m_W$.

We compute the branching ratio for the SM-like Higgs boson decays to the BSM Higgs states for the type-II model as a function of $m_{A,H}$ for the parameter space that explains $m_{W, \text{CDF-II}}$.
We find the resulting branching ratio to lie between $80\%-97\%$ if the decay modes are kinematically accessible ($m_{A,H} < m_h/2$). Such a large branching ratio for the exotic decays is in tension with the Higgs precision measurements from the LHC. For example, the ATLAS experiment places a 95\% CL constraint of $\text{Br}(h\to\text{BSM}) = \text{Br}(h\to\text{inv}) + \text{Br}(h\to\text{undetected}) < 49\%$~\cite{ATLAS:2019nkf}. A similar constraint has also been placed by CMS experiment~\cite{CMS:2018uag}. These constraints exclude the type-II model if neutral states are lighter than $m_h/2$.


\section{Direct searches at the LHC}
\label{sec:direct_searches}

In this Section, we study potential LHC signals of the new scalars with a spectrum preferred by $m_{W,{\rm CDF-II}}$. As we have demonstrated in the previous sections, an explanation of the $W$ mass deviation in the context of the HTM points to new Higgs bosons below a TeV which makes them targets for direct searches at the LHC. As a guide for the dedicated experimental searches in the future,  the main goal of this section is to highlight the promising search channels with distinct signatures. Of course, some of the LHC searches designed to look for different signal processes would also have sensitivity to the signals considered here. To this end, we recast some of the most relevant searches. Instead of providing detailed limits on the model, our focus is to obtain an indication whether the parameter space has been thoroughly covered. As we will show later in this section, most of the parameter space remains open. We expect dedicated searches designed specifically for the signature described in this section will be much more sensitive. For the rest of this section, we start by discussing the various production channels for the BSM states. We then differentiate between three situations for the decays of the BSM Higgs states resulting in distinct collider signatures: a promptly-decaying lightest BSM state, a detector-stable lightest BSM state, a long-lived lightest BSM state.


\subsection{Production}

\begin{figure*}[t]
    \centering
    \includegraphics[width=1\textwidth]{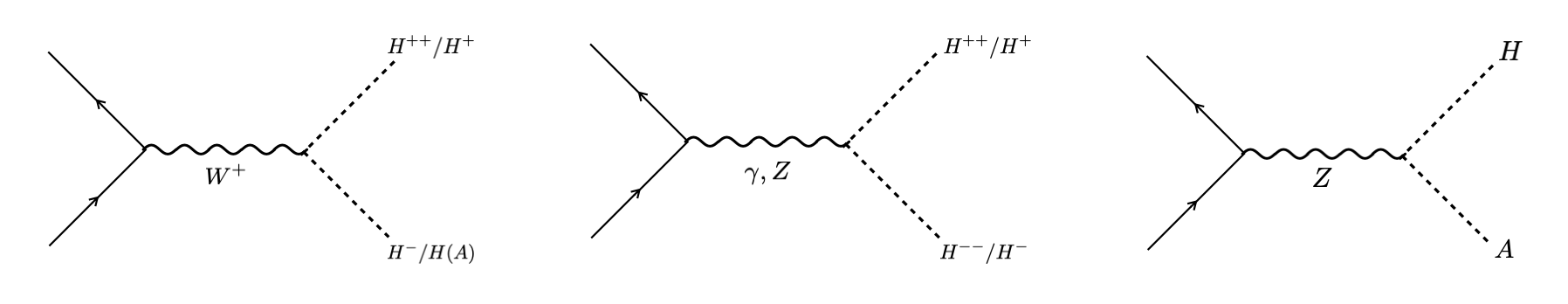}\\
        \includegraphics[width=1\textwidth]{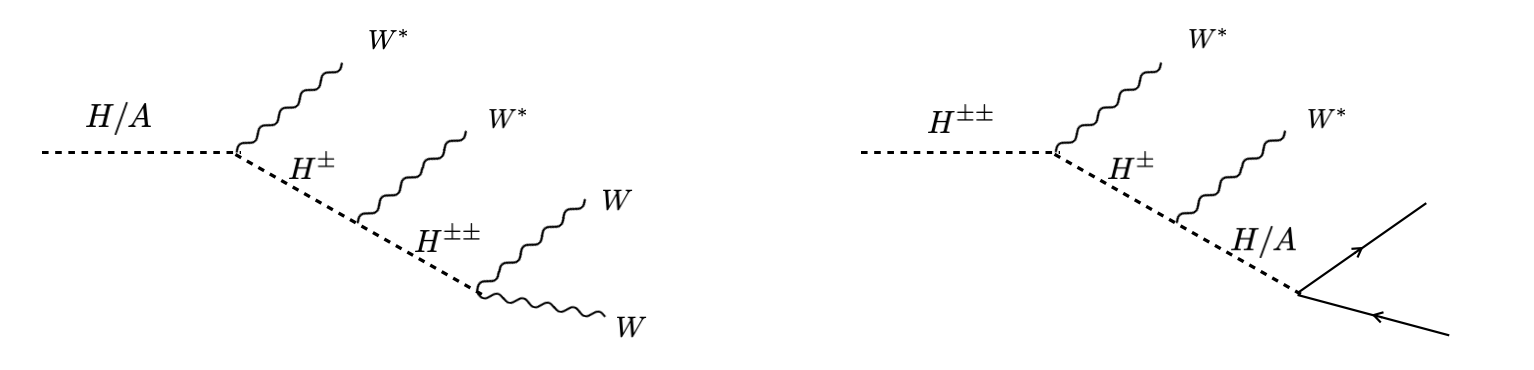}
    \caption{(Upper) Main production channels of the exotic Higgs states at the LHC. (Lower) Decay chains for the heaviest new Higgs state in type I (left) and type II (right). In type II, the lightest state $H/A$ can decay to both SM fermions (as shown on the right) and the SM-like Higgs and gauge bosons.}
    \label{fig:productiondecay}
\end{figure*}

In the absence of additional Yukawa-type interaction terms and for $v_\Delta \ll v$, the exotic Higgs states are dominantly produced via electroweak pair production as shown in the upper row of~\figref{productiondecay}.

\begin{figure*}
    \centering
    \includegraphics[width=0.96\textwidth]{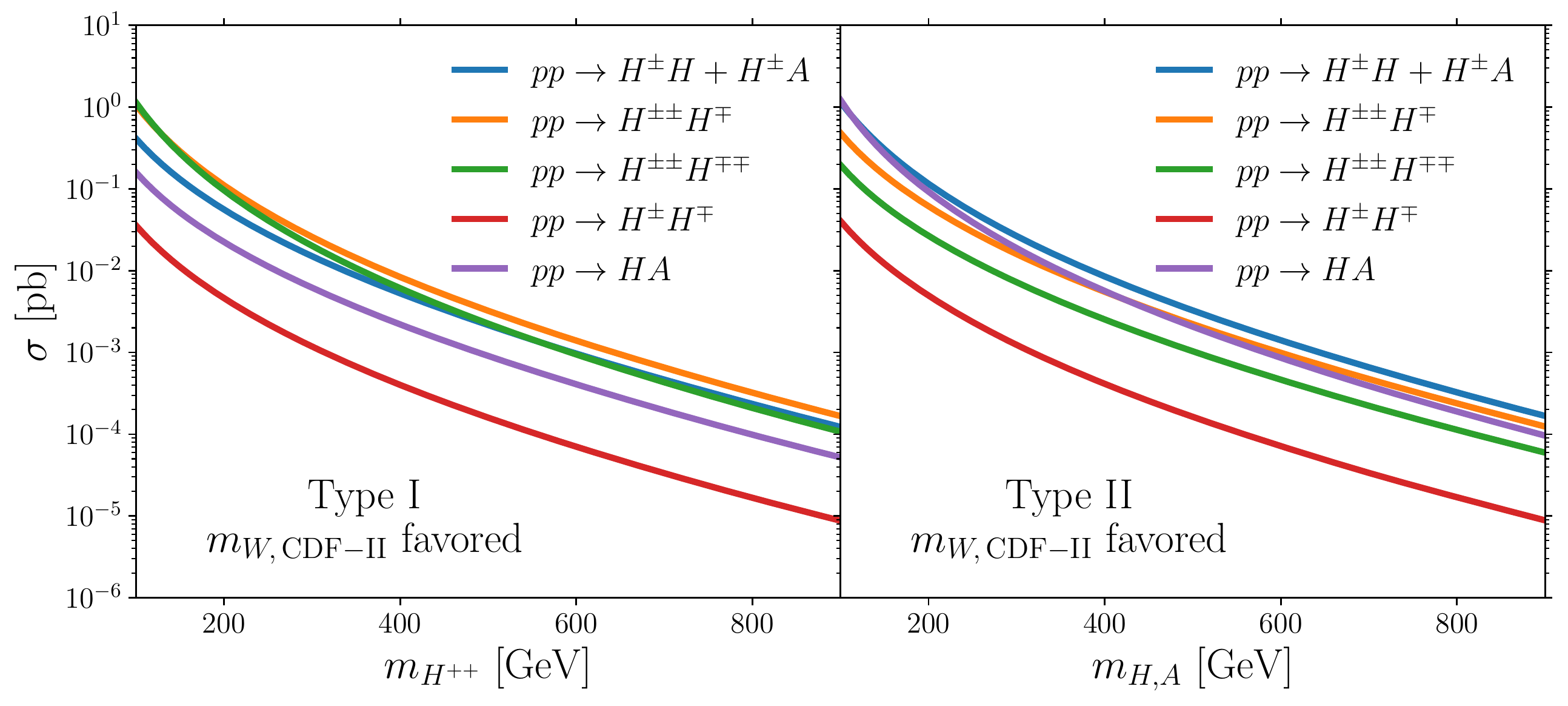}
    \caption{NLO pair production cross sections for Type-I (left) and Type-II (right) as a function of $m_\text{lightest}$ at the 14 TeV run of the LHC.}
    \label{fig:xsec}
\end{figure*} 

In order to obtain an overview of the rate of the various production channels, we computed the next-leading-order (NLO) pair production cross sections for both mass hierarchies using a modified version of the \textsc{Type-II Seesaw} model file~\cite{Fuks:2019clu} (derived using \texttt{FeynRules 2.3}~\cite{Alloul:2013bka}) and \texttt{MG5aMC@NLO v2.9.10}~\cite{Alwall:2014hca}. The dependence of the production cross sections on the lightest BSM state mass in the respective model type is shown in \figref{xsec}. Here, we have chosen the mass spectrum such that we can reproduce the CDF-II central value for $m_W$ as shown in~\figref{fig3}.

For type I (see left panel of \figref{xsec}), the $pp\to H^{\pm\pm}H^{\mp}$ channel mediated by a $W$ boson has the largest cross section of up to $\sim 1$~pb for $m_{H^{++}}\sim 100\gev$. The $pp\to H^{\pm\pm}H^{\mp\mp}$ production cross section is of similar size (especially for lower mass values). Less important are the $pp\to H^\pm H + H^\pm A$, $pp\to HA$, and the $pp\to H^\pm H^\mp$ production channels.

The overall behavior is similar for type II (see right panel of \figref{xsec}). As a consequence of $H$ and $A$ being the lightest BSM Higgs bosons, the $pp\to H^\pm H + H^\pm A$ and $pp\to HA$ channels have, however, now the largest cross sections given their larger phase spaces. Their cross sections reach $\sim 1$~pb for $m_{H,A}\sim 100\gev$.

In our discussion of potential search strategies at the LHC below, we will only focus on the production channels with the largest cross sections.

\subsection{Detection signatures}
\label{sec:detection}

In order to correctly reproduce the $W$ mass measured by CDF-II, the triplet vev $v_\Delta$ generically needs to be small. Given the size of $v_\Delta$ is controlled by the amount of soft breaking, a small value can be naturally achieved.
If the triplet vev is exactly zero, the lightest triplet state is stable. This implies that the choice of $v_\Delta$ directly affects the lifetime of the lightest state, thus affecting the detection signature at the LHC. We discuss the cosmological implications in Sec.~\ref{sec:darkmatter}.

\begin{figure*}[t]
    \centering
    \includegraphics[width=0.48\textwidth]{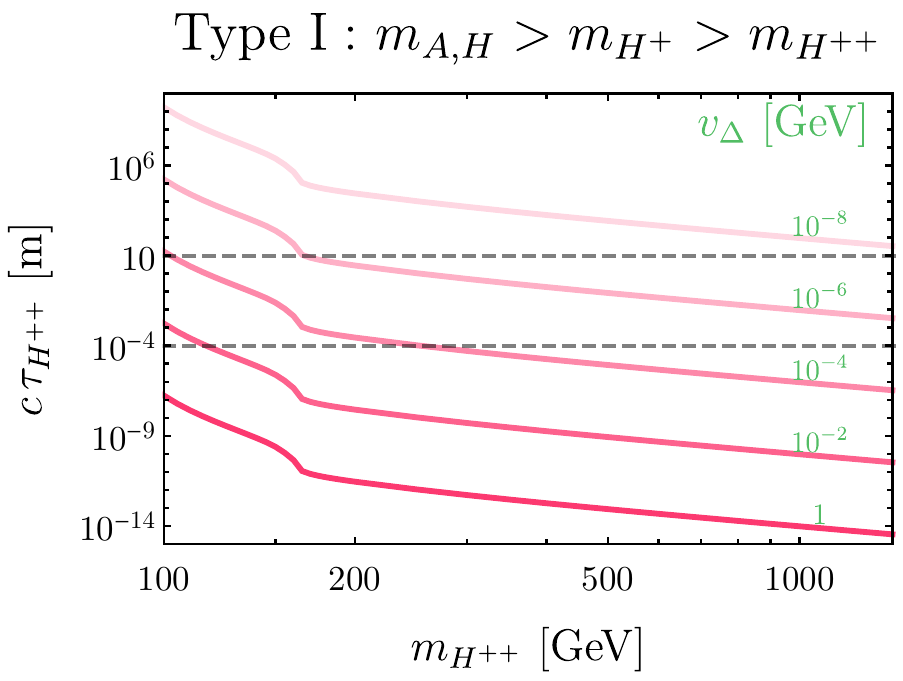}\hfill
    \includegraphics[width=0.48\textwidth]{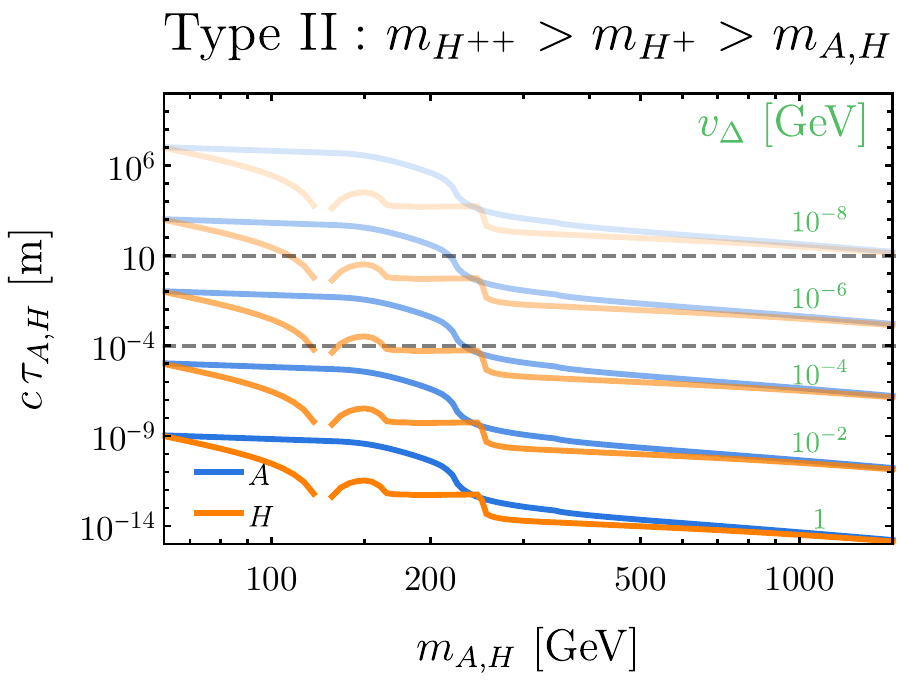}
    \caption{Lifetime (times the speed of light) of the lightest exotic Higgs state for type I (left) and type II (right) as a function of mass and $v_\Delta$ (setting $\lambda_2=\lambda_3=\lambda_4=0$). The remaining parameters are chosen such that CDF-II $m_{W}$ value is explained. 
    The sharp drop at around 160 GeV correspond to the threshold at which the $WW$ decay becomes on shell. For type II (right), additional sharp drops occur at around 250 (215) GeV where $H\to hh$ ($A\to Zh$) becomes on shell; furthermore, $m_H$ is restricted to be $\notin(120,130)$ GeV since $H$ maximally mixes with $h$ there (see App.~\ref{app:soft_Z2_breaking} for more details). For reference, we have drawn dashed lines representing $c\tau = 10^{-4}$ meter (corresponding to $\sim10^{-12}$ sec, which is the typical B meson lifetime) and 10 meter. This is the range in which long lived particle searches at the LHC could be sensitive. 
    }
    \label{fig:lifetime}
\end{figure*}
 
In~\figref{lifetime}, we show this lifetime of the lightest state for different choices of $v_\Delta$ ranging from $10^{-8}\,{\rm GeV}$ to 1 GeV for the type-I/II HTM. For $v_\Delta\sim 10^{-4}$ GeV, the lifetime of the lightest state is generically of the order of the $B$-meson lifetime. As such, any decay products of the lightest state will be tagged as displaced. For $v_\Delta\sim10^{-8}$ GeV, the lifetime is generically orders of magnitude greater than the radius of the detector. In this case, the lightest state is unlikely to decay within the detector volume.

\begin{figure*}[t]
    \centering
    \includegraphics[width=0.96\textwidth]{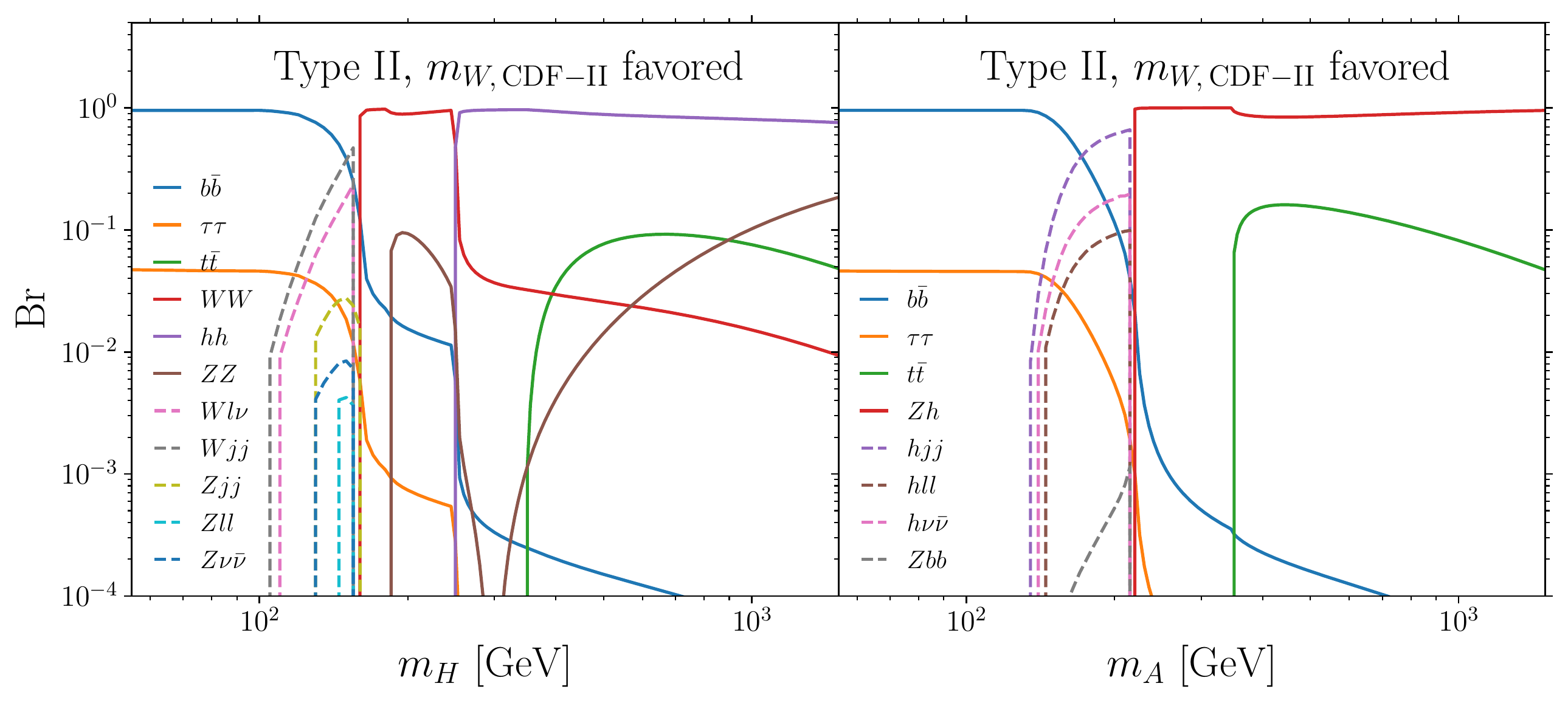}
    \caption{Branching ratios for $H$ (left) and $A$ (right) in the type-II HTM for $\lambda_2=\lambda_3=\lambda_4=0$,  $v_\Delta=1$ GeV, with a mass spectrum explaining the CDF-II $m_W$ measurement. Branching ratios for two-body (three-body) decays are shown as solid (dashed) lines. Since all decay widths are proportional to $v_\Delta^2$, the branching ratios do not depend on the choice of $v_\Delta$ provided that it is nonzero.}
    \label{fig:br}
\end{figure*} 

We further show the decay branching ratio of $H$ and $A$ for type-II model in~\figref{br} (assuming a small but finite $v_\Delta$).\footnote{Branching ratios for the lightest state in the type-I HTM, $H^{++}$, can be found in e.g. in Ref.~\cite{Kanemura:2014goa}.} Generically, the branching ratio of the dominant decay mode is always very close to one. This dominant decay mode depends on whether or not the preferred final state is kinematically accessible. For the \CP-even BSM Higgs boson, $H$, the important thresholds are the $hh$ and $WW$ mass thresholds. For the \CP-odd Higgs boson, $A$, the important threshold is the $Zh$ mass threshold. Below the lowest mass threshold, they both predominantly decay to $b\bar{b}$ due to the bottom Yukawa inherited from its mixing with the SM Higgs doublet.   

In the remainder of the section, we discuss the qualitatively different LHC signatures for the three different lifetime domains: prompt decay of the lightest state, detector-stable lightest state, long-lived lightest state.

\subsubsection{The lightest state promptly decays}

An overview of the main LHC search channels for a promptly decaying lightest BSM state for the type-I and type-II HTM can be found in Tab.~\ref{table:typeI}.

\begin{table}[ht!]
\centering
\begin{tabular}{|c| c|} 
\hline
\multicolumn{2}{|c|}{Type I, Prompt}\\
 \hline
  Main Channels & Example Signature \\ [0.5ex] 
 \hline
 $pp\to H^{\pm\pm}H^\mp$  $\to H^{\pm\pm} (H^{\mp\mp}W^\pm)\to 5W^{(*)}$ & 5~$\ell+\slashed{E}_T$ \\ [0.5ex]\hline
 {$pp\to H^{++}H^{--}\to (W^+W^+) (W^-W^+)$ }
 & 4~$\ell+\slashed{E}_T$   \\
[0.5ex]\hline
$pp\to H^\pm H(A)$  $\to H^\pm (H^\pm W^\mp) \to (H^{\pm\pm}W^\mp)(H^{\pm\pm}W^\mp W^\mp)\to 7W^{(*)}$ & $7 \ell+\slashed{E}_T$ \\ 
[0.5ex]

 \hline
\end{tabular}\\
\vspace{.2cm} 
\begin{tabular}{|c |c |c|} 
\hline
\multicolumn{3}{|c|}{Type II, Prompt}\\
 \hline
 $m_{H,A}$&  Main Channels & Example Signature \\ [0.5ex]\hline
\multirow{2}{*}{ $\lesssim 250 \gev$}& $pp\to H^\pm H/A\to W^{\pm(*)} b\bar{b}b\bar{b}$&monolepton+up to 4 $b$-jets+$\slashed{E}_T$ \\ \cline{2-2}\cline{2-3}
& $pp\to HA\rightarrow b\bar{b}b\bar{b}$ & up to 4 $b$-jets\\[0.5ex]
\hline
\multirow{2}{*}{$\gtrsim 250 \gev$}& $pp\to H^\pm H/A\to (W^\pm H/A)H/A$, $H/A\to hh/Z$ &multi $b$-jets + leptons+$\slashed{E}_T$\\ \cline{2-2}\cline{2-3}
& $pp\to HA\rightarrow Zhhh$ & up to 6 $b$-jets + leptonic $Z$\\[1ex]
\hline
 \end{tabular}
\caption{Summary of main channels and example search signatures for additional Higgs bosons of the HTM that promptly decay at the LHC. The upper and lower tables consider the type-I and II HTM, respectively. See text for more details.}
\label{table:typeI}
\end{table}

For type-I HTM, the production process with the largest cross section is $pp\rightarrow H^{\pm\pm}H^\mp$. The singly-charged Higgs boson then decays to a doubly-charged Higgs boson via emission of a $W$ boson, $H^{\mp}\to H^{\mp\mp} W^{\pm}$. All doubly-charged Higgs bosons will then promptly decay into a pair of $W$ bosons, $H^{\pm\pm} \to W^{\pm} W^{\pm}$, with branching ratio $\approx 1$. (c.f., the lower left diagram of~\figref{productiondecay}.) As such, the corresponding search channel will be a final state of five $W$ bosons. These $W$ bosons could be off-shell depending on the masses. 

No dedicated searches for this channel exist so far. To nevertheless gain an estimate for the LHC sensitivity for this signature, we use \texttt{CheckMATE 2.2}~\cite{Dercks:2016npn,Sjostrand:2014zea, deFavereau:2013fsa, Cacciari:2011ma, Cacciari:2005hq, Cacciari:2008gp, Read:2002hq} to recast a large set of existing searches on a set of benchmark points. \texttt{CheckMATE} will generically summarize the result with $r=S/S_{95}$, where $S$ is the number of signal events and $S_{95}$ is the 95\% C.L. limit on the number of signal events for the given analysis. For statistically limited searches, one would expect $r$ to scale as $\sqrt{\int \mathcal{L}dt}$. We will use this naive scaling to make statements about potential reach with searches involving more data.

We find that $m_{H^{++}}=150$ GeV can be excluded by recasting the multi-lepton final state search of Ref.~\cite{CMS:2017moi} (i.e., by the B02 signal region).
Based on this channel, one could potentially expect to fully close the gap of $84\, \text{GeV}\leq m_{H^{++}}\leq 200 \,\text{GeV}$ between the searches for doubly-charged Higgs boson pair production based, as described below.
We also checked a benchmark point of $m_{H^{++}}=350$ GeV. Here, we expect four on-shell $W$ bosons and one off-shell $W$ boson in the final state. This benchmark is not constrained, for example,  
by using the search of Ref.~\cite{CMS:2017moi} in the G05 signal region. Applying the naive integrated luminosity based rescaling indicates that the full high-luminosity (HL)-LHC dataset (3 ab$^{-1}$) can exclude this mass point; albeit with an analysis that is not dedicated to searching for a doubly-charged Higgs.

In the type-I HTM, the process with the second largest cross section is doubly-charged Higgs boson pair production, $pp\rightarrow H^{\pm\pm}H^{\mp\mp}$.  The corresponding search channel involves a final state of four $W$ bosons. A dedicated search for this signature has been performed by ATLAS using 13 TeV data~\cite{ATLAS:2018ceg,ATLAS:2021jol} . Their search excludes doubly charged Higgs promptly decaying into $W$ bosons with masses $200\, \text{GeV}\leq m_{H^{++}}\leq 350 \,\text{GeV}$. Studies recasting 8~TeV ATLAS data excludes the mass range $m_{H^{++}} < 84\,\text{GeV}$~\cite{Kanemura:2014goa,Kanemura:2014ipa}.

Another significant production process for the type-I HTM is $pp\rightarrow H^\pm H/A$. The neutral Higgs boson in the type-I HTM decays to a singly-charged Higgs boson via $W^{(*)}$ emission with a branching ratio close to one. ($H/A \to W^{\mp} H^{\pm}$, c.f., the lower left diagram of~\figref{productiondecay}.) Fully decaying all of the extra Higgs bosons will generate a final state of seven $W^{(*)}$ bosons. 
The corresponding experimental final state will contain various jets, leptons and missing transverse energy. We have checked a benchmark point with $m_{H^{++}}=350$ GeV using the search in Ref. \cite{ATLAS:2021fbt} in signal region SR12, and found that it is not sensitive to this point. 

For the type-II HTM, the production process with the largest cross section is $pp\rightarrow H^\pm H/A$. The singly-charged Higgs boson will decay to $H/A$ via $W$ boson emission, $H^\pm\to W^\pm H/A$; both $H$ and $A$ have roughly the same probability of being produced. From \figref{br}, the neutral Higgs boson will likely decay to either to a heavy fermion pair or a pair of SM bosons. As before, all of these SM bosons could be off-shell. For this scenario, we ran \texttt{CheckMATE} for both $m_H=100$ GeV and $m_H=300$ GeV. We find both benchmark values to be allowed using the built-in 13 TeV run analyses. The $m_H=100$ GeV benchmark point scenario yielded $r\approx0.6$ using the search of Ref.~\cite{ATLAS:2018zdn} in the $3b1j$ signal region. As this study only used 3.2~fb$^{-1}$ of 13 TeV data, one could potentially exclude the benchmark (i.e., cases where di-boson decays are kinematically forbidden) at $2\sigma$ using a dedicated search with existing data. The $m_H=300$ GeV benchmark point yielded $r\approx0.1$ using the search of Ref.~\cite{ATLAS:2021fbt}. Even with the full HL-LHC dataset, it seems unlikely that a re-analysis could exclude this parameter point based on naive rescaling alone. This analysis is not dedicated to this particular search. It does not make use of the $h$ or $Z$ in the final state.

For the type-II HTM with a light $H/A$, $pp\to HA$ production can be sizable. For $m_H=100$~GeV, recasting Ref.~\cite{ATLAS:2018zdn} in the same signal regions as the previous production mode yielded $r\approx0.4$. Once again, naive luminosity based scaling indicates that existing data is potentially sufficient to exclude this. For $m_H=300$~GeV, we obtained $r\approx0.02$ using \cite{ATLAS:2018zdn} in the $4b1j$ signal region. Accounting for the differences in integrated luminosity used in this and Ref.~\cite{ATLAS:2021fbt}, the exclusion reach comparable to the previous production mode. It should be noted that this production mode ensures a $Z$ boson in the final state. Reconstructing it can potentially reduce the background.

\subsubsection{The lightest state is detector stable}
In this section, we consider the case in which the lightest member of the Higgs triplet is stable on detector timescales. This can be achieve with a small $v_\Delta\lesssim10^{-8}\text{ GeV}$.

In type I, if the lightest state is detector stable, charged tracks in multiple subsystems of the detector are a generic signature. ATLAS presented a search for such tracks excluding doubly-charged particles masses below 1050~GeV~\cite{ATLAS:2022cob}. The unexcluded mass regions will typically require very large values of $\lambda_5$ to give the desired shift in the $W$ mass as shown in~\figref{fig3}. 

In type II, starting with the $pp\to H^\pm H/A$ production, a generic final state consists of $W^*$ and missing transverse energy (MET). As such, the search channels are either monolepton + MET or dijet + MET. Recasting existing searches using \texttt{CheckMATE} did not yield any exclusions for the $m_H=100$~GeV benchmark point.
$pp\to H^{\pm\pm}H^\mp$ production leads to a different final state with more visible particles. The final state consists now out of three $W^{(*)}$ boson. The final state signature could be three charged leptons + MET, two charged leptons + jets + MET, monolepton + jets + MET, or jets + MET. We will focus on the three charged lepton signature. Our recasting with this benchmark show that current searches, such as the one in Ref.~\cite{CMS:2017moi}, is not yet sensitive. Naively rescaling based on the full HL-LHC integrated luminosity shows that this analysis barely misses the exclusion.
Lastly, for $pp\to HA$ production, the main search channel is a mono-jet or mono-photon + MET signature (with the jet or photon originating from initial-state radiation). Current available searches, such as Ref.~\cite{ATLAS:2018zdn} in the MET1$j$ signal region, are not sensitive. This scenario can potentially be excluded using the full HL-LHC dataset.

\subsubsection{The lightest state is long-lived}

If the charged particle decays before reaching the muon spectrometer, the previously mentioned ATLAS charged track search~\cite{ATLAS:2022cob} is not sensitive to it. If the particle decays in the inner tracker, the signal caused by doubly-charged Higgs bosons will be disappearing tracks plus delayed multi-lepton/multi-jet final states. Depending on the initial state, one may also expect prompt off-shell $W$ bosons. These prompt jets/leptons could be used to tag the events provided that the intrinsic jet time spread is sufficiently low~\cite{Chiu:2021sgs}. It should also be noted that recently ATLAS found am anomalously large ionization energy loss~\cite{ATLAS:2022pib}. A highly boosted, long-lived, doubly-charged particle is a potential explanation to explain this excess~\cite{Giudice:2022bpq} suggesting that $H^{\pm \pm}$ could be a good candidate. A large partonic center-of-mass energy could provide the desired boost. A detailed study should be performed to determine the viability of the HTM as an explanation for the $dE/dx$ anomaly. 

For the neutral Higgs states, Ref.~\cite{ATLAS:2022zhj} could be recasted for pair production of the neutral Higgs. However, the only hard objects in this production mode are delayed objects. Generically, we expect a search strategy involving prompt jets/lepton tagging + delayed jets/leptons to be better. Furthermore, for $m_A>215$ GeV, the dominant decay mode involves an on-shell $Z$ boson. Reconstructing a delayed $Z$ boson will be a good signal to search for.


\section{Cosmological implications}
\label{sec:darkmatter}

For sufficiently small $v_\Delta$, the lifetime of the lightest states in type-II, $H$ and $A$, could be longer than the age of the Universe. Given $H$ and $A$ are electrically neutral, they could provide a good candidate for dark matter or a massive relic. To explain the $m_W$ value measured by CDF-II, a large $|\lambda_5|$ is needed. This requires $H$ and $A$ to strongly couple to $h$. Such strong couplings yield a small relic density for $H/A$ if they are produced through the standard thermal freeze-out. The large couplings also lead to large scattering cross sections between $H/A$ and nucleons as well as the production of significant amounts of electromagnetic or hadronic energy  if they are not cosmologically stable.

\begin{figure}[t]
    \centering
    \includegraphics[width=0.49\textwidth]{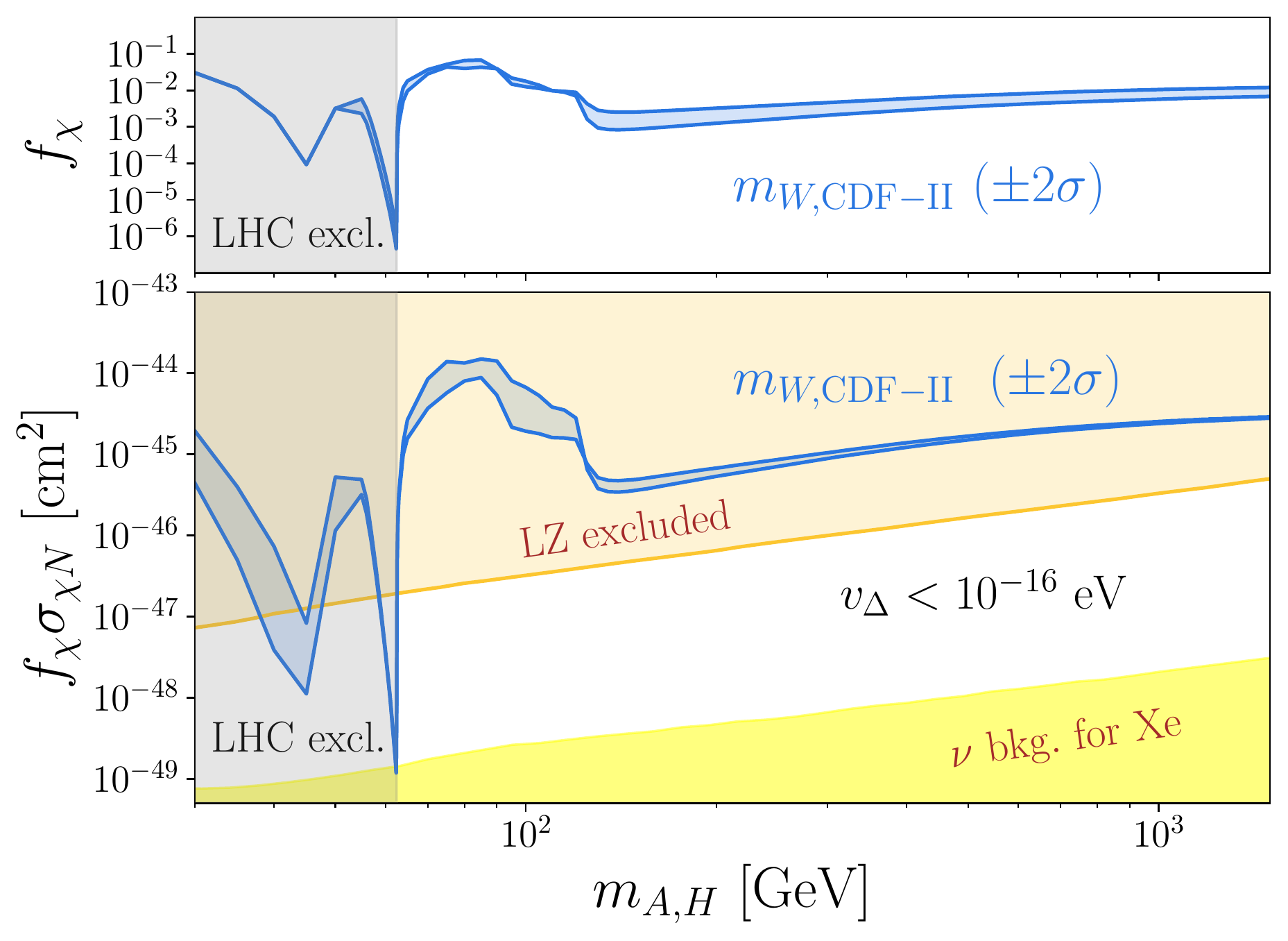}~\includegraphics[width=0.49\textwidth]{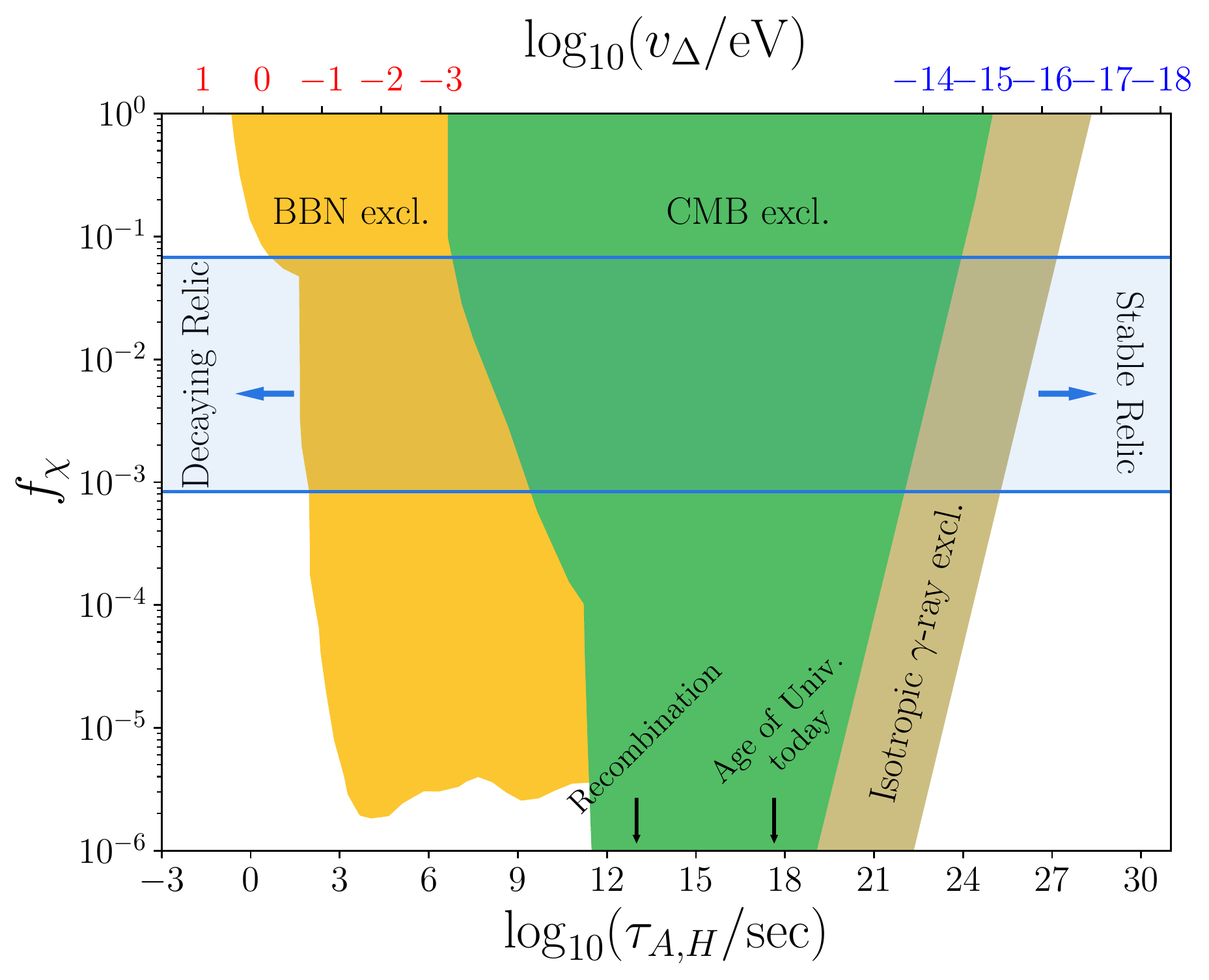}
    \caption{(Upper left) Sum of the relative relic abundances for $H$ and $A$ in the type-II HTM  with respect to that of cold dark matter $f_\chi = \Omega_{H+A}/\Omega_c$ as a function of $m_{H,A}=m_H=m_A$ for parameters that explain the measured $m_{W,\text{CDF-II}}$ within $2\sigma$. (Lower left) Sum of the direct detection cross sections times the relative abundance for cosmologically stable $H$ and $A$ for parameters that explain the measured $m_W$ by CDF-II within $2\sigma$ (blue band) (assuming $v_\Delta < 10^{-16}\,\text{eV}$). 95\% CL constraints from the LZ experiment with 5.5 ton$\cdot$ 60 day exposure~\cite{LZ}  are shown as the orange shaded region. The neutrino background for a xenon target~\cite{Ruppin:2014bra} is shown as the yellow shaded region. For both panels, we added constraints from the exotic decays of the SM-like Higgs as the gray shaded region. (Right panel) Constraints on the relative abundance of visibly decaying relic with respect to cold dark matter as a function of their lifetime. 95\% CL constraints from BBN~\cite{Kawasaki:2017bqm}, CMB~\cite{Acharya:2019uba, Acharya:2019owx}, and isotropic $\gamma$-ray backgrounds~\cite{Blanco:2018esa} are shown as yellow, green, and brown shaded regions. We highlighted the range of $f_\chi$ that explains $m_{W,\text{CDF-II}}$ as the blue band with arrows indicating the allowed lifetimes for stable massive relic and decaying massive relic. In the lifetime axes, we indicate the age of Universe at recombination and today with black arrows. In the upper axes, we show the corresponding values of $v_\Delta$ for $m_A = 65\,\text{GeV}$ and $m_A=1.5\,\text{TeV}$ as red and blue ticks, respectively. See text for more details.}
    \label{fig:figdm}
\end{figure}

We first compute the thermal relic density for $A$ and $H$ using \texttt{MadDM 3.2}~\cite{Arina:2021gfn}. The upper panel of~\figref{figdm} shows the resulting sum of the relative relic abundances for $H$ and $A$ with respect to that of cold dark matter, $f_\chi \equiv \Omega_{H+A}/\Omega_c$ as a function of $m_{A,H}\equiv m_H\simeq m_A$\footnote{The mass splitting between $H$ and $A$ is  at $\mathcal{O}(v_\Delta^2/v^2)$. It is negligible for the value of $v_\Delta$ we are interested in.}  for model parameters that explains the CDF-II measured $m_W$ within $2\sigma$.  The relative relic abundance of $A$ and $H$ ranges from $10^{-7}$ to $10^{-1}$ of the total dark matter abundance. It reaches a maximum of $7\%$ around $m_{A,H}\approx m_W$ and converges to $\sim 1\%$ for $m_{A,H} >  700\,\text{GeV}$. The dips around $m_{A,H} =m_Z/2$ and $m_{A,H} =m_h/2$ correspond to the resonant enhancement of the annihilation cross sections. Note that the parameter space below $m_{A,H} < m_h/2$ (shaded in gray) is excluded by Higgs precision measurements at the LHC (see Sec.~\ref{sec:exotichiggsdecays}). If we restrict $m_{H,A}$ to be away from the resonant region of $m_h/2$, i.e.\ $m_{H,A} > 63\, \text{GeV}$, the relative abundance $f_\chi$ varies from $0.08\%$ to $7\%$.

To discuss the observational signatures of the massive relic, we consider two scenarios according to the lifetime of $H$ and $A$: (i) $H$ and $A$ are cosmologically stable and (ii) $H$ and $A$ are cosmologically unstable.\footnote{We do not consider the scenario where $A$ is stable and $H$ is unstable given the small difference in their lifetimes for a fixed $v_\Delta$ compared to the cosmological timescales.} For the parameters that explain $m_{W,\text{CDF-II}}$, the lifetime of $H$ and $A$ mostly depends on the size of $v_\Delta$ and weakly depends on $m_{H,A}$. Besides the two parameters, the observational signature of the massive relic additionally depends on $f_\chi$, which is determined by $m_{A,H}$ as shown in the upper left panel of~\figref{figdm}.

\subsection{Stable massive relic}

For scenario (i), $H/A$ could enter dark matter direct detection experiments on Earth and leave imprints even if they are subdominant components of dark matter. Given $H/A$ are thermally produced in the early universe, their lifetime coincides with the age of the Universe. To realize the stable relic scenario, the lifetime for $H/A$ needs to be longer than the age of universe today $\tau_{H,A} \gtrsim \tau_U=10^{18}\,\text{sec}$. A stronger constraints on $\tau_{A,H}$ comes from the observations of the diffused $\gamma$-ray backgrounds~\cite{Ando:2015qda, Cohen:2016uyg, Liu:2016ngs, Blanco:2018esa}. Observations from Fermi-LAT telescope restrict the decaying time of dark matter $\tau_\chi \gtrsim 10^{28}\,\text{sec}$ if it consists all the dark matter~\cite{Blanco:2018esa}. We translate this bound into $\tau_{H,A} \gtrsim 10^{28} f_\chi\,\text{sec}$ if only an   $f_\chi$ fraction of dark matter decays visibly. This is shown as the brown shaded region in the right panel of~\figref{figdm}. To satisfy the constraint, $v_\Delta$ needs to be small. In the right panel of~\figref{figdm}, we explicitly show the value of $v_\Delta$ for a given lifetime for $A$ with mass $m_A=1.5\,\text{TeV}$ as the blue upper ticks. For the $m_{A,H}$ parameter space we consider, we find that setting $$v_\Delta \lesssim 10^{-16}\,\text{eV}$$ guarantees the cosmological stability.
 
We use \texttt{MadDM 3.2}~\cite{Arina:2021gfn} to compute the spin-independent direct detection cross section for $A$ and $H$. The lower panel of~\figref{figdm} shows the corresponding sum of the spin-independent direct detection cross section between the nucleon and $H/A$, weighted by the relative abundance. In the computation, we assume the relative abundance $f_\chi$ between the massive relic and cold dark matter stays the same for the local dark matter environment (with cold dark matter density $\rho_{\text{local}} = 0.3\,\text{GeV}/\text{cm}^3$). Besides, the two share the same velocity distribution. The resulting weighted cross section (blue band), which is favored to explain $m_{W,\text{CDF-II}}$, is ranging from $10^{-49}\,\text{cm}^2$ to $10^{-44}\,\text{cm}^2$ for $m_{A,H}$ ranging from 30 GeV to 1.5 TeV. In the same panel, we also show 95\% CL constraint from the LUX-ZEPLIN (LZ) experiment with 5.5 ton$\cdot$60 day exposure, where we scale up the cross section by 1.96/1.64 to estimate 95\% CL limit based on the 90\% CL limit reported in~\cite{LZ}. Note that most of the parameter space to explain $m_{W,\text{CDF-II}}$ is excluded by the LZ experiment together with the Higgs precision measurement. One exception is a fine-tuned parameter space with $m_{A,H}$ slightly above $m_h/2$, which could be excluded by future direct detection experiments or Higgs precision measurements. Otherwise, an additional mechanism is needed to further deplete its relic abundance to make this case viable.

\subsection{Decaying massive relic}

 If $H$ and $A$ are not cosmologically stable, they could decay into the Standard Model particles through their couplings to the SM-like Higgs boson. The decays could inject significant amount of  electromagnetic or hadronic energy into the Standard Model plasma in the early universe or intergalactic medium in the late universe, depending on the their lifetimes. This could lead to various observational signature in astrophysics and cosmology, such as those from Big Bang Nucleosynthesis (BBN)~\cite{Poulin:2015opa,Kawasaki:2017bqm,Acharya:2019uba}, Cosmic Microwave Background (CMB)~\cite{Slatyer:2016qyl, Poulin:2016anj,Chluba:2020oip, Acharya:2019uba, Acharya:2019owx}, and galactic and extragalactic diffuse $\gamma$-ray background observations~\cite{Ando:2015qda, Cohen:2016uyg, Liu:2016ngs, Blanco:2018esa}, even if they are subdominant components of dark matter.

In the right panel of~\figref{figdm}, we summarize current cosmological constraints on visibly-decaying massive relic from BBN~\cite{Kawasaki:2017bqm}, CMB (combining constraints from anisotropy~\cite{Acharya:2019uba} from Planck 2018 and spectra distortion~\cite{Acharya:2019owx} from COBE/FIRAS), and isotropic $\gamma$-ray background~\cite{Blanco:2018esa} as yellow, green, and brown shaded regions, respectively. To get the BBN constraints, we take the constraints on massive relic $\chi$ with $m_\chi = 1\,\text{TeV}$ that decaying to $b\bar{b}$ from Ref.~\cite{Kawasaki:2017bqm}~\footnote{The original constraints are expressed in the variable $m_\chi n_\chi/s$ where $m_\chi n_\chi$ is the density of the massive relic and $s$ is the entropy density. We translate the constraints into those on $f_\chi$.}. This constraints are representative for massive relic that mostly decaying to hadronic energy. As shown in Ref.~\cite{Kawasaki:2017bqm}, lighter relic ($m_\chi = 30 \,\text{GeV}$ and $m_\chi=100\,\text{GeV}$) or other hadronic energy-dominant decay channels ($\chi \to \bar u u, \bar t t, gg, WW$) share similar constraints. Constraints for massive relics decaying to electromagnetic energy, e.g. $\chi \to e^+e^-$, are generically weaker than those for relics decaying to hadronic energy. In our scenario explaining the CDF-II $m_W$ measurement, the dominant decay channels of $H$ ($A$) are $b\bar{b}$, $WW$, and $hh$ ($bb$ and $Zh$), depending on the kinematic accessibility (c.f.~\figref{br}). All these decay channels generate significant amount of hadronic energy. Hence the BBN constraint we quoted are applicable.

In the same panel, we highlight a light blue band to show the range of the relative abundance for $m_{H,A} > 63\,\text{GeV}$ (away from the fine-tuned mass region) whose corresponding parameters explain $m_{W,\text{CDF-II}}$. For such an abundance range (0.08\%--7\%), the strongest constraints for the decaying relic come from BBN, which restrict $\tau_{A,H} \lesssim 50\,\text{sec}$. To satisfy this constraint, the value of $v_\Delta$ needs to be large. In the right panel of~\figref{figdm}, we explicitly show the value of $v_\Delta$ for a given lifetime of $A$ with mass $m_A=65\,\text{GeV}$ as the red upper ticks. For the $m_{A,H}$ parameter space we consider, we find that
$$v_\Delta \gtrsim 1\,\text{eV}$$
guarantees that $A$ and $H$ evade all the cosmological constraints for a visibly-decaying massive relic in the scenario which explains the CDF-II $m_W$ measurement. Note that $v_\Delta \geq 1\,\text{eV}$ corresponds to $c\tau_{H,A} \lesssim 1\,\text{km}$. Such decay signal could be searched at the long-lived particle search facilities at the LHC.


\section{Summary}
\label{sec:summary}

In this work, we studied the HTM with hypercharge $Y=1$ in light of the recent  CDF-II $W$ mass measurement. 
The HTM can be realized with two distinct types of spectra: type I for which $m_{H^{++}}<m_{H^+}<m_{H,A}$, and type II for which $m_{H^{++}}>m_{H^+}>m_{H,A}$. First, we derived the mass spectrum of the additional Higgs bosons (for both type I and type II) preferred by the CDF-II $m_{W}$ measurement. For this mass spectra, we then checked the compatibility with experimental measurements of the effective weak mixing angle and Higgs precision data (i.e., measurements of the Higgs di-photon rate, constraints on the Higgs trilinear coupling and constraints on exotic decay channels of the SM-like Higgs boson). For the type-I HTM, we find that mass spectra (as shown in the first, third, and fifth panel of~\figref{fig3}) with the lightest state mass $m_{H^{++}}\gtrsim 250\,\text{GeV}$ explain the observed $m_{W,\text{CDF-II}}$ while being consistent with the measurements of the effective weak mixing angle and Higgs precision measurements, while also satisfying the theoretical constraints of perturbative unitarity and vacuum stability. For the type-II HTM, we find that mass spectra (as shown in the second, fourth, and sixth panel of~\figref{fig3}) with the lightest state mass $62.5\,\text{GeV}\lesssim m_{H,A}\lesssim 350\,\text{GeV}$ explain the observed $m_{W,\text{CDF-II}}$ while being consistent with the Higgs precision measurements, perturbative unitarity, and vacuum stability. For type II, we, however, find a mild tension with the world average measurement of $\sin^2_{\theta_{\rm eff}}$ at the $2\sigma$ level, while still being well consistent with the single-most precise measurement of the effective weak mixing angle by the SLD collaboration.

Direct searches at the LHC provide stronger yet model-dependent constraints on the HTM. The model dependence mainly originates from the decay length of the lightest state, which is mostly controlled by the value of $v_\Delta$ and $m_\text{lightest}$ (c.f.~\figref{lifetime}). We classified the LHC signatures according to if the lightest state promptly decays, if it is detector-stable, or it is long-lived. We investigated the collider phenomenology for each of these cases and pointed out a number of promising discovery channels that the LHC could be sensitive to (summaries of those channels can be found in~Tabs.~\ref{table:typeI}). Current LHC searches are most sensitive to the type-I HTM with a prompt decay of the lightest state (excluding $200\,\text{GeV} < m_{H^{++}} < 350\,\text{GeV}$) or detector-stable lightest state (excluding $m_{H^{++}} < 1050\,\text{GeV}$). A dedicated analysis using current data can also exclude a promptly decaying doubly-charged Higgs with $m_{H^{++}}\sim 150$ GeV by studying the $H^{\pm\pm}H^\mp$ production channel. The case of a long-lived lightest state is so far largely unconstrained for type I. Dedicated searches with the existing data could effectively cover the parameter space for the type-II HTM, especially given the constrained mass range for which the CDF-II measurement can be explained while evading other constraints (see above).

Furthermore, we explored the scenario that the new Higgs triplet is approximately inert. In this case, its lightest neutral state can be a candidate for a sub-dominate fraction of stable dark matter if $v_\Delta \lesssim 10^{-16}\, \text{eV}$ or decaying dark matter if $v_\Delta \gtrsim 1\,\text{eV}$. The former scenario is almost fully constrained by current dark matter direct detection experiments such as the LZ experiment. The later scenario remains possible.


\begin{acknowledgments}
We thank Joaquim Iguaz, Seth Koren, Ying-Ying Li, Zhen Liu, and Anastasia Sokolenko for helpful discussions. CG and YZ acknowledge the Aspen Center for Physics for its hospitality during the final phase of this study, which is supported by National Science Foundation grant PHY-1607611. HB acknowledges support by the Alexander von Humboldt foundation. CG is supported by the DOE QuantISED program through the
theory consortium “Intersections of QIS and Theoretical Particle Physics” at Fermilab.  LTW is supported by the DOE grant DE-SC0013642. YZ is supported by the Kavli Institute for Cosmological Physics at the University of Chicago through an endowment from the Kavli Foundation and its founder Fred Kavli. 
\end{acknowledgments}


\appendix


\section{Self-energy corrections}
\label{app:self_energies}

We take the one-loop contributions to self energies from Ref.~\cite{Aoki:2012jj}, setting $v_\Delta=0$, in the computation of $m_W$ and $\sin^2\theta_\text{eff}$. We listed all the relevant formula here for readers' convenience. 
The corrections are parameterized in terms of $g^2 = e^2/s_W^2$ and $g_Z^2 = e^2/(s_W^2 c_W^2)$.

The BSM contributions to the vector-boson self energies are given by
\begin{align}
\Pi_{WW}^\text{1PI, BSM}(p^2) &=\frac{g^2}{16\pi^2}\Big( B_5(p^2, m_{H^{++}}^2, m_{H^+}^2) +\frac{1}{2} B_5(p^2, m_{H^+}^2, m_H^2)+\frac{1}{2} B_5(p^2, m_{H^+}^2, m_A^2)\Big), \\
\Pi_{ZZ}^\text{1PI, BSM}(p^2)&=\frac{g_Z^2}{16\pi^2} \Big((c_W^2-s_W^2)^2 B_5(p^2, m_{H^{++}}^2, m_{H^{++}}^2)
\nonumber\\
&\hspace{1.7cm}+ s_W^4 B_5(p^2, m_{H^{+}}^2, m_{H^{+}}^2) + B_5(p^2, m_H^2, m_A^2)\Big),\\
\Pi_{\gamma \gamma}^\text{1PI, BSM}(p^2)&=\frac{e^2}{16\pi^2}\big(4 B_5(p^2, m_{H^{++}}^2, m_{H^{++}}^2)+B_5(p^2, m_{H^{+}}^2, m_{H^{+}}^2)\big),\\
\Pi_{\gamma \gamma}^{\prime\text{1PI, BSM}}(p^2)&=\frac{e^2}{16\pi^2}\big(4 B'_5(p^2, m_{H^{++}}^2, m_{H^{++}}^2)+B'_5(p^2, m_{H^{+}}^2, m_{H^{+}}^2)\big), \\
\Pi_{Z \gamma}^\text{1PI, BSM}(p^2)&=-\frac{e g_Z}{16 \pi^2}\Big(2(c_W^2-s_W^2)B_5(p^2, m_{H^{++}}^2, m_{H^{++}}^2)
\nonumber\\
&\hspace{2.cm}+\frac{1}{2}(c_W^2-s_W^2-1)B_5(p^2, m_{H^{+}}^2, m_{H^{+}}^2)\Big)
\end{align}
where $B_{0,1,00,11}(p^2, m_1^2,m_2^2)$  are the Passarino-Veltman two-point functions, which we evaluate using \texttt{LoopTools 2.16}~\cite{Hahn:1998yk}. The remaining loop-functions are given by
\begin{align}
B_3(p^2, m_1^2, m_2^2) ={}& -B_1(p^2, m_1^2, m_2^2) - B_{11}(p^2, m_1^2, m_2^2),\\
B_4(p^2, m_1^2, m_2^2) ={}& -m_1^2 B_1(p^2, m_2^2, m_1^2) - m_2^2 B_{1}(p^2, m_1^2, m_2^2)\\
B_5(p^2, m_1^2, m_2^2) ={}& A_0(m_1^2)+ A_0(m_2^2)-4B_{00}(p^2, m_1^2, m_2^2),
\end{align}
where $A_0(m^2)$ is the Passarino-Veltman one-point function. 

The SM contributions to $\Delta r$ (and $\sin^2 \theta_\text{eff}$) can be separated into three classes: those from scalar bosons, fermions, and gauge bosons --- i.e., $\Pi_{i}^\text{1PI, SM}(p^2) = \Pi_{i,S}^\text{1PI, SM}(p^2) +\Pi_{i,F}^\text{1PI, SM}(p^2) + \Pi_{i,V}^\text{1PI, SM}(p^2)$ where $i= WW, ZZ, \gamma\gamma, Z\gamma$.  The scalar contributions are given by
\begin{align}
    \Pi_{WW}^\text{1PI, SM}(p^2)_S&=\frac{g^2}{64\pi^2}\left(B_5(p^2,m_W^2, m_h^2)+ B_5(p^2,m_W^2, m_Z^2)\right),\\
    \Pi_{ZZ}^\text{1PI, SM}(p^2)_S &=\frac{g_Z^2}{64\pi^2}\left((c_W^2-s_W^2)^2 B_5 (p^2, m_W^2, m_W^2)+ B_5(p^2, m_h^2, m_Z^2) \right),\\
    \Pi_{Z\gamma}^\text{1PI, SM}(p^2)_S &= -\frac{e g_Z}{32\pi^2}(c_W^2-s_W^2) B_5(p^2, m_W^2, m_W^2),\\
    \Pi_{\gamma\gamma}^\text{1PI, SM}(p^2)_S &=\frac{e^2}{16\pi^2} B_5(p^2, m_W^2, m_W^2),\\
    {\Pi'}_{\gamma\gamma}^\text{1PI, SM}(p^2)_S &=\frac{e^2}{16\pi^2} B'_5(p^2, m_W^2, m_W^2).
\end{align}
The fermionic contributions are given by
\begin{align}
\Pi_{WW}^\text{1PI, SM}(p^2)_F &= \frac{g^2}{16\pi^2} N_c^f (2 p^2 B_3(p^2, m_f^2, m_{f'}^2)-B_4(p^2, m_f^2, m_{f'}^2)),\\
\Pi_{ZZ}^\text{1PI, SM}(p^2)_F &= \frac{g_Z^2}{8\pi^2}N_c^f \left(2 p^2 (2 s_W^4 Q_f^2- 2 s_W^2 Q_f I_f+ I_f^2)B_3(p^2, m_f^2, m_f^2)\right. \nonumber\\
&\left.\hspace{2cm}-  I_f^2 m_f^2 B_0(p^2, m_f^2, m_f^2)\right),\\
\Pi_{Z\gamma}^\text{1PI, SM}(p^2)_F &= \frac{e g_Z}{4\pi^2} N_c^f p^2 (2 s_W^2 Q_f^2 -I_f Q_f) B_3 (p^2, m_f, m_f),\\
\Pi_{\gamma \gamma}^\text{1PI, SM}(p^2)_F &= \frac{e^2}{2\pi^2} N_c^f Q_f^2 p^2 B_3(p^2, m_f^2, m_f^2),\\
\Pi_{\gamma \gamma}^{\prime\text{1PI, SM}}(p^2)_F &= \frac{e^2}{2\pi^2} N_c^f Q_f^2 \left(B_3(p^2, m_f^2, m_f^2)+ p^2 B'_3(p^2, m_f^2, m_f^2)\right)
\end{align}
where $m_f$, $Q_f$, $I_f$, and $N_c^f$ are the mass, electric charge, isospin, and color numbers of the SM fermion $f$, respectively. Here, we sum over all the SM quarks and leptons.

Finally, the gauge boson contributions are given by
\begin{align}
    \Pi_{WW}^\text{1PI, SM}(p^2)_V={}& \frac{g^2}{16\pi^2}\bigg(m_W^2 \left(B_0(p^2, m_h^2, m_W^2)+s_W^2 B_0(p^2, m_W^2, 0)+\frac{s_W^4}{c_W^2}B_0(p^2, m_W^2, m_Z^2)\right) \nonumber\\
   & -c_W^2(6 D-8) B_{00}(p^2, m_Z^2, m_W^2)
    - 2 p^2 c_W^2  B_{11}(p^2, m_Z^2, m_W^2)
     \nonumber\\
   & - 2 p^2 c_W^2  B_{1}(p^2, m_Z^2, m_W^2) - 5 p^2 c_W^2  B_{0}(p^2, m_Z^2, m_W^2) +c_W^2 (D-1) A_0(m_Z^2)\nonumber\\
   &+(D-1) A_0(m_W^2)-s_W^2 (6 D-8) B_{00}(p^2, 0, m_W^2)-2 p^2 s_W^2  B_{11}(p^2, 0, m_W^2) \nonumber\\
   & - 2 p^2 s_W^2  B_1(p^2, 0, m_W^2)- 5 p^2 s_W^2  B_0(p^2, 0, m_W^2)
    \bigg) \nonumber\\
    &-\frac{g^2}{4 \pi^2}(p^2 -m_W^2)\left(c_W^2 B_0 (p^2, m_Z^2, m_W^2)+ s_W^2 B_0(p^2, 0, m_W^2)\right),\\
\Pi_{ZZ}^\text{1PI, SM}(p^2)_V={}&\frac{g_Z^2}{16 \pi^2}\bigg(m_Z^2 B_0(p^2, m_h^2, m_Z^2) + 2m_W^2 s_W^4 B_0(p^2, m_W^2, m_W^2)\nonumber\\
&-c_W^4(6D-8)B_{00}(p^2, m_W^2, m_W^2)- 2 p^2 c_W^4  B_{11}(p^2, m_W^2, m_W^2)  \nonumber\\
&- 2 p^2 c_W^4  B_1(p^2, m_W^2, m_W^2) - 5 p^2 c_W^4  B_0(p^2, m_W^2, m_W^2)+2(D-1)c_W^4 A_0(m_W^2)\bigg) \nonumber\\
&-\frac{g_Z^2}{4\pi^2}(p^2-m_Z^2)c_W^4 B_0(p^2, m_W^2, m_W^2),\\
\Pi_{Z\gamma}^\text{1PI, SM}(p^2)_V={}& \frac{e g_Z}{16\pi^2}\bigg(c_W^2(6D-8)B_{00}(p^2, m_W^2, m_W^2)+2 p^2 c_W^2  B_{11}(p^2, m_W^2, m_W^2)\nonumber\\
&+2 p^2 c_W^2  B_1(p^2, m_W^2, m_W^2)+ 5 p^2 c_W^2  B_0(p^2, m_W^2, m_W^2)-2 c_W^2(D-1)A_0(m_W^2) \nonumber\\
&+2 m_W^2 s_W^2 B_0(p^2, m_W^2, m_W^2)\bigg)+\frac{e g_Z}{8\pi^2} \left(2p^2-m_Z^2\right)c_W^2 B_0(p^2, m_W^2, m_W^2),\\
\Pi_{\gamma\gamma}^\text{1PI, SM}(p^2)_V ={}&-\frac{e^2}{16 \pi^2}\bigg((6D-8) B_{00}(p^2, m_W^2, m_W^2)+ 2 p^2 B_{11}(p^2, m_W^2, m_W^2) + 2 p^2 B_1(p^2, m_W^2, m_W^2)\nonumber\\
&+ 5 p^2 B_0(p^2, m_W^2, m_W^2)- 2(D-1)A_0(m_W^2)- 2 m_W^2 B_0(p^2, m_W^2, m_W^2)\bigg) \nonumber\\
&-\frac{ e^2}{4\pi^2} p^2 B_0(p^2, m_W^2, m_W^2),\\
\Pi_{\gamma\gamma}^{\prime\text{1PI, SM}}(p^2)_V ={}&-\frac{e^2}{16 \pi^2}\bigg((6D-8) B'_{00}(p^2, m_W^2, m_W^2)+ 2 B_{11}(p^2, m_W^2, m_W^2) + 2 p^2 B'_{11}(p^2, m_W^2, m_W^2) \nonumber\\ 
&+ 2  B_1(p^2, m_W^2, m_W^2) + 2 p^2 B'_1(p^2, m_W^2, m_W^2)
+ 5 B_0(p^2, m_W^2, m_W^2) \nonumber\\
& + 5 p^2 B'_0(p^2, m_W^2, m_W^2) - 2 m_W^2 B'_0(p^2, m_W^2, m_W^2)\bigg) \nonumber\\
& -\frac{ e^2}{4\pi^2} \left(B_0(p^2, m_W^2, m_W^2)+ p^2 B'_0(p^2, m_W^2, m_W^2)\right),
\end{align}
where $D=4-2\epsilon$ with $\epsilon$ being the dimensional regulator. The divergences of the loop functions are given by
\begin{align}
A_0(m^2)_\text{div} ={}& m^2 \Delta, \quad    B_0(p^2, m_1^2, m_2^2)_\text{div} =\Delta \nonumber \\
B_1(p^2, m_1^2, m_2^2)_\text{div} ={}& -\frac{\Delta}{2}, \quad B_{11}(p^2, m_1^2, m_2^2)_\text{div} =\frac{\Delta}{3} \nonumber\\
B_{00}(p^2, m_1^2, m_2^2)_\text{div} ={}& \left(\frac{m_1^2+m_2^2}{4}-\frac{p^2}{12}\right)\Delta, \nonumber\\
B'_{00}(p^2, m_1^2, m_2^2)_\text{div} ={}& -\frac{\Delta}{12}, \nonumber\\
B_3(p^2, m_1^2, m_2^2)_\text{div} ={}&\frac{\Delta}{6}, B_4(p^2, m_1^2, m_2^2)_\text{div} =\frac{m_1^2 +m_2^2}{2} \Delta  \nonumber\\
B_5(p^2, m_1^2, m_2^2)_\text{div} = {}&\frac{p^2}{3} \Delta, \quad B'_5 (p^2, m_1^2, m_2^2)_\text{div} =\frac{\Delta}{3},
\label{eq:div}
\end{align}
where $\Delta = \frac{1}{\epsilon} + \ln \mu^2$ with $\mu$ being the renormalization scale.


\section{The SM fitting formula}
\label{sec:fitting}

The SM prediction for the $W$ boson mass, $m_W$, and the leptonic effective mixing angle, $\sin^2 \theta_{W}$, are parameterized by the fitting formula in Ref.~\cite{Awramik:2003rn} and Ref.~\cite{Awramik:2006uz}, respectively. We list them here for readers' convenience. The fitting formula for $m_{W, \text{SM}}$ is given by
\begin{align}
m_{W,\text{SM}} ={}& m_{W}^0 - c_1 dH - c_2 dH^2 +c_3 dH^4 +c_4 (dh-1) -c_5 d\alpha + c_6 dt \nonumber\\
& -c_7 dt^2 - c_8 dH dt +c_9 dh dt - c_{10} d\alpha_s + c_{11} dZ,
\end{align}
where
\begin{alignat}{2}
    dH={}& \ln\left(\frac{m_h}{100~\text{GeV}}\right),\quad&& dh =\left(\frac{m_h}{100~\text{GeV}}\right)^2, \nonumber\\
    dt ={}&\left(\frac{m_t}{174.3~\text{GeV}}\right)^2-1, \quad && dZ = \frac{m_Z}{91.1875~\text{GeV}} -1, \nonumber\\
    d\alpha ={}& \frac{\Delta \alpha}{0.05907}-1, \quad && d\alpha_s = \frac{\alpha_s(m_Z^2)}{0.119}-1
\end{alignat}
and the coefficients are given by
\begin{alignat}{3}
& m_{W}^0 = 80.3799 \gev,  \quad&& c_1 = 0.05263 \gev, \quad&& c_2 = 0.010239 \gev, \nonumber\\
& c_3 = 0.000954 \gev, \quad&& c_4 = -0.000054 \gev, \quad&& c_5 = 1.077 \gev, \nonumber\\
& c_6 = 0.5252 \gev, \quad&& c_7 = 0.0700 \gev, \quad&& c_8 = 0.004102 \gev, \nonumber\\
& c_9 = 0.000111 \gev, \quad&& c_{10} = 0.0774 \gev, \quad&& c_{11} = 115.0 \gev.
\end{alignat}
This fitting formula includes the complete one-loop and two-loop results~\cite{Sirlin:1980nh,Marciano:1980pb,Djouadi:1987gn,Djouadi:1987di,Kniehl:1989yc,Halzen:1990je,Kniehl:1991gu,Kniehl:1992dx,Halzen:1991ik,Freitas:2000gg,Freitas:2002ja,Awramik:2002wn,Awramik:2003ee,Onishchenko:2002ve,Awramik:2002vu,Bauberger:1996ix,Bauberger:1997ey,Awramik:2006uz}. Moreover, partial higher-order corrections up to four-loop order are included~\cite{Avdeev:1994db,Chetyrkin:1995ix,Chetyrkin:1995js,Chetyrkin:1996cf,Faisst:2003px,vanderBij:2000cg,Boughezal:2004ef,Schroder:2005db,Chetyrkin:2006bj,Boughezal:2006xk}.

The fitting formula for $\sin^2 \theta_{\text{eff, SM}}^{\ell}$ is given by
\begin{align}
\sin^2 \theta_{\text{eff, SM}}^{\ell} ={}& s_0 + d_1 L_H + d_2 L_H^2 + d_3 L_H^4 +d_4 (\Delta_H^2-1)+ d_5 \Delta_\alpha +d_6 \Delta_t \nonumber\\
& + d_7 \Delta_t^2 + d_8 \Delta_t (\Delta_H-1)+ d_9 \Delta_{\alpha_s} + d_{10} \Delta_Z,
\end{align}
where
\begin{align}
    L_H={}& \ln\left(\frac{m_h}{100~\text{GeV}}\right),\quad \Delta_H = \frac{m_h}{100~\text{GeV}}, \nonumber\\
    \Delta_\alpha ={}&\frac{\Delta \alpha}{0.05907}-1,\quad \Delta_t = \left(\frac{m_t}{178.0~\text{GeV}}\right)^2-1, \nonumber\\
    \Delta_{\alpha_s} ={}&\frac{\alpha_s (m_Z^2)}{0.117}-1, \quad  \Delta_Z = \frac{m_Z}{91.1876~\text{GeV}}-1
\end{align}
and the coefficients
\begin{alignat}{3}
& s_0 = 0.2312527, \quad&& d_1 = 4.729\times 10^{-4},  \quad&& d_2 = 2.07\times 10^{-5},\nonumber\\
& d_3 = 3.85 \times 10^{-6}, \quad&& d_4 = -1.85\times 10^{-6}, 
\quad&& d_5 = 2.07\times 10^{-2}, \nonumber \\
& d_6 = -2.851\times 10^{-3}, \quad&& d_7 = 1.82\times 10^{-4}, \quad&& d_8 = -9.74 \times 10^{-6}, \nonumber \\
& d_9 =  3.98\times 10^{-4}, \quad&& d_{10} = -0.655. \quad&&
\end{alignat}
This fitting formula is based on the full one-, and two-loop corrections as well as the leading three- and four-loop corrections computed in Refs.~\cite{Avdeev:1994db,Chetyrkin:1995ix,Faisst:2003px,vanderBij:2000cg,Djouadi:1987gn,Djouadi:1987di,Kniehl:1989yc,Halzen:1990je,Kniehl:1992dx,Djouadi:1993ss,Chetyrkin:1995js,Kniehl:1991gu}. The corresponding $\Delta \kappa_\text{SM}$ can then be derived by
\begin{equation}
    \Delta \kappa_\text{SM} = \frac{\sin^2 \theta_{\text{eff, SM}}^{\ell}}{1-m_{W,\text{SM}}^2/m_Z^2}-1.
\end{equation}


\section{Soft \texorpdfstring{$\mathbb{Z}_2$}{Z2} breaking}
\label{app:soft_Z2_breaking}

Here we introduce a soft $\mathbb{Z}_2$ breaking term in the Higgs potential
\begin{equation}
\begin{split}
    \Delta V=\mu\Phi^{\rm T}i\sigma_2\Delta^\dagger \Phi+h.c.
    =-\sqrt{2}\mu Hh^2+\mu(-\sqrt2H^-G^+G^0+H^{--}G^+G^++h.c.)
\end{split}
\end{equation}
where $\mu$ is assumed to be real. 
Such a term breaks the degeneracy between the two neutral states $H,A$, therefore it is expected that $m_H^2-m_A^2\propto \mu^2$. $\Delta V$ generates a non-zero $v_\Delta$:
\begin{equation}
\begin{split}
    v_\Delta\approx& \frac{\sqrt2}{2M^2/v^2+\lambda_{4}+\lambda_5}\mu+\mathcal{O}(\mu^3),\\
v^2_\phi\approx& -\frac{m^2}{\lambda_1}+\mathcal{O}(\mu^2)=v^2-2v_\Delta^2.
\end{split}
\end{equation}
Defining $\epsilon\equiv \sqrt{2}v_\Delta/v\ll1$, the explicit $\mathbb{Z}_2$ breaking mixes the states of $\Phi$ and $\Delta$ with the same quantum numbers at $\mathcal{O}(\epsilon)$. To avoid confusion, we write the weak eigenstates as 
\begin{equation}
\begin{split}
    \Phi'= \left(\begin{array}{c}
         H_1^+ \\
         \frac {v_{\phi}+H_1+iA_1}{\sqrt2}
    \end{array}
    \right)~,~
    \Delta'={}& \left(\begin{array}{c c}
        \frac{H_2^+}{\sqrt2}&H^{++} \\
        \frac{v_\Delta+H_2+iA_2}{\sqrt 2}&-\frac{H_2^+}{\sqrt2}
    \end{array}
    \right)
    \end{split}
\end{equation}
The physical 125~GeV Higgs boson, and the Goldstone bosons that would become the longitudinal $W$ and $Z$ all have small mixtures of the corresponding component of the triplet:
\begin{equation}\label{eq:mixing}
\begin{split}
    h\approx H_1 +\frac{M^2}{M_\Delta^2-2\lambda_1v^2}\sqrt{2}\epsilon H_2~,~
    G^+\approx H_1^++\epsilon H_2^+~,~
    G^0\approx A_1 +\sqrt2\epsilon A_2~,
    \end{split}
\end{equation}
where $M_\Delta^2=M^2+\frac{1}{2}(\lambda_{4}+\lambda_5)v^2$. This mixture allows the mass eigenstates $H$ and $A$ to decay to fermions even though no Yukawa interactions are explicitly introduced in the triplet model.  Up to quadratic order in $\epsilon$, the physical states have masses given by
\begin{equation}
\begin{split}
m_h^2&=2\lambda_1v^2-\left(2\lambda_1v^2+2M^2\frac{M^2}{M_\Delta^2-2\lambda_1v^2}\right)\epsilon^2,\\
m_H^2&=M_\Delta^2+\left(v^2(\lambda_{2}+\lambda_3)+M^2\frac{M^2}{M_\Delta^2-2\lambda_1v^2}\right)\epsilon^2,\\
    m_A^2&=M_\Delta^2+2M_\Delta^2\epsilon^2,\\
    m_{H^+}^2&=M_\Delta^2-\frac{\lambda_5}{4}v^2+M_\Delta^2 \epsilon^2,\\
    m_{H^{++}}^2&=M_\Delta^2-\frac{\lambda_5}{2}v^2+\frac{\lambda_5-\lambda_3}2v^2\epsilon^2.
\end{split}
\end{equation}
From \eqref{eq:mixing}, when $m_H^2\to m_h^2$, the mixing parameter between $H_1$ and $H_2$ diverges. This means that $h$ and $H$ can be maximally mixed even when $\epsilon\ll1$. In this limit, the $H-h$ mixing depends on the details of the Higgs potential parameters.


\section{Unitarity and Vacuum Stability Bounds}
\label{sec:unitarity_stability}

We follow the analysis of vacuum stability and unitarity constraints as given by \cite{Arhrib:2011uy,Aoki:2012jj}.

Demanding perturbative unitarity impose an upper bound on the eigenvalues of the $2\to 2$ scattering matrix
\begin{equation}
    |x_i|< 8\pi,\quad i=1,2,3
\end{equation}
where
\begin{equation}
\begin{split}
    x_1=3\lambda_1+7\lambda_{\Delta}+\sqrt{(3\lambda_1-7\lambda_\Delta)^2+\frac32(2\lambda_4+\lambda_5)^2}~,~
    x_2=\frac12(2\lambda_4+3\lambda_5)~,~
    x_3=\frac12(2\lambda_4-\lambda_5)
    \end{split}.
\end{equation}
Taking $\lambda_\Delta\equiv\lambda_2=\lambda_3>0$, the necessary and sufficient condition for the Higgs potential to be bounded from below is 
\begin{equation}
\begin{split}
    \lambda_1>0,~\lambda_\Delta>0,~
    2\sqrt{2\lambda_1\lambda_\Delta}+\lambda_4+{\rm min}(0,\lambda_5)>0
    \end{split}
\end{equation}
The first two conditions are trivially satisfied. 
In terms of masses of physical states and $M^2$, the last condition can be written as
\begin{equation}
    \frac{m_h}{v}\sqrt{\lambda_\Delta}+\frac{m_{H^{++}}^2-M^2}{v^2}+{\rm min}\left(0,2\frac{m_{H^+}^2-m_{H^{++}}^2}{v^2}\right)>0
\end{equation}
This can be easily satisfied if $\lambda_4\geq 0$ and the mass spectrum is that $m_{A(H)}>m_{H^+}>m_{H^{++}}$. If the spectrum is $m_{A(H)}<m_{H^+}<m_{H^{++}}$, the vacuum stability condition places an upper bound on the mass of $H^{++}$:  
\begin{equation}\label{eq:BFB}
    m_h v\sqrt{\lambda_\Delta}+2m_{H^+}^2-m_{H^{++}}^2>M^2>0
\end{equation}
While boundedness-from-below is only a necessary condition for vacuum stability, we do not expect a second minimum deeper than the electroweak vacuum to exists since we always assume that $v_\Delta \ll v$ recovering approximately the SM vacuum structure.

The conditions in Eq.~\ref{eq:BFB} imposes absolute stability. One can slightly relax this assumption by demanding metastability with a lifetime longer than the age of the Universe. For large field values, the quartic term dominates allowing to solve analytical for the bounce action $B$~\cite{Lee:1985uv}. Following Ref.~\cite{Hollik:2018wrr,Wittbrodt:2019bsu}, we then demand that $B < 440$ resulting in the condition
\begin{align}
    \frac{1}{4}\lambda_1 \cos^4\varphi + (\lambda_2 + \lambda_3)\sin^4\varphi + \frac{1}{8}(\lambda_4 + \lambda_5)\cos^2\varphi\sin^2\varphi < - \frac{\pi^2}{165}\simeq -0.06
\end{align}
for any $\varphi\in[0,2\pi]$.


\section{Landau Pole}
\label{app:landau}

From Sec.~\ref{sec:oneloop}, we see that a larger choice of $m_\text{lightest}$ generically requires a larger value of $\lambda_5$. This generally tells us that the Landau pole could potentially be very close to $m_\text{lightest}$. For our purposes, we will denote the Landau pole as the scale at which the running coupling $\lambda_5(\mu)$ grows to $4\pi$.

Here, we will compute the one-loop beta function for $\lambda_5$. For simplicity, we will only compute the leading $\lambda_5^2$ term. The one-loop counterterm for $\lambda_5$ in $d=4-2\epsilon$ is given by
\begin{equation}
    \delta^{(1)}\lambda_5=\frac{\lambda_5^2}{8\pi^2}\left(\frac{1}{\epsilon}+\text{finite}\right).
\end{equation}
At leading order in $\lambda_5$, the wave functions of $\Phi$ and $\Delta$ are not renormalized at the one-loop level. As such, the one-loop beta function for $\lambda_5$ is simply
\begin{equation}
    \beta_{\lambda_5}= \frac{d\lambda_5}{d\ln\mu}=\frac{\lambda_5^2}{4\pi^2}
\end{equation}
Solving the RGE yields
\begin{equation}\label{eq:landau_pole}
    \Lambda_\text{Landau-pole}=m_\text{lightest}\exp\left(\frac{4\pi^2}{\lambda_5(m_\text{lightest})}-\pi\right).
\end{equation}
The curve for $m_\text{lightest} = 1$~TeV in \figref{fig1} crosses the CDF-II band at $|\lambda_5| \sim 7$. Inserting this value into Eq.~(\ref{eq:landau_pole}) implies a Landau pole at $\sim 12 m_\text{lightest} = 12$~TeV. In this situation, additional BSM physics preventing the appearance of the Landau pole should appear in the multi-TeV range. For a more precise estimate also subleading RGE effects would need to be taken into account.


\section{Decoupling}
\label{app:decoupling}

\begin{figure*}[t]
    \centering
    \includegraphics[width=0.48\textwidth]{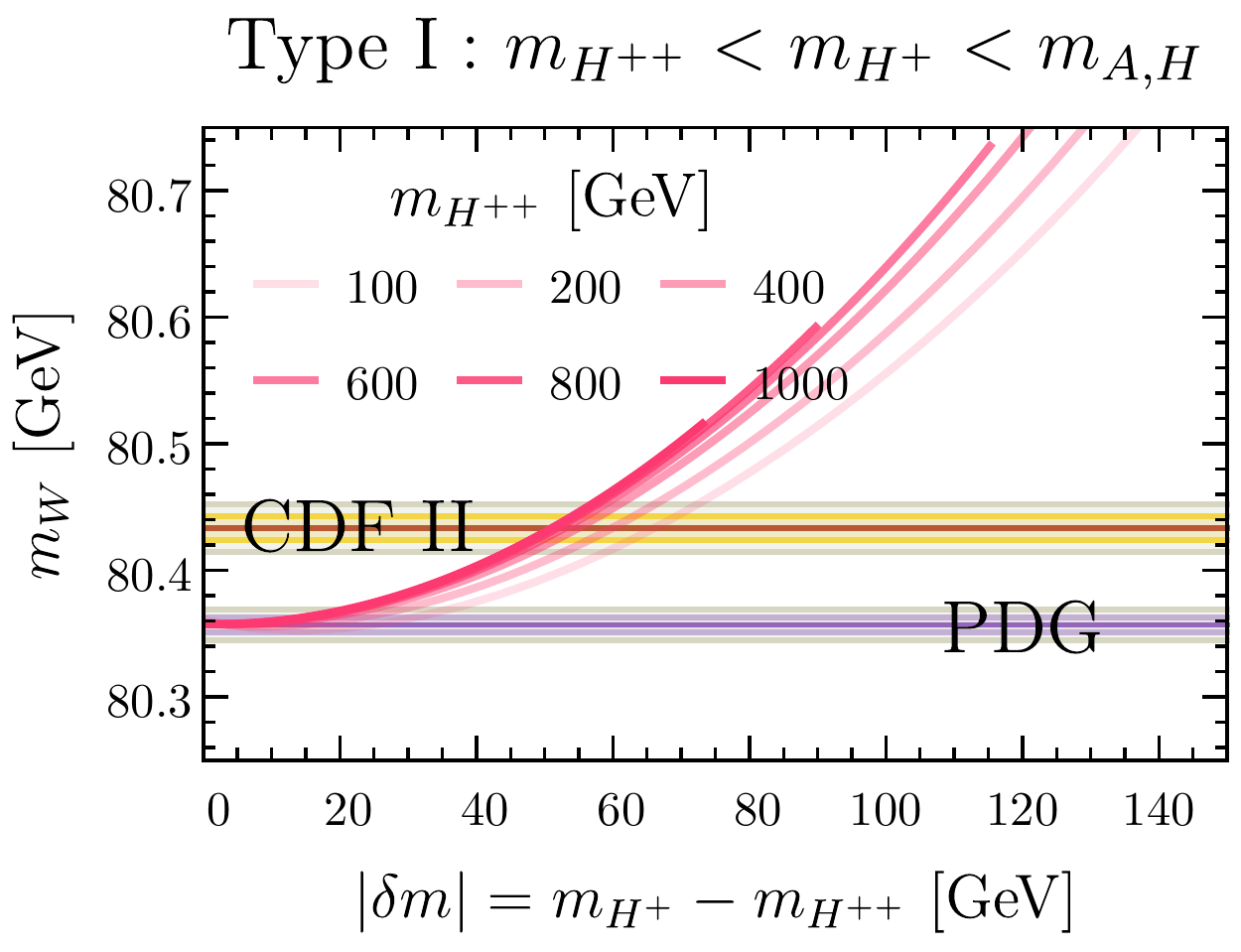}\quad\includegraphics[width=0.48\textwidth]{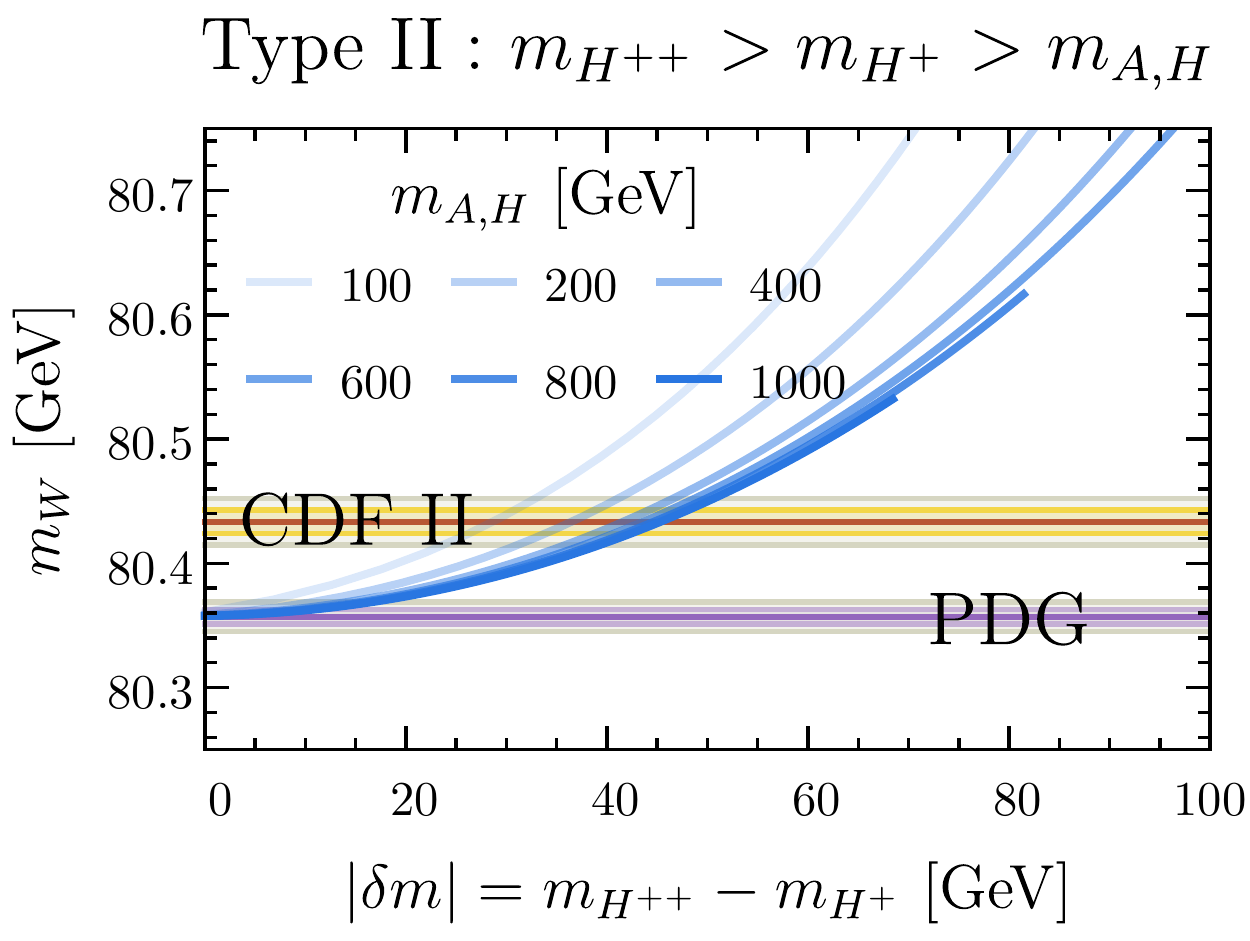}
    \caption{1-loop corrected $W$ boson mass $m_W$ as a function of $|\delta m| = |m_{H^+}- m_{H^{++}}| $ for various masses of the lightest state in the HTM. We assume the mass hierarchy of the new states following $m_{H^{++}}< m_{H^+}< m_A$ ($m_{H^{++}}> m_{H^+}> m_A$) in the left (right) panel. We do not show the parameters for $|\lambda_5|>10$ in drawing the curves. The color bands are the same as those in~\figref{fig1}.}
    \label{fig:figold}
\end{figure*}

In the literature, the one-loop corrected $m_W$ is often presented as a function of $|\delta m| \equiv |m_{H^+} - m_{H^{++}}|$, as shown in~\figref{figold}. At first sight, it is confusing that to reach a given amount of $m_W$ increment, $|\delta m|$ stays almost the same as $m_\text{lightest}$ increases (for the type-I case, it even decreases). Naturally, one would expect that the BSM corrections go to zero in the limit $M\to\infty$. 

To understand such behavior, we should notice that decoupling behavior is only manifest if $\Delta m^2 \equiv m_{H^{\pm}}^2 - m_{H^{\pm\pm}}^2$ or $\lambda_5=4 \Delta m^2 /v^2$ is fixed. If, however, $|\delta m|$ is fixed and $M$ is increased, no decoupling occurs. This happens because
\begin{align}
 |\delta m| = |m_{H^{\pm}} - m_{H^{\pm\pm}}| = 
\left|\sqrt{M^2 - \frac{1}{4}\lambda_5 v^2} - \sqrt{M^2 - \frac{1}{2}\lambda_5 v^2}\right| = \frac{1}{8}|\lambda_5| \frac{v^2}{M} \left(1 + \mathcal{O}\left(\frac{v^2}{M^2}\right)\right)
\end{align}
and therefore $|\lambda_5| \sim M \delta m/v^2$ leads to $|\lambda_5|\to\infty$ in the limit $M\to\infty$. Consequently, this limit will unavoidably violate perturbative unitarity. (We truncated all the curves once $|\lambda_5|$ grows to 10 in~\figref{figold}.)


\bibliographystyle{utphys}
\bibliography{main}

\providecommand{\href}[2]{#2}\begingroup\raggedright\begin{thebibliography}{100}

\bibitem{CDF:2022hxs}
{\bfseries CDF} Collaboration, T.~Aaltonen {\em et~al.}, ``{High-precision
  measurement of the W boson mass with the CDF II detector},''
  \href{http://dx.doi.org/10.1126/science.abk1781}{{\em Science} {\bfseries
  376} no.~6589, (2022) 170--176}.

\bibitem{ParticleDataGroup:2020ssz}
{\bfseries Particle Data Group} Collaboration, P.~A. Zyla {\em et~al.},
  ``{Review of Particle Physics},''
  \href{http://dx.doi.org/10.1093/ptep/ptaa104}{{\em PTEP} {\bfseries 2020}
  no.~8, (2020) 083C01}.

\bibitem{Dorigo}
{Tommaso Dorigo}.
\newblock
  \url{https://www.science20.com/tommaso_dorigo/how_inconsistent_really_are_the_w_mass_measurements-256071}.

\bibitem{Lu:2022bgw}
C.-T. Lu, L.~Wu, Y.~Wu, and B.~Zhu, ``{Electroweak Precision Fit and New
  Physics in light of $W$ Boson Mass},''
  \href{http://arxiv.org/abs/2204.03796}{{\ttfamily arXiv:2204.03796
  [hep-ph]}}.

\bibitem{DiLuzio:2022xns}
L.~Di~Luzio, R.~Gr\"ober, and P.~Paradisi, ``{Higgs physics confronts the $M_W$
  anomaly},'' \href{http://arxiv.org/abs/2204.05284}{{\ttfamily
  arXiv:2204.05284 [hep-ph]}}.

\bibitem{Song:2022xts}
H.~Song, W.~Su, and M.~Zhang, ``{Electroweak Phase Transition in 2HDM under
  Higgs, Z-pole, and W precision measurements},''
  \href{http://arxiv.org/abs/2204.05085}{{\ttfamily arXiv:2204.05085
  [hep-ph]}}.

\bibitem{Sakurai:2022hwh}
K.~Sakurai, F.~Takahashi, and W.~Yin, ``{Singlet extensions and W boson mass in
  the light of the CDF II result},''
  \href{http://arxiv.org/abs/2204.04770}{{\ttfamily arXiv:2204.04770
  [hep-ph]}}.

\bibitem{Cheng:2022jyi}
Y.~Cheng, X.-G. He, Z.-L. Huang, and M.-W. Li, ``{Type-II seesaw triplet scalar
  effects on neutrino trident scattering},''
  \href{http://dx.doi.org/10.1016/j.physletb.2022.137218}{{\em Phys. Lett. B}
  {\bfseries 831} (2022) 137218},
  \href{http://arxiv.org/abs/2204.05031}{{\ttfamily arXiv:2204.05031
  [hep-ph]}}.

\bibitem{Bahl:2022xzi}
H.~Bahl, J.~Braathen, and G.~Weiglein, ``{New physics effects on the $W$-boson
  mass from a doublet extension of the SM Higgs sector},''
  \href{http://arxiv.org/abs/2204.05269}{{\ttfamily arXiv:2204.05269
  [hep-ph]}}.

\bibitem{Heo:2022dey}
Y.~Heo, D.-W. Jung, and J.~S. Lee, ``{Impact of the CDF $W$-mass anomaly on two
  Higgs doublet model},'' \href{http://arxiv.org/abs/2204.05728}{{\ttfamily
  arXiv:2204.05728 [hep-ph]}}.

\bibitem{Biekotter:2022abc}
T.~Biek\"otter, S.~Heinemeyer, and G.~Weiglein, ``{Excesses in the low-mass
  Higgs-boson search and the $W$-boson mass measurement},''
  \href{http://arxiv.org/abs/2204.05975}{{\ttfamily arXiv:2204.05975
  [hep-ph]}}.

\bibitem{Du:2022brr}
X.~K. Du, Z.~Li, F.~Wang, and Y.~K. Zhang, ``{Explaining The New CDF II W-Boson
  Mass Data In The Georgi-Machacek Extension Models},''
  \href{http://arxiv.org/abs/2204.05760}{{\ttfamily arXiv:2204.05760
  [hep-ph]}}.

\bibitem{Han:2022juu}
X.-F. Han, F.~Wang, L.~Wang, J.~M. Yang, and Y.~Zhang, ``{A joint explanation
  of W-mass and muon g-2 in 2HDM},''
  \href{http://arxiv.org/abs/2204.06505}{{\ttfamily arXiv:2204.06505
  [hep-ph]}}.

\bibitem{Ahn:2022xeq}
Y.~H. Ahn, S.~K. Kang, and R.~Ramos, ``{Implications of New CDF-II $W$ Boson
  Mass on Two Higgs Doublet Model},''
  \href{http://arxiv.org/abs/2204.06485}{{\ttfamily arXiv:2204.06485
  [hep-ph]}}.

\bibitem{FileviezPerez:2022lxp}
P.~Fileviez~Perez, H.~H. Patel, and A.~D. Plascencia, ``{On the $W$-mass and
  New Higgs Bosons},'' \href{http://arxiv.org/abs/2204.07144}{{\ttfamily
  arXiv:2204.07144 [hep-ph]}}.

\bibitem{Ghoshal:2022vzo}
A.~Ghoshal, N.~Okada, S.~Okada, D.~Raut, Q.~Shafi, and A.~Thapa, ``{Type III
  seesaw with R-parity violation in light of $m_W$ (CDF)},''
  \href{http://arxiv.org/abs/2204.07138}{{\ttfamily arXiv:2204.07138
  [hep-ph]}}.

\bibitem{Kanemura:2022ahw}
S.~Kanemura and K.~Yagyu, ``{Implication of the W boson mass anomaly at CDF II
  in the Higgs triplet model with a mass difference},''
  \href{http://dx.doi.org/10.1016/j.physletb.2022.137217}{{\em Phys. Lett. B}
  {\bfseries 831} (2022) 137217},
  \href{http://arxiv.org/abs/2204.07511}{{\ttfamily arXiv:2204.07511
  [hep-ph]}}.

\bibitem{Popov:2022ldh}
O.~Popov and R.~Srivastava, ``{The Triplet Dirac Seesaw in the View of the
  Recent CDF-II W Mass Anomaly},''
  \href{http://arxiv.org/abs/2204.08568}{{\ttfamily arXiv:2204.08568
  [hep-ph]}}.

\bibitem{Arcadi:2022dmt}
G.~Arcadi and A.~Djouadi, ``{The 2HD+a model for a combined explanation of the
  possible excesses in the CDF $\mathbf{M_W}$ measurement and
  $\mathbf{(g-2)_\mu}$ with Dark Matter},''
  \href{http://arxiv.org/abs/2204.08406}{{\ttfamily arXiv:2204.08406
  [hep-ph]}}.

\bibitem{Ghorbani:2022vtv}
K.~Ghorbani and P.~Ghorbani, ``{$W$-Boson Mass Anomaly from Scale Invariant
  2HDM},'' \href{http://arxiv.org/abs/2204.09001}{{\ttfamily arXiv:2204.09001
  [hep-ph]}}.

\bibitem{Lee:2022gyf}
S.~Lee, K.~Cheung, J.~Kim, C.-T. Lu, and J.~Song, ``{Status of the
  two-Higgs-doublet model in light of the CDF $m_W$ measurement},''
  \href{http://arxiv.org/abs/2204.10338}{{\ttfamily arXiv:2204.10338
  [hep-ph]}}.

\bibitem{Heeck:2022fvl}
J.~Heeck, ``{W-boson mass in the triplet seesaw model},''
  \href{http://arxiv.org/abs/2204.10274}{{\ttfamily arXiv:2204.10274
  [hep-ph]}}.

\bibitem{Abouabid:2022lpg}
H.~Abouabid, A.~Arhrib, R.~Benbrik, M.~Krab, and M.~Ouchemhou, ``{Is the new
  CDF $M_W$ measurement consistent with the two higgs doublet model?},''
  \href{http://arxiv.org/abs/2204.12018}{{\ttfamily arXiv:2204.12018
  [hep-ph]}}.

\bibitem{Benbrik:2022dja}
R.~Benbrik, M.~Boukidi, and B.~Manaut, ``{$W$-mass and 96 GeV excess in
  type-III 2HDM},'' \href{http://arxiv.org/abs/2204.11755}{{\ttfamily
  arXiv:2204.11755 [hep-ph]}}.

\bibitem{Kim:2022hvh}
J.~Kim, S.~Lee, P.~Sanyal, and J.~Song, ``{CDF $W$ boson mass and muon $g-2$ in
  type-X two-Higgs-doublet model with a Higgs-phobic light pseudoscalar},''
  \href{http://arxiv.org/abs/2205.01701}{{\ttfamily arXiv:2205.01701
  [hep-ph]}}.

\bibitem{Atkinson:2022qnl}
O.~Atkinson, M.~Black, C.~Englert, A.~Lenz, and A.~Rusov, ``{MUonE, muon $g-2$
  and electroweak precision constraints within 2HDMs},''
  \href{http://arxiv.org/abs/2207.02789}{{\ttfamily arXiv:2207.02789
  [hep-ph]}}.

\bibitem{Strumia:2022qkt}
A.~Strumia, ``{Interpreting electroweak precision data including the $W$-mass
  CDF anomaly},'' \href{http://arxiv.org/abs/2204.04191}{{\ttfamily
  arXiv:2204.04191 [hep-ph]}}.

\bibitem{deBlas:2022hdk}
J.~de~Blas, M.~Pierini, L.~Reina, and L.~Silvestrini, ``{Impact of the recent
  measurements of the top-quark and W-boson masses on electroweak precision
  fits},'' \href{http://arxiv.org/abs/2204.04204}{{\ttfamily arXiv:2204.04204
  [hep-ph]}}.

\bibitem{Yang:2022gvz}
J.~M. Yang and Y.~Zhang, ``{Low energy SUSY confronted with new measurements of
  W-boson mass and muon g-2},''
  \href{http://arxiv.org/abs/2204.04202}{{\ttfamily arXiv:2204.04202
  [hep-ph]}}.

\bibitem{Yuan:2022cpw}
G.-W. Yuan, L.~Zu, L.~Feng, Y.-F. Cai, and Y.-Z. Fan, ``{Hint on new physics
  from the $W$-boson mass excess$-$axion-like particle, dark photon or
  Chameleon dark energy},'' \href{http://arxiv.org/abs/2204.04183}{{\ttfamily
  arXiv:2204.04183 [hep-ph]}}.

\bibitem{Athron:2022qpo}
P.~Athron, A.~Fowlie, C.-T. Lu, L.~Wu, Y.~Wu, and B.~Zhu, ``{The $W$ boson Mass
  and Muon $g-2$: Hadronic Uncertainties or New Physics?},''
  \href{http://arxiv.org/abs/2204.03996}{{\ttfamily arXiv:2204.03996
  [hep-ph]}}.

\bibitem{Fan:2022dck}
Y.-Z. Fan, T.-P. Tang, Y.-L.~S. Tsai, and L.~Wu, ``{Inert Higgs Dark Matter for
  New CDF W-boson Mass and Detection Prospects},''
  \href{http://arxiv.org/abs/2204.03693}{{\ttfamily arXiv:2204.03693
  [hep-ph]}}.

\bibitem{Babu:2022pdn}
K.~S. Babu, S.~Jana, and V.~P. K., ``{Correlating $W$-Boson Mass Shift with
  Muon ${g-2}$ in the 2HDM},''
  \href{http://arxiv.org/abs/2204.05303}{{\ttfamily arXiv:2204.05303
  [hep-ph]}}.

\bibitem{Heckman:2022the}
J.~J. Heckman, ``{Extra $W$-Boson Mass from a D3-Brane},''
  \href{http://arxiv.org/abs/2204.05302}{{\ttfamily arXiv:2204.05302
  [hep-ph]}}.

\bibitem{Gu:2022htv}
J.~Gu, Z.~Liu, T.~Ma, and J.~Shu, ``{Speculations on the W-Mass Measurement at
  CDF},'' \href{http://arxiv.org/abs/2204.05296}{{\ttfamily arXiv:2204.05296
  [hep-ph]}}.

\bibitem{Athron:2022isz}
P.~Athron, M.~Bach, D.~H.~J. Jacob, W.~Kotlarski, D.~St\"ockinger, and
  A.~Voigt, ``{Precise calculation of the W boson pole mass beyond the Standard
  Model with FlexibleSUSY},'' \href{http://arxiv.org/abs/2204.05285}{{\ttfamily
  arXiv:2204.05285 [hep-ph]}}.

\bibitem{Asadi:2022xiy}
P.~Asadi, C.~Cesarotti, K.~Fraser, S.~Homiller, and A.~Parikh, ``{Oblique
  Lessons from the $W$ Mass Measurement at CDF II},''
  \href{http://arxiv.org/abs/2204.05283}{{\ttfamily arXiv:2204.05283
  [hep-ph]}}.

\bibitem{Paul:2022dds}
A.~Paul and M.~Valli, ``{Violation of custodial symmetry from W-boson mass
  measurements},'' \href{http://arxiv.org/abs/2204.05267}{{\ttfamily
  arXiv:2204.05267 [hep-ph]}}.

\bibitem{Bagnaschi:2022whn}
E.~Bagnaschi, J.~Ellis, M.~Madigan, K.~Mimasu, V.~Sanz, and T.~You, ``{SMEFT
  Analysis of $m_{W}$},'' \href{http://arxiv.org/abs/2204.05260}{{\ttfamily
  arXiv:2204.05260 [hep-ph]}}.

\bibitem{Lee:2022nqz}
H.~M. Lee and K.~Yamashita, ``{A Model of Vector-like Leptons for the Muon
  $g-2$ and the $W$ Boson Mass},''
  \href{http://arxiv.org/abs/2204.05024}{{\ttfamily arXiv:2204.05024
  [hep-ph]}}.

\bibitem{Liu:2022jdq}
X.~Liu, S.-Y. Guo, B.~Zhu, and Y.~Li, ``{Unifying gravitational waves with $W$
  boson, FIMP dark matter, and Majorana Seesaw mechanism},''
  \href{http://arxiv.org/abs/2204.04834}{{\ttfamily arXiv:2204.04834
  [hep-ph]}}.

\bibitem{Fan:2022yly}
J.~Fan, L.~Li, T.~Liu, and K.-F. Lyu, ``{$W$-Boson Mass, Electroweak Precision
  Tests and SMEFT},'' \href{http://arxiv.org/abs/2204.04805}{{\ttfamily
  arXiv:2204.04805 [hep-ph]}}.

\bibitem{Balkin:2022glu}
R.~Balkin, E.~Madge, T.~Menzo, G.~Perez, Y.~Soreq, and J.~Zupan, ``{On the
  implications of positive W mass shift},''
  \href{http://dx.doi.org/10.1007/JHEP05(2022)133}{{\em JHEP} {\bfseries 05}
  (2022) 133}, \href{http://arxiv.org/abs/2204.05992}{{\ttfamily
  arXiv:2204.05992 [hep-ph]}}.

\bibitem{Endo:2022kiw}
M.~Endo and S.~Mishima, ``{New physics interpretation of $W$-boson mass
  anomaly},'' \href{http://arxiv.org/abs/2204.05965}{{\ttfamily
  arXiv:2204.05965 [hep-ph]}}.

\bibitem{Crivellin:2022fdf}
A.~Crivellin, M.~Kirk, T.~Kitahara, and F.~Mescia, ``{Correlating $t\to cZ$ to
  the $W$ Mass and $B$ Physics with Vector-Like Quarks},''
  \href{http://arxiv.org/abs/2204.05962}{{\ttfamily arXiv:2204.05962
  [hep-ph]}}.

\bibitem{Blennow:2022yfm}
M.~Blennow, P.~Coloma, E.~Fern\'andez-Mart\'\i{}nez, and M.~Gonz\'alez-L\'opez,
  ``{Right-handed neutrinos and the CDF II anomaly},''
  \href{http://arxiv.org/abs/2204.04559}{{\ttfamily arXiv:2204.04559
  [hep-ph]}}.

\bibitem{Cacciapaglia:2022xih}
G.~Cacciapaglia and F.~Sannino, ``{The W boson mass weighs in on the
  non-standard Higgs},''
  \href{http://dx.doi.org/10.1016/j.physletb.2022.137232}{{\em Phys. Lett. B}
  {\bfseries 832} (2022) 137232},
  \href{http://arxiv.org/abs/2204.04514}{{\ttfamily arXiv:2204.04514
  [hep-ph]}}.

\bibitem{Tang:2022pxh}
T.-P. Tang, M.~Abdughani, L.~Feng, Y.-L.~S. Tsai, J.~Wu, and Y.-Z. Fan,
  ``{NMSSM neutralino dark matter for $W$-boson mass and muon $g-2$ and the
  promising prospect of direct detection},''
  \href{http://arxiv.org/abs/2204.04356}{{\ttfamily arXiv:2204.04356
  [hep-ph]}}.

\bibitem{Zhu:2022tpr}
C.-R. Zhu, M.-Y. Cui, Z.-Q. Xia, Z.-H. Yu, X.~Huang, Q.~Yuan, and Y.~Z. Fan,
  ``{GeV antiproton/gamma-ray excesses and the $W$-boson mass anomaly: three
  faces of $\sim 60-70$ GeV dark matter particle?},''
  \href{http://arxiv.org/abs/2204.03767}{{\ttfamily arXiv:2204.03767
  [astro-ph.HE]}}.

\bibitem{Zheng:2022irz}
M.-D. Zheng, F.-Z. Chen, and H.-H. Zhang, ``{The $W\ell\nu$-vertex corrections
  to W-boson mass in the R-parity violating MSSM},''
  \href{http://arxiv.org/abs/2204.06541}{{\ttfamily arXiv:2204.06541
  [hep-ph]}}.

\bibitem{Krasnikov:2022xsi}
N.~V. Krasnikov, ``{Nonlocal generalization of the SM as an explanation of
  recent CDF result},'' \href{http://arxiv.org/abs/2204.06327}{{\ttfamily
  arXiv:2204.06327 [hep-ph]}}.

\bibitem{Arias-Aragon:2022ats}
F.~Arias-Arag\'on, E.~Fern\'andez-Mart\'\i{}nez, M.~Gonz\'alez-L\'opez, and
  L.~Merlo, ``{Dynamical Minimal Flavour Violating Inverse Seesaw},''
  \href{http://arxiv.org/abs/2204.04672}{{\ttfamily arXiv:2204.04672
  [hep-ph]}}.

\bibitem{Du:2022pbp}
X.~K. Du, Z.~Li, F.~Wang, and Y.~K. Zhang, ``{Explaining The Muon $g-2$ Anomaly
  and New CDF II W-Boson Mass in the Framework of (Extra)Ordinary Gauge
  Mediation},'' \href{http://arxiv.org/abs/2204.04286}{{\ttfamily
  arXiv:2204.04286 [hep-ph]}}.

\bibitem{Kawamura:2022uft}
J.~Kawamura, S.~Okawa, and Y.~Omura, ``{$W$ boson mass and muon $g-2$ in a
  lepton portal dark matter model},''
  \href{http://arxiv.org/abs/2204.07022}{{\ttfamily arXiv:2204.07022
  [hep-ph]}}.

\bibitem{Nagao:2022oin}
K.~I. Nagao, T.~Nomura, and H.~Okada, ``{A model explaining the new CDF II W
  boson mass linking to muon $g-2$ and dark matter},''
  \href{http://arxiv.org/abs/2204.07411}{{\ttfamily arXiv:2204.07411
  [hep-ph]}}.

\bibitem{Zhang:2022nnh}
K.-Y. Zhang and W.-Z. Feng, ``{Explaining $W$ boson mass anomaly and dark
  matter with a $U(1)$ dark sector},''
  \href{http://arxiv.org/abs/2204.08067}{{\ttfamily arXiv:2204.08067
  [hep-ph]}}.

\bibitem{Carpenter:2022oyg}
L.~M. Carpenter, T.~Murphy, and M.~J. Smylie, ``{Changing patterns in
  electroweak precision with new color-charged states: Oblique corrections and
  the $W$ boson mass},'' \href{http://arxiv.org/abs/2204.08546}{{\ttfamily
  arXiv:2204.08546 [hep-ph]}}.

\bibitem{Senjanovic:2022zwy}
G.~Senjanovi\'c and M.~Zantedeschi, ``{$SU(5)$ grand unification and $W$-boson
  mass},'' \href{http://arxiv.org/abs/2205.05022}{{\ttfamily arXiv:2205.05022
  [hep-ph]}}.

\bibitem{Chowdhury:2022moc}
T.~A. Chowdhury, J.~Heeck, S.~Saad, and A.~Thapa, ``{$W$ boson mass shift and
  muon magnetic moment in the Zee model},''
  \href{http://arxiv.org/abs/2204.08390}{{\ttfamily arXiv:2204.08390
  [hep-ph]}}.

\bibitem{Borah:2022obi}
D.~Borah, S.~Mahapatra, D.~Nanda, and N.~Sahu, ``{Type II Dirac Seesaw with
  Observable $\Delta N_{eff}$ in the light of W-mass Anomaly},''
  \href{http://arxiv.org/abs/2204.08266}{{\ttfamily arXiv:2204.08266
  [hep-ph]}}.

\bibitem{Zeng:2022lkk}
Y.-P. Zeng, C.~Cai, Y.-H. Su, and H.-H. Zhang, ``{Extra boson mix with Z boson
  explaining the mass of W boson},''
  \href{http://arxiv.org/abs/2204.09487}{{\ttfamily arXiv:2204.09487
  [hep-ph]}}.

\bibitem{Du:2022fqv}
M.~Du, Z.~Liu, and P.~Nath, ``{CDF W mass anomaly from a dark sector with a
  Stueckelberg-Higgs portal},''
  \href{http://arxiv.org/abs/2204.09024}{{\ttfamily arXiv:2204.09024
  [hep-ph]}}.

\bibitem{Bhaskar:2022vgk}
A.~Bhaskar, A.~A. Madathil, T.~Mandal, and S.~Mitra, ``{Combined explanation of
  $W$-mass, muon $g-2$, $R_{K^{(*)}}$ and $R_{D^{(*)}}$ anomalies in a
  singlet-triplet scalar leptoquark model},''
  \href{http://arxiv.org/abs/2204.09031}{{\ttfamily arXiv:2204.09031
  [hep-ph]}}.

\bibitem{Baek:2022agi}
S.~Baek, ``{Implications of CDF $W$-mass and $(g-2)_\mu$ on
  $U(1)_{L_\mu-L_\tau}$ model},''
  \href{http://arxiv.org/abs/2204.09585}{{\ttfamily arXiv:2204.09585
  [hep-ph]}}.

\bibitem{Cao:2022mif}
J.~Cao, L.~Meng, L.~Shang, S.~Wang, and B.~Yang, ``{Interpreting the $W$ mass
  anomaly in the vectorlike quark models},''
  \href{http://arxiv.org/abs/2204.09477}{{\ttfamily arXiv:2204.09477
  [hep-ph]}}.

\bibitem{Borah:2022zim}
D.~Borah, S.~Mahapatra, and N.~Sahu, ``{Singlet-doublet fermion origin of dark
  matter, neutrino mass and W-mass anomaly},''
  \href{http://dx.doi.org/10.1016/j.physletb.2022.137196}{{\em Phys. Lett. B}
  {\bfseries 831} (2022) 137196},
  \href{http://arxiv.org/abs/2204.09671}{{\ttfamily arXiv:2204.09671
  [hep-ph]}}.

\bibitem{Batra:2022org}
A.~Batra, S.~K.~A., S.~Mandal, and R.~Srivastava, ``{W boson mass in
  Singlet-Triplet Scotogenic dark matter model},''
  \href{http://arxiv.org/abs/2204.09376}{{\ttfamily arXiv:2204.09376
  [hep-ph]}}.

\bibitem{Almeida:2022lcs}
E.~d.~S. Almeida, A.~Alves, O.~J.~P. Eboli, and M.~C. Gonzalez-Garcia,
  ``{Impact of CDF-II measurement of $M_W$ on the electroweak legacy of the LHC
  Run II},'' \href{http://arxiv.org/abs/2204.10130}{{\ttfamily arXiv:2204.10130
  [hep-ph]}}.

\bibitem{Cheng:2022aau}
Y.~Cheng, X.-G. He, F.~Huang, J.~Sun, and Z.-P. Xing, ``{Dark photon kinetic
  mixing effects for CDF W mass excess},''
  \href{http://arxiv.org/abs/2204.10156}{{\ttfamily arXiv:2204.10156
  [hep-ph]}}.

\bibitem{Batra:2022pej}
A.~Batra, S.~K. A, S.~Mandal, H.~Prajapati, and R.~Srivastava, ``{CDF-II $W$
  Boson Mass Anomaly in the Canonical Scotogenic Neutrino-Dark Matter Model},''
  \href{http://arxiv.org/abs/2204.11945}{{\ttfamily arXiv:2204.11945
  [hep-ph]}}.

\bibitem{Cai:2022cti}
C.~Cai, D.~Qiu, Y.-L. Tang, Z.-H. Yu, and H.-H. Zhang, ``{Corrections to
  electroweak precision observables from mixings of an exotic vector boson in
  light of the CDF $W$-mass anomaly},''
  \href{http://arxiv.org/abs/2204.11570}{{\ttfamily arXiv:2204.11570
  [hep-ph]}}.

\bibitem{Zhou:2022cql}
Q.~Zhou and X.-F. Han, ``{The CDF W-mass, muon g-2, and dark matter in a
  $U(1)_{L_\mu-L_\tau}$ model with vector-like leptons},''
  \href{http://arxiv.org/abs/2204.13027}{{\ttfamily arXiv:2204.13027
  [hep-ph]}}.

\bibitem{Gupta:2022lrt}
R.~S. Gupta, ``{Running away from the T-parameter solution to the W mass
  anomaly},'' \href{http://arxiv.org/abs/2204.13690}{{\ttfamily
  arXiv:2204.13690 [hep-ph]}}.

\bibitem{Wang:2022dte}
J.-W. Wang, X.-J. Bi, P.-F. Yin, and Z.-H. Yu, ``{Electroweak dark matter model
  accounting for the CDF $W$-mass anomaly},''
  \href{http://arxiv.org/abs/2205.00783}{{\ttfamily arXiv:2205.00783
  [hep-ph]}}.

\bibitem{Barman:2022qix}
B.~Barman, A.~Das, and S.~Sengupta, ``{New $W$-Boson mass in the light of
  doubly warped braneworld model},''
  \href{http://arxiv.org/abs/2205.01699}{{\ttfamily arXiv:2205.01699
  [hep-ph]}}.

\bibitem{Kim:2022xuo}
J.~Kim, ``{Compatibility of muon g-2, W mass anomaly in type-X 2HDM},''
  \href{http://dx.doi.org/10.1016/j.physletb.2022.137220}{{\em Phys. Lett. B}
  {\bfseries 832} (2022) 137220},
  \href{http://arxiv.org/abs/2205.01437}{{\ttfamily arXiv:2205.01437
  [hep-ph]}}.

\bibitem{Dcruz:2022dao}
R.~Dcruz and A.~Thapa, ``{$W$ boson mass, dark matter and $(g-2)_\ell$ in
  ScotoZee neutrino mass model},''
  \href{http://arxiv.org/abs/2205.02217}{{\ttfamily arXiv:2205.02217
  [hep-ph]}}.

\bibitem{Isaacson:2022rts}
J.~Isaacson, Y.~Fu, and C.~P. Yuan, ``{ResBos2 and the CDF W Mass
  Measurement},'' \href{http://arxiv.org/abs/2205.02788}{{\ttfamily
  arXiv:2205.02788 [hep-ph]}}.

\bibitem{Chowdhury:2022dps}
T.~A. Chowdhury and S.~Saad, ``{Leptoquark-vectorlike quark model for $m_W$
  (CDF), $(g-2)_\mu$, $R_{K^{(\ast)}}$ anomalies and neutrino mass},''
  \href{http://arxiv.org/abs/2205.03917}{{\ttfamily arXiv:2205.03917
  [hep-ph]}}.

\bibitem{Kim:2022zhj}
S.-S. Kim, H.~M. Lee, A.~G. Menkara, and K.~Yamashita, ``{The $SU(2)_D$ lepton
  portals for muon $g-2$, $W$ boson mass and dark matter},''
  \href{http://arxiv.org/abs/2205.04016}{{\ttfamily arXiv:2205.04016
  [hep-ph]}}.

\bibitem{Gao:2022wxk}
J.~Gao, D.~Liu, and K.~Xie, ``{Understanding PDF uncertainty on the $W$ boson
  mass measurements in CT18 global analysis},''
  \href{http://arxiv.org/abs/2205.03942}{{\ttfamily arXiv:2205.03942
  [hep-ph]}}.

\bibitem{Lazarides:2022spe}
G.~Lazarides, R.~Maji, R.~Roshan, and Q.~Shafi, ``{Heavier $W$-boson, dark
  matter and gravitational waves from strings in an SO(10) axion model},''
  \href{http://arxiv.org/abs/2205.04824}{{\ttfamily arXiv:2205.04824
  [hep-ph]}}.

\bibitem{Rizzo:2022jti}
T.~G. Rizzo, ``{Kinetic Mixing, Dark Higgs Triplets, $M_W$ and All That},''
  \href{http://arxiv.org/abs/2206.09814}{{\ttfamily arXiv:2206.09814
  [hep-ph]}}.

\bibitem{VanLoi:2022eir}
D.~Van~Loi and P.~Van~Dong, ``{Novel effects of the $W$-boson mass shift in the
  3-3-1 model},'' \href{http://arxiv.org/abs/2206.10100}{{\ttfamily
  arXiv:2206.10100 [hep-ph]}}.

\bibitem{YaserAyazi:2022tbn}
S.~Yaser~Ayazi and M.~Hosseini, ``{W boson mass anomaly and vacuum structure in
  vector dark matter model with a singlet scalar mediator},''
  \href{http://arxiv.org/abs/2206.11041}{{\ttfamily arXiv:2206.11041
  [hep-ph]}}.

\bibitem{Chakrabarty:2022voz}
N.~Chakrabarty, ``{The muon $g-2$ and $W$-mass anomalies explained and the
  electroweak vacuum stabilised by extending the minimal Type-II seesaw},''
  \href{http://arxiv.org/abs/2206.11771}{{\ttfamily arXiv:2206.11771
  [hep-ph]}}.

\bibitem{CentellesChulia:2022vpz}
S.~Centelles~Chuli\'a, R.~Srivastava, and S.~Yadav, ``{CDF-II W boson mass in
  the Dirac Scotogenic model},''
  \href{http://arxiv.org/abs/2206.11903}{{\ttfamily arXiv:2206.11903
  [hep-ph]}}.

\bibitem{Nagao:2022dgl}
K.~I. Nagao, T.~Nomura, and H.~Okada, ``{An alternative gauged $U(1)_R$
  symmetric model in light of the CDF II $W$ boson mass anomaly},''
  \href{http://arxiv.org/abs/2206.15256}{{\ttfamily arXiv:2206.15256
  [hep-ph]}}.

\bibitem{Aoki:2012jj}
M.~Aoki, S.~Kanemura, M.~Kikuchi, and K.~Yagyu, ``{Radiative corrections to the
  Higgs boson couplings in the triplet model},''
  \href{http://dx.doi.org/10.1103/PhysRevD.87.015012}{{\em Phys. Rev. D}
  {\bfseries 87} no.~1, (2013) 015012},
  \href{http://arxiv.org/abs/1211.6029}{{\ttfamily arXiv:1211.6029 [hep-ph]}}.

\bibitem{Steinhauser:1998rq}
M.~Steinhauser, ``{Leptonic contribution to the effective electromagnetic
  coupling constant up to three loops},''
  \href{http://dx.doi.org/10.1016/S0370-2693(98)00503-6}{{\em Phys. Lett. B}
  {\bfseries 429} (1998) 158--161},
  \href{http://arxiv.org/abs/hep-ph/9803313}{{\ttfamily arXiv:hep-ph/9803313}}.

\bibitem{Bohm:2001yx}
M.~Bohm, A.~Denner, and H.~Joos,
  \href{http://dx.doi.org/10.1007/978-3-322-80160-9}{{\em {Gauge theories of
  the strong and electroweak interaction}}}.
\newblock 2001.

\bibitem{Hessenberger:2018xzo}
S.~Hessenberger, {\em {Two-loop corrections to electroweak precision
  observables in Two-Higgs-Doublet-Models}}.
\newblock PhD thesis, Munich, Tech. U., 2018.

\bibitem{Awramik:2003rn}
M.~Awramik, M.~Czakon, A.~Freitas, and G.~Weiglein, ``{Precise prediction for
  the W boson mass in the standard model},''
  \href{http://dx.doi.org/10.1103/PhysRevD.69.053006}{{\em Phys. Rev. D}
  {\bfseries 69} (2004) 053006},
  \href{http://arxiv.org/abs/hep-ph/0311148}{{\ttfamily arXiv:hep-ph/0311148}}.

\bibitem{Kanemura:2014goa}
S.~Kanemura, M.~Kikuchi, K.~Yagyu, and H.~Yokoya, ``{Bounds on the mass of
  doubly-charged Higgs bosons in the same-sign diboson decay scenario},''
  \href{http://dx.doi.org/10.1103/PhysRevD.90.115018}{{\em Phys. Rev. D}
  {\bfseries 90} no.~11, (2014) 115018},
  \href{http://arxiv.org/abs/1407.6547}{{\ttfamily arXiv:1407.6547 [hep-ph]}}.

\bibitem{ALEPH:2005ab}
{\bfseries ALEPH, DELPHI, L3, OPAL, SLD, LEP Electroweak Working Group, SLD
  Electroweak Group, SLD Heavy Flavour Group} Collaboration, S.~Schael {\em
  et~al.}, ``{Precision electroweak measurements on the $Z$ resonance},''
  \href{http://dx.doi.org/10.1016/j.physrep.2005.12.006}{{\em Phys. Rept.}
  {\bfseries 427} (2006) 257--454},
  \href{http://arxiv.org/abs/hep-ex/0509008}{{\ttfamily arXiv:hep-ex/0509008}}.

\bibitem{Awramik:2006uz}
M.~Awramik, M.~Czakon, and A.~Freitas, ``{Electroweak two-loop corrections to
  the effective weak mixing angle},''
  \href{http://dx.doi.org/10.1088/1126-6708/2006/11/048}{{\em JHEP} {\bfseries
  11} (2006) 048}, \href{http://arxiv.org/abs/hep-ph/0608099}{{\ttfamily
  arXiv:hep-ph/0608099}}.

\bibitem{Bechtle:2013xfa}
P.~Bechtle, S.~Heinemeyer, O.~St\r{a}l, T.~Stefaniak, and G.~Weiglein,
  ``{$HiggsSignals$: Confronting arbitrary Higgs sectors with measurements at
  the Tevatron and the LHC},''
  \href{http://dx.doi.org/10.1140/epjc/s10052-013-2711-4}{{\em Eur. Phys. J. C}
  {\bfseries 74} no.~2, (2014) 2711},
  \href{http://arxiv.org/abs/1305.1933}{{\ttfamily arXiv:1305.1933 [hep-ph]}}.

\bibitem{Bechtle:2020uwn}
P.~Bechtle, S.~Heinemeyer, T.~Klingl, T.~Stefaniak, G.~Weiglein, and
  J.~Wittbrodt, ``{HiggsSignals-2: Probing new physics with precision Higgs
  measurements in the LHC 13 TeV era},''
  \href{http://dx.doi.org/10.1140/epjc/s10052-021-08942-y}{{\em Eur. Phys. J.
  C} {\bfseries 81} no.~2, (2021) 145},
  \href{http://arxiv.org/abs/2012.09197}{{\ttfamily arXiv:2012.09197
  [hep-ph]}}.

\bibitem{Bahl:2022jnx}
H.~Bahl, J.~Braathen, and G.~Weiglein, ``{New constraints on extended Higgs
  sectors from the trilinear Higgs coupling},''
  \href{http://arxiv.org/abs/2202.03453}{{\ttfamily arXiv:2202.03453
  [hep-ph]}}.

\bibitem{Hahn:2000kx}
T.~Hahn, ``{Generating Feynman diagrams and amplitudes with FeynArts 3},''
  \href{http://dx.doi.org/10.1016/S0010-4655(01)00290-9}{{\em Comput. Phys.
  Commun.} {\bfseries 140} (2001) 418--431},
  \href{http://arxiv.org/abs/hep-ph/0012260}{{\ttfamily arXiv:hep-ph/0012260}}.

\bibitem{Hahn:1998yk}
T.~Hahn and M.~Perez-Victoria, ``{Automatized one loop calculations in
  four-dimensions and D-dimensions},''
  \href{http://dx.doi.org/10.1016/S0010-4655(98)00173-8}{{\em Comput. Phys.
  Commun.} {\bfseries 118} (1999) 153--165},
  \href{http://arxiv.org/abs/hep-ph/9807565}{{\ttfamily arXiv:hep-ph/9807565}}.

\bibitem{Christensen:2008py}
N.~D. Christensen and C.~Duhr, ``{FeynRules - Feynman rules made easy},''
  \href{http://dx.doi.org/10.1016/j.cpc.2009.02.018}{{\em Comput. Phys.
  Commun.} {\bfseries 180} (2009) 1614--1641},
  \href{http://arxiv.org/abs/0806.4194}{{\ttfamily arXiv:0806.4194 [hep-ph]}}.

\bibitem{Alloul:2013bka}
A.~Alloul, N.~D. Christensen, C.~Degrande, C.~Duhr, and B.~Fuks, ``{FeynRules
  2.0 - A complete toolbox for tree-level phenomenology},''
  \href{http://dx.doi.org/10.1016/j.cpc.2014.04.012}{{\em Comput. Phys.
  Commun.} {\bfseries 185} (2014) 2250--2300},
  \href{http://arxiv.org/abs/1310.1921}{{\ttfamily arXiv:1310.1921 [hep-ph]}}.

\bibitem{ATLAS:2021tyg}
{\bfseries ATLAS} Collaboration, ``{Combination of searches for non-resonant
  and resonant Higgs boson pair production in the $b\bar{b}\gamma\gamma$,
  $b\bar{b}\tau^{+}\tau^{-}$ and $b\bar{b}b\bar{b}$ decay channels using $pp$
  collisions at $\sqrt{s}$ = 13 TeV with the ATLAS detector},''.

\bibitem{ATLAS:2019nkf}
{\bfseries ATLAS} Collaboration, G.~Aad {\em et~al.}, ``{Combined measurements
  of Higgs boson production and decay using up to $80$ fb$^{-1}$ of
  proton-proton collision data at $\sqrt{s}=$ 13 TeV collected with the ATLAS
  experiment},'' \href{http://dx.doi.org/10.1103/PhysRevD.101.012002}{{\em
  Phys. Rev. D} {\bfseries 101} no.~1, (2020) 012002},
  \href{http://arxiv.org/abs/1909.02845}{{\ttfamily arXiv:1909.02845
  [hep-ex]}}.

\bibitem{CMS:2018uag}
{\bfseries CMS} Collaboration, A.~M. Sirunyan {\em et~al.}, ``{Combined
  measurements of Higgs boson couplings in proton\textendash{}proton collisions
  at $\sqrt{s}=13\,\text {Te}\text {V} $},''
  \href{http://dx.doi.org/10.1140/epjc/s10052-019-6909-y}{{\em Eur. Phys. J. C}
  {\bfseries 79} no.~5, (2019) 421},
  \href{http://arxiv.org/abs/1809.10733}{{\ttfamily arXiv:1809.10733
  [hep-ex]}}.

\bibitem{Fuks:2019clu}
B.~Fuks, M.~Nemev\v{s}ek, and R.~Ruiz, ``{Doubly Charged Higgs Boson Production
  at Hadron Colliders},''
  \href{http://dx.doi.org/10.1103/PhysRevD.101.075022}{{\em Phys. Rev. D}
  {\bfseries 101} no.~7, (2020) 075022},
  \href{http://arxiv.org/abs/1912.08975}{{\ttfamily arXiv:1912.08975
  [hep-ph]}}.

\bibitem{Alwall:2014hca}
J.~Alwall, R.~Frederix, S.~Frixione, V.~Hirschi, F.~Maltoni, O.~Mattelaer,
  H.~S. Shao, T.~Stelzer, P.~Torrielli, and M.~Zaro, ``{The automated
  computation of tree-level and next-to-leading order differential cross
  sections, and their matching to parton shower simulations},''
  \href{http://dx.doi.org/10.1007/JHEP07(2014)079}{{\em JHEP} {\bfseries 07}
  (2014) 079}, \href{http://arxiv.org/abs/1405.0301}{{\ttfamily arXiv:1405.0301
  [hep-ph]}}.

\bibitem{Dercks:2016npn}
D.~Dercks, N.~Desai, J.~S. Kim, K.~Rolbiecki, J.~Tattersall, and T.~Weber,
  ``{CheckMATE 2: From the model to the limit},''
  \href{http://dx.doi.org/10.1016/j.cpc.2017.08.021}{{\em Comput. Phys.
  Commun.} {\bfseries 221} (2017) 383--418},
  \href{http://arxiv.org/abs/1611.09856}{{\ttfamily arXiv:1611.09856
  [hep-ph]}}.

\bibitem{Sjostrand:2014zea}
T.~Sj\"ostrand, S.~Ask, J.~R. Christiansen, R.~Corke, N.~Desai, P.~Ilten,
  S.~Mrenna, S.~Prestel, C.~O. Rasmussen, and P.~Z. Skands, ``{An introduction
  to PYTHIA 8.2}'' \href{http://dx.doi.org/10.1016/j.cpc.2015.01.024}{{\em
  Comput. Phys. Commun.} {\bfseries 191} (2015) 159--177},
  \href{http://arxiv.org/abs/1410.3012}{{\ttfamily arXiv:1410.3012 [hep-ph]}}.

\bibitem{deFavereau:2013fsa}
{\bfseries DELPHES 3} Collaboration, J.~de~Favereau, C.~Delaere, P.~Demin,
  A.~Giammanco, V.~Lema\^\i{}tre, A.~Mertens, and M.~Selvaggi, ``{DELPHES 3, A
  modular framework for fast simulation of a generic collider experiment},''
  \href{http://dx.doi.org/10.1007/JHEP02(2014)057}{{\em JHEP} {\bfseries 02}
  (2014) 057}, \href{http://arxiv.org/abs/1307.6346}{{\ttfamily arXiv:1307.6346
  [hep-ex]}}.

\bibitem{Cacciari:2011ma}
M.~Cacciari, G.~P. Salam, and G.~Soyez, ``{FastJet User Manual},''
  \href{http://dx.doi.org/10.1140/epjc/s10052-012-1896-2}{{\em Eur. Phys. J. C}
  {\bfseries 72} (2012) 1896}, \href{http://arxiv.org/abs/1111.6097}{{\ttfamily
  arXiv:1111.6097 [hep-ph]}}.

\bibitem{Cacciari:2005hq}
M.~Cacciari and G.~P. Salam, ``{Dispelling the $N^{3}$ myth for the $k_t$
  jet-finder},'' \href{http://dx.doi.org/10.1016/j.physletb.2006.08.037}{{\em
  Phys. Lett. B} {\bfseries 641} (2006) 57--61},
  \href{http://arxiv.org/abs/hep-ph/0512210}{{\ttfamily arXiv:hep-ph/0512210}}.

\bibitem{Cacciari:2008gp}
M.~Cacciari, G.~P. Salam, and G.~Soyez, ``{The anti-$k_t$ jet clustering
  algorithm},'' \href{http://dx.doi.org/10.1088/1126-6708/2008/04/063}{{\em
  JHEP} {\bfseries 04} (2008) 063},
  \href{http://arxiv.org/abs/0802.1189}{{\ttfamily arXiv:0802.1189 [hep-ph]}}.

\bibitem{Read:2002hq}
A.~L. Read, ``{Presentation of search results: The CL(s) technique},''
  \href{http://dx.doi.org/10.1088/0954-3899/28/10/313}{{\em J. Phys. G}
  {\bfseries 28} (2002) 2693--2704}.

\bibitem{CMS:2017moi}
{\bfseries CMS} Collaboration, A.~M. Sirunyan {\em et~al.}, ``{Search for
  electroweak production of charginos and neutralinos in multilepton final
  states in proton-proton collisions at $\sqrt{s}=$ 13 TeV},''
  \href{http://dx.doi.org/10.1007/JHEP03(2018)166}{{\em JHEP} {\bfseries 03}
  (2018) 166}, \href{http://arxiv.org/abs/1709.05406}{{\ttfamily
  arXiv:1709.05406 [hep-ex]}}.

\bibitem{ATLAS:2018ceg}
{\bfseries ATLAS} Collaboration, M.~Aaboud {\em et~al.}, ``{Search for doubly
  charged scalar bosons decaying into same-sign $W$ boson pairs with the ATLAS
  detector},'' \href{http://dx.doi.org/10.1140/epjc/s10052-018-6500-y}{{\em
  Eur. Phys. J. C} {\bfseries 79} no.~1, (2019) 58},
  \href{http://arxiv.org/abs/1808.01899}{{\ttfamily arXiv:1808.01899
  [hep-ex]}}.

\bibitem{ATLAS:2021jol}
{\bfseries ATLAS} Collaboration, G.~Aad {\em et~al.}, ``{Search for doubly and
  singly charged Higgs bosons decaying into vector bosons in multi-lepton final
  states with the ATLAS detector using proton-proton collisions at $
  \sqrt{\mathrm{s}} $ = 13 TeV},''
  \href{http://dx.doi.org/10.1007/JHEP06(2021)146}{{\em JHEP} {\bfseries 06}
  (2021) 146}, \href{http://arxiv.org/abs/2101.11961}{{\ttfamily
  arXiv:2101.11961 [hep-ex]}}.

\bibitem{Kanemura:2014ipa}
S.~Kanemura, M.~Kikuchi, H.~Yokoya, and K.~Yagyu, ``{LHC Run-I constraint on
  the mass of doubly charged Higgs bosons in the same-sign diboson decay
  scenario},'' \href{http://dx.doi.org/10.1093/ptep/ptv071}{{\em PTEP}
  {\bfseries 2015} (2015) 051B02},
  \href{http://arxiv.org/abs/1412.7603}{{\ttfamily arXiv:1412.7603 [hep-ph]}}.

\bibitem{ATLAS:2021fbt}
{\bfseries ATLAS} Collaboration, G.~Aad {\em et~al.}, ``{Search for
  R-parity-violating supersymmetry in a final state containing leptons and many
  jets with the ATLAS experiment using $\sqrt{s} = 13 { TeV}$
  proton\textendash{}proton collision data},''
  \href{http://dx.doi.org/10.1140/epjc/s10052-021-09761-x}{{\em Eur. Phys. J.
  C} {\bfseries 81} no.~11, (2021) 1023},
  \href{http://arxiv.org/abs/2106.09609}{{\ttfamily arXiv:2106.09609
  [hep-ex]}}.

\bibitem{ATLAS:2018zdn}
{\bfseries ATLAS} Collaboration, M.~Aaboud {\em et~al.}, ``{A strategy for a
  general search for new phenomena using data-derived signal regions and its
  application within the ATLAS experiment},''
  \href{http://dx.doi.org/10.1140/epjc/s10052-019-6540-y}{{\em Eur. Phys. J. C}
  {\bfseries 79} no.~2, (2019) 120},
  \href{http://arxiv.org/abs/1807.07447}{{\ttfamily arXiv:1807.07447
  [hep-ex]}}.

\bibitem{ATLAS:2022cob}
{\bfseries ATLAS} Collaboration, ``{Search for heavy long-lived multi-charged
  particles in the full Run-II $pp$ collision data at $\sqrt{s}$ = 13 TeV using
  the ATLAS detector},''.

\bibitem{Chiu:2021sgs}
W.~H. Chiu, Z.~Liu, M.~Low, and L.-T. Wang, ``{Jet timing},''
  \href{http://dx.doi.org/10.1007/JHEP01(2022)014}{{\em JHEP} {\bfseries 01}
  (2022) 014}, \href{http://arxiv.org/abs/2109.01682}{{\ttfamily
  arXiv:2109.01682 [hep-ph]}}.

\bibitem{ATLAS:2022pib}
{\bfseries ATLAS} Collaboration, ``{Search for heavy, long-lived, charged
  particles with large ionisation energy loss in $pp$ collisions at $\sqrt{s} =
  13~\text{TeV}$ using the ATLAS experiment and the full Run 2 dataset},''
  \href{http://arxiv.org/abs/2205.06013}{{\ttfamily arXiv:2205.06013
  [hep-ex]}}.

\bibitem{Giudice:2022bpq}
G.~F. Giudice, M.~McCullough, and D.~Teresi, ``{dE/dx from boosted long-lived
  particles},'' \href{http://arxiv.org/abs/2205.04473}{{\ttfamily
  arXiv:2205.04473 [hep-ph]}}.

\bibitem{ATLAS:2022zhj}
{\bfseries ATLAS} Collaboration, G.~Aad {\em et~al.}, ``{Search for neutral
  long-lived particles in $pp$ collisions at $ \sqrt{s} $ = 13 TeV that decay
  into displaced hadronic jets in the ATLAS calorimeter},''
  \href{http://dx.doi.org/10.1007/JHEP06(2022)005}{{\em JHEP} {\bfseries 06}
  (2022) 005}, \href{http://arxiv.org/abs/2203.01009}{{\ttfamily
  arXiv:2203.01009 [hep-ex]}}.

\bibitem{LZ}
J.~Aalbers {\em et~al.}, ``{First Dark Matter Search Results from the
  LUX-ZEPLIN (LZ) Experiment},''.
  \url{https://lz.lbl.gov/wp-content/uploads/sites/6/2022/07/LZ_SR1_Paper_7July2022.pdf}.

\bibitem{Ruppin:2014bra}
F.~Ruppin, J.~Billard, E.~Figueroa-Feliciano, and L.~Strigari,
  ``{Complementarity of dark matter detectors in light of the neutrino
  background},'' \href{http://dx.doi.org/10.1103/PhysRevD.90.083510}{{\em Phys.
  Rev. D} {\bfseries 90} no.~8, (2014) 083510},
  \href{http://arxiv.org/abs/1408.3581}{{\ttfamily arXiv:1408.3581 [hep-ph]}}.

\bibitem{Kawasaki:2017bqm}
M.~Kawasaki, K.~Kohri, T.~Moroi, and Y.~Takaesu, ``{Revisiting Big-Bang
  Nucleosynthesis Constraints on Long-Lived Decaying Particles},''
  \href{http://dx.doi.org/10.1103/PhysRevD.97.023502}{{\em Phys. Rev. D}
  {\bfseries 97} no.~2, (2018) 023502},
  \href{http://arxiv.org/abs/1709.01211}{{\ttfamily arXiv:1709.01211
  [hep-ph]}}.

\bibitem{Acharya:2019uba}
S.~K. Acharya and R.~Khatri, ``{CMB anisotropy and BBN constraints on
  pre-recombination decay of dark matter to visible particles},''
  \href{http://dx.doi.org/10.1088/1475-7516/2019/12/046}{{\em JCAP} {\bfseries
  12} (2019) 046}, \href{http://arxiv.org/abs/1910.06272}{{\ttfamily
  arXiv:1910.06272 [astro-ph.CO]}}.

\bibitem{Acharya:2019owx}
S.~K. Acharya and R.~Khatri, ``{New CMB spectral distortion constraints on
  decaying dark matter with full evolution of electromagnetic cascades before
  recombination},'' \href{http://dx.doi.org/10.1103/PhysRevD.99.123510}{{\em
  Phys. Rev. D} {\bfseries 99} no.~12, (2019) 123510},
  \href{http://arxiv.org/abs/1903.04503}{{\ttfamily arXiv:1903.04503
  [astro-ph.CO]}}.

\bibitem{Blanco:2018esa}
C.~Blanco and D.~Hooper, ``{Constraints on Decaying Dark Matter from the
  Isotropic Gamma-Ray Background},''
  \href{http://dx.doi.org/10.1088/1475-7516/2019/03/019}{{\em JCAP} {\bfseries
  03} (2019) 019}, \href{http://arxiv.org/abs/1811.05988}{{\ttfamily
  arXiv:1811.05988 [astro-ph.HE]}}.

\bibitem{Arina:2021gfn}
C.~Arina, J.~Heisig, F.~Maltoni, D.~Massaro, and O.~Mattelaer, ``{Indirect
  dark-matter detection with MadDM v3.2 -- Lines and Loops},''
  \href{http://arxiv.org/abs/2107.04598}{{\ttfamily arXiv:2107.04598
  [hep-ph]}}.

\bibitem{Ando:2015qda}
S.~Ando and K.~Ishiwata, ``{Constraints on decaying dark matter from the
  extragalactic gamma-ray background},''
  \href{http://dx.doi.org/10.1088/1475-7516/2015/05/024}{{\em JCAP} {\bfseries
  05} (2015) 024}, \href{http://arxiv.org/abs/1502.02007}{{\ttfamily
  arXiv:1502.02007 [astro-ph.CO]}}.

\bibitem{Cohen:2016uyg}
T.~Cohen, K.~Murase, N.~L. Rodd, B.~R. Safdi, and Y.~Soreq,
  ``{\ensuremath{\gamma} -ray Constraints on Decaying Dark Matter and
  Implications for IceCube},''
  \href{http://dx.doi.org/10.1103/PhysRevLett.119.021102}{{\em Phys. Rev.
  Lett.} {\bfseries 119} no.~2, (2017) 021102},
  \href{http://arxiv.org/abs/1612.05638}{{\ttfamily arXiv:1612.05638
  [hep-ph]}}.

\bibitem{Liu:2016ngs}
W.~Liu, X.-J. Bi, S.-J. Lin, and P.-F. Yin, ``{Constraints on dark matter
  annihilation and decay from the isotropic gamma-ray background},''
  \href{http://dx.doi.org/10.1088/1674-1137/41/4/045104}{{\em Chin. Phys. C}
  {\bfseries 41} no.~4, (2017) 045104},
  \href{http://arxiv.org/abs/1602.01012}{{\ttfamily arXiv:1602.01012
  [astro-ph.CO]}}.

\bibitem{Poulin:2015opa}
V.~Poulin and P.~D. Serpico, ``{Nonuniversal BBN bounds on electromagnetically
  decaying particles},''
  \href{http://dx.doi.org/10.1103/PhysRevD.91.103007}{{\em Phys. Rev. D}
  {\bfseries 91} no.~10, (2015) 103007},
  \href{http://arxiv.org/abs/1503.04852}{{\ttfamily arXiv:1503.04852
  [astro-ph.CO]}}.

\bibitem{Slatyer:2016qyl}
T.~R. Slatyer and C.-L. Wu, ``{General Constraints on Dark Matter Decay from
  the Cosmic Microwave Background},''
  \href{http://dx.doi.org/10.1103/PhysRevD.95.023010}{{\em Phys. Rev. D}
  {\bfseries 95} no.~2, (2017) 023010},
  \href{http://arxiv.org/abs/1610.06933}{{\ttfamily arXiv:1610.06933
  [astro-ph.CO]}}.

\bibitem{Poulin:2016anj}
V.~Poulin, J.~Lesgourgues, and P.~D. Serpico, ``{Cosmological constraints on
  exotic injection of electromagnetic energy},''
  \href{http://dx.doi.org/10.1088/1475-7516/2017/03/043}{{\em JCAP} {\bfseries
  03} (2017) 043}, \href{http://arxiv.org/abs/1610.10051}{{\ttfamily
  arXiv:1610.10051 [astro-ph.CO]}}.

\bibitem{Chluba:2020oip}
J.~Chluba, A.~Ravenni, and S.~K. Acharya, ``{Thermalization of large energy
  release in the early Universe},''
  \href{http://dx.doi.org/10.1093/mnras/staa2131}{{\em Mon. Not. Roy. Astron.
  Soc.} {\bfseries 498} no.~1, (2020) 959--980},
  \href{http://arxiv.org/abs/2005.11325}{{\ttfamily arXiv:2005.11325
  [astro-ph.CO]}}.

\bibitem{Sirlin:1980nh}
A.~Sirlin, ``{Radiative Corrections in the SU(2)-L x U(1) Theory: A Simple
  Renormalization Framework},''
  \href{http://dx.doi.org/10.1103/PhysRevD.22.971}{{\em Phys. Rev. D}
  {\bfseries 22} (1980) 971--981}.

\bibitem{Marciano:1980pb}
W.~J. Marciano and A.~Sirlin, ``{Radiative Corrections to Neutrino Induced
  Neutral Current Phenomena in the SU(2)-L x U(1) Theory},''
  \href{http://dx.doi.org/10.1103/PhysRevD.22.2695}{{\em Phys. Rev. D}
  {\bfseries 22} (1980) 2695}. [Erratum: Phys.Rev.D 31, 213 (1985)].

\bibitem{Djouadi:1987gn}
A.~Djouadi and C.~Verzegnassi, ``{Virtual Very Heavy Top Effects in LEP / SLC
  Precision Measurements},''
  \href{http://dx.doi.org/10.1016/0370-2693(87)91206-8}{{\em Phys. Lett. B}
  {\bfseries 195} (1987) 265--271}.

\bibitem{Djouadi:1987di}
A.~Djouadi, ``{O(alpha alpha-s) Vacuum Polarization Functions of the Standard
  Model Gauge Bosons},'' \href{http://dx.doi.org/10.1007/BF02812964}{{\em Nuovo
  Cim. A} {\bfseries 100} (1988) 357}.

\bibitem{Kniehl:1989yc}
B.~A. Kniehl, ``{Two Loop Corrections to the Vacuum Polarizations in
  Perturbative QCD},''
  \href{http://dx.doi.org/10.1016/0550-3213(90)90552-O}{{\em Nucl. Phys. B}
  {\bfseries 347} (1990) 86--104}.

\bibitem{Halzen:1990je}
F.~Halzen and B.~A. Kniehl, ``{$\Delta$ r beyond one loop},''
  \href{http://dx.doi.org/10.1016/0550-3213(91)90319-S}{{\em Nucl. Phys. B}
  {\bfseries 353} (1991) 567--590}.

\bibitem{Kniehl:1991gu}
B.~A. Kniehl and A.~Sirlin, ``{Dispersion relations for vacuum polarization
  functions in electroweak physics},''
  \href{http://dx.doi.org/10.1016/0550-3213(92)90232-Z}{{\em Nucl. Phys. B}
  {\bfseries 371} (1992) 141--148}.

\bibitem{Kniehl:1992dx}
B.~A. Kniehl and A.~Sirlin, ``{On the effect of the $t \bar{t}$ threshold on
  electroweak parameters},''
  \href{http://dx.doi.org/10.1103/PhysRevD.47.883}{{\em Phys. Rev. D}
  {\bfseries 47} (1993) 883--893}.

\bibitem{Halzen:1991ik}
F.~Halzen, B.~A. Kniehl, and M.~L. Stong, ``{Two loop electroweak
  parameters},'' \href{http://dx.doi.org/10.1007/BF01554085}{{\em Z. Phys. C}
  {\bfseries 58} (1993) 119--132}.

\bibitem{Freitas:2000gg}
A.~Freitas, W.~Hollik, W.~Walter, and G.~Weiglein, ``{Complete fermionic two
  loop results for the M(W) - M(Z) interdependence},''
  \href{http://dx.doi.org/10.1016/S0370-2693(00)01263-6}{{\em Phys. Lett. B}
  {\bfseries 495} (2000) 338--346},
  \href{http://arxiv.org/abs/hep-ph/0007091}{{\ttfamily arXiv:hep-ph/0007091}}.
  [Erratum: Phys.Lett.B 570, 265 (2003)].

\bibitem{Freitas:2002ja}
A.~Freitas, W.~Hollik, W.~Walter, and G.~Weiglein, ``{Electroweak two loop
  corrections to the $M_W-M_Z$ mass correlation in the standard model},''
  \href{http://dx.doi.org/10.1016/S0550-3213(02)00243-2}{{\em Nucl. Phys. B}
  {\bfseries 632} (2002) 189--218},
  \href{http://arxiv.org/abs/hep-ph/0202131}{{\ttfamily arXiv:hep-ph/0202131}}.
  [Erratum: Nucl.Phys.B 666, 305--307 (2003)].

\bibitem{Awramik:2002wn}
M.~Awramik and M.~Czakon, ``{Complete two loop bosonic contributions to the
  muon lifetime in the standard model},''
  \href{http://dx.doi.org/10.1103/PhysRevLett.89.241801}{{\em Phys. Rev. Lett.}
  {\bfseries 89} (2002) 241801},
  \href{http://arxiv.org/abs/hep-ph/0208113}{{\ttfamily arXiv:hep-ph/0208113}}.

\bibitem{Awramik:2003ee}
M.~Awramik and M.~Czakon, ``{Complete two loop electroweak contributions to the
  muon lifetime in the standard model},''
  \href{http://dx.doi.org/10.1016/j.physletb.2003.06.007}{{\em Phys. Lett. B}
  {\bfseries 568} (2003) 48--54},
  \href{http://arxiv.org/abs/hep-ph/0305248}{{\ttfamily arXiv:hep-ph/0305248}}.

\bibitem{Onishchenko:2002ve}
A.~Onishchenko and O.~Veretin, ``{Two loop bosonic electroweak corrections to
  the muon lifetime and M(Z) - M(W) interdependence},''
  \href{http://dx.doi.org/10.1016/S0370-2693(02)03004-6}{{\em Phys. Lett. B}
  {\bfseries 551} (2003) 111--114},
  \href{http://arxiv.org/abs/hep-ph/0209010}{{\ttfamily arXiv:hep-ph/0209010}}.

\bibitem{Awramik:2002vu}
M.~Awramik, M.~Czakon, A.~Onishchenko, and O.~Veretin, ``{Bosonic corrections
  to Delta r at the two loop level},''
  \href{http://dx.doi.org/10.1103/PhysRevD.68.053004}{{\em Phys. Rev. D}
  {\bfseries 68} (2003) 053004},
  \href{http://arxiv.org/abs/hep-ph/0209084}{{\ttfamily arXiv:hep-ph/0209084}}.

\bibitem{Bauberger:1996ix}
S.~Bauberger and G.~Weiglein, ``{Calculation of two loop top quark and Higgs
  boson corrections in the electroweak standard model},''
  \href{http://dx.doi.org/10.1016/S0168-9002(97)00116-2}{{\em Nucl. Instrum.
  Meth. A} {\bfseries 389} (1997) 318--322},
  \href{http://arxiv.org/abs/hep-ph/9611445}{{\ttfamily arXiv:hep-ph/9611445}}.

\bibitem{Bauberger:1997ey}
S.~Bauberger and G.~Weiglein, ``{Higgs mass dependence of two loop corrections
  to $\delta$ r},'' \href{http://dx.doi.org/10.1016/S0370-2693(97)01458-5}{{\em
  Phys. Lett. B} {\bfseries 419} (1998) 333--339},
  \href{http://arxiv.org/abs/hep-ph/9707510}{{\ttfamily arXiv:hep-ph/9707510}}.

\bibitem{Avdeev:1994db}
L.~Avdeev, J.~Fleischer, S.~Mikhailov, and O.~Tarasov, ``{$0(\alpha
  \alpha_s^2)$ correction to the electroweak $\rho$ parameter},''
  \href{http://dx.doi.org/10.1016/0370-2693(94)90573-8}{{\em Phys. Lett. B}
  {\bfseries 336} (1994) 560--566},
  \href{http://arxiv.org/abs/hep-ph/9406363}{{\ttfamily arXiv:hep-ph/9406363}}.
  [Erratum: Phys.Lett.B 349, 597--598 (1995)].

\bibitem{Chetyrkin:1995ix}
K.~G. Chetyrkin, J.~H. Kuhn, and M.~Steinhauser, ``{Corrections of order ${\cal
  O}(G_F M_t^2 \alpha_s^2)$ to the $\rho$ parameter},''
  \href{http://dx.doi.org/10.1016/0370-2693(95)00380-4}{{\em Phys. Lett. B}
  {\bfseries 351} (1995) 331--338},
  \href{http://arxiv.org/abs/hep-ph/9502291}{{\ttfamily arXiv:hep-ph/9502291}}.

\bibitem{Chetyrkin:1995js}
K.~G. Chetyrkin, J.~H. Kuhn, and M.~Steinhauser, ``{QCD corrections from top
  quark to relations between electroweak parameters to order alpha-s**2},''
  \href{http://dx.doi.org/10.1103/PhysRevLett.75.3394}{{\em Phys. Rev. Lett.}
  {\bfseries 75} (1995) 3394--3397},
  \href{http://arxiv.org/abs/hep-ph/9504413}{{\ttfamily arXiv:hep-ph/9504413}}.

\bibitem{Chetyrkin:1996cf}
K.~G. Chetyrkin, J.~H. Kuhn, and M.~Steinhauser, ``{Three loop polarization
  function and O (alpha-s**2) corrections to the production of heavy quarks},''
  \href{http://dx.doi.org/10.1016/S0550-3213(96)00534-2}{{\em Nucl. Phys. B}
  {\bfseries 482} (1996) 213--240},
  \href{http://arxiv.org/abs/hep-ph/9606230}{{\ttfamily arXiv:hep-ph/9606230}}.

\bibitem{Faisst:2003px}
M.~Faisst, J.~H. Kuhn, T.~Seidensticker, and O.~Veretin, ``{Three loop top
  quark contributions to the rho parameter},''
  \href{http://dx.doi.org/10.1016/S0550-3213(03)00450-4}{{\em Nucl. Phys. B}
  {\bfseries 665} (2003) 649--662},
  \href{http://arxiv.org/abs/hep-ph/0302275}{{\ttfamily arXiv:hep-ph/0302275}}.

\bibitem{vanderBij:2000cg}
J.~J. van~der Bij, K.~G. Chetyrkin, M.~Faisst, G.~Jikia, and T.~Seidensticker,
  ``{Three loop leading top mass contributions to the rho parameter},''
  \href{http://dx.doi.org/10.1016/S0370-2693(01)00002-8}{{\em Phys. Lett. B}
  {\bfseries 498} (2001) 156--162},
  \href{http://arxiv.org/abs/hep-ph/0011373}{{\ttfamily arXiv:hep-ph/0011373}}.

\bibitem{Boughezal:2004ef}
R.~Boughezal, J.~B. Tausk, and J.~J. van~der Bij, ``{Three-loop electroweak
  correction to the Rho parameter in the large Higgs mass limit},''
  \href{http://dx.doi.org/10.1016/j.nuclphysb.2005.02.020}{{\em Nucl. Phys. B}
  {\bfseries 713} (2005) 278--290},
  \href{http://arxiv.org/abs/hep-ph/0410216}{{\ttfamily arXiv:hep-ph/0410216}}.

\bibitem{Schroder:2005db}
Y.~Schroder and M.~Steinhauser, ``{Four-loop singlet contribution to the rho
  parameter},'' \href{http://dx.doi.org/10.1016/j.physletb.2005.06.085}{{\em
  Phys. Lett. B} {\bfseries 622} (2005) 124--130},
  \href{http://arxiv.org/abs/hep-ph/0504055}{{\ttfamily arXiv:hep-ph/0504055}}.

\bibitem{Chetyrkin:2006bj}
K.~G. Chetyrkin, M.~Faisst, J.~H. Kuhn, P.~Maierhofer, and C.~Sturm,
  ``{Four-Loop QCD Corrections to the Rho Parameter},''
  \href{http://dx.doi.org/10.1103/PhysRevLett.97.102003}{{\em Phys. Rev. Lett.}
  {\bfseries 97} (2006) 102003},
  \href{http://arxiv.org/abs/hep-ph/0605201}{{\ttfamily arXiv:hep-ph/0605201}}.

\bibitem{Boughezal:2006xk}
R.~Boughezal and M.~Czakon, ``{Single scale tadpoles and O(G(F m(t)**2
  alpha(s)**3)) corrections to the rho parameter},''
  \href{http://dx.doi.org/10.1016/j.nuclphysb.2006.08.007}{{\em Nucl. Phys. B}
  {\bfseries 755} (2006) 221--238},
  \href{http://arxiv.org/abs/hep-ph/0606232}{{\ttfamily arXiv:hep-ph/0606232}}.

\bibitem{Djouadi:1993ss}
A.~Djouadi and P.~Gambino, ``{Electroweak gauge bosons selfenergies: Complete
  QCD corrections},'' \href{http://dx.doi.org/10.1103/PhysRevD.49.3499}{{\em
  Phys. Rev. D} {\bfseries 49} (1994) 3499--3511},
  \href{http://arxiv.org/abs/hep-ph/9309298}{{\ttfamily arXiv:hep-ph/9309298}}.
  [Erratum: Phys.Rev.D 53, 4111 (1996)].

\bibitem{Arhrib:2011uy}
A.~Arhrib, R.~Benbrik, M.~Chabab, G.~Moultaka, M.~C. Peyranere, L.~Rahili, and
  J.~Ramadan, ``{The Higgs Potential in the Type II Seesaw Model},''
  \href{http://dx.doi.org/10.1103/PhysRevD.84.095005}{{\em Phys. Rev. D}
  {\bfseries 84} (2011) 095005},
  \href{http://arxiv.org/abs/1105.1925}{{\ttfamily arXiv:1105.1925 [hep-ph]}}.

\bibitem{Lee:1985uv}
K.-M. Lee and E.~J. Weinberg, ``{TUNNELING WITHOUT BARRIERS},''
  \href{http://dx.doi.org/10.1016/0550-3213(86)90150-1}{{\em Nucl. Phys. B}
  {\bfseries 267} (1986) 181--202}.

\bibitem{Hollik:2018wrr}
W.~G. Hollik, G.~Weiglein, and J.~Wittbrodt, ``{Impact of Vacuum Stability
  Constraints on the Phenomenology of Supersymmetric Models},''
  \href{http://dx.doi.org/10.1007/JHEP03(2019)109}{{\em JHEP} {\bfseries 03}
  (2019) 109}, \href{http://arxiv.org/abs/1812.04644}{{\ttfamily
  arXiv:1812.04644 [hep-ph]}}.

\bibitem{Wittbrodt:2019bsu}
J.~Wittbrodt, \href{http://dx.doi.org/10.3204/PUBDB-2019-03809}{{\em {Exploring
  Models of Electroweak Symmetry Breaking at the LHC and Beyond}}}.
\newblock PhD thesis, Hamburg U., Hamburg, 2019.

\end{thebibliography}\endgroup

\end{document}